%% file: main.tex
\documentclass[a4paper]{article}

\usepackage{graphicx}
\usepackage{multirow}
\usepackage{amsmath,amssymb,amsfonts}
\usepackage{amsthm}
\usepackage{mathrsfs}
\usepackage[title]{appendix}
\usepackage{xcolor}
\usepackage{textcomp}
\usepackage{manyfoot}
\usepackage{booktabs}
\usepackage{algorithm}
\usepackage{algorithmicx}
\usepackage{algpseudocode}
\usepackage{listings}
\usepackage{comment, soul}
\usepackage{boxedminipage}
\usepackage[T1]{fontenc}
\usepackage{lmodern}
\usepackage{subcaption}
\usepackage{enumitem}
\setlist{itemsep=1pt, parsep=0pt, topsep=3pt}

\makeatletter
\renewcommand\subparagraph{\@startsection{subparagraph}{5}{\parindent}%
  {1.2ex \@plus .3ex \@minus .2ex}{-1em}{\normalfont\normalsize\bfseries}}
\makeatother

\usepackage{fancyhdr}
\usepackage[affil-it]{authblk}
\usepackage[numbers]{natbib}
\usepackage[colorlinks=true]{hyperref}
\usepackage[a4paper]{geometry}

\definecolor{injected}{RGB}{20,90,170}

\lstdefinestyle{promptbox}{
    basicstyle=\ttfamily\fontsize{6.5}{7.5}\selectfont,
    breaklines=true,
    breakindent=0pt,
    columns=fullflexible,
    keepspaces=true,
    lineskip=-1pt,
    aboveskip=2pt,
    belowskip=2pt,
    literate={—}{{---}}1 {–}{{--}}1 {→}{{$\rightarrow$}}1 {•}{{\textbullet}}1
             {"}{{``}}1 {"}{{''}}1 {'}{{'}}1 {…}{{...}}1 {─}{{-}}1,
    frame=none,
    xleftmargin=0pt,
    moredelim=[is][\color{injected}]{@}{@},
}

\newcommand{\tabref}[1]{Table~\ref{#1}}
\newcommand{\figref}[1]{Fig.~\ref{#1}}
\newcommand{\figsref}[1]{Figs.~\ref{#1}}
\newcommand{\secref}[1]{\nameref{#1}}
\newcommand{\eqnref}[1]{Eqn.~\ref{#1}}

\newcommand{\ie}{\textit{i.e.},~}

\date{}

\begin{document}

\title{Empirical Evidence of Large Language Model's Influence on Human Spoken Communication}

\author[1]{Hiromu Yakura$^{\ast\dagger}$}
\author[2,3]{Ezequiel Lopez-Lopez$^{\ast\dagger}$}
\author[1]{Levin Brinkmann$^{\ast\dagger}$}
\author[1]{Ignacio de la Serna}
\author[1]{Lara Kirfel}
\author[1]{Prateek Gupta}
\author[1]{Ivan Soraperra}
\author[1]{Thomas F. Eisenmann}
\author[2,4]{Dirk U. Wulff}
\author[1]{Iyad Rahwan$^{\ast}$}

\affil[1]{Center for Humans and Machines, Max-Planck Institute for Human Development, Berlin, Germany}
\affil[2]{Center for Adaptive Rationality, Max-Planck Institute for Human Development, Berlin, Germany}
\affil[3]{Center Synergy of Systems, TUD Dresden University of Technology, Dresden, Germany}
\affil[4]{Department of Business Analytics and Decision Science, Vienna University of Economics and Business, Vienna, Austria}

\maketitle
{\let\thefootnote\relax\footnotetext{$^{\ast}$Corresponding authors: \texttt{\{yakura, lopez, brinkmann, rahwan\}@mpib-berlin.mpg.de}.\\\hspace*{1.8em}$^{\dagger}$These authors contributed equally to this work.}}

\begin{abstract}
\input{text/abstract}
\end{abstract}

\input{text/intro}
\input{text/results}
\input{text/discussion}
\input{text/methods}

\input{text/acknowledgments}

\begin{appendices}
\input{text/appendix}

\end{appendices}

\bibliographystyle{unsrtnat}
\bibliography{reference}

\newpage


\renewcommand{\thefigure}{S\arabic{figure}}
\renewcommand{\thetable}{S\arabic{table}}
\renewcommand{\theequation}{S\arabic{equation}}
\renewcommand{\thepage}{S\arabic{page}}
\setcounter{figure}{0}
\setcounter{table}{0}
\setcounter{equation}{0}
\setcounter{page}{1}

\begin{center}
\section*{Supplementary Materials}

Hiromu Yakura$^{1\ast\dagger}$,
Ezequiel Lopez-Lopez$^{2,3\dagger}$,
Levin Brinkmann$^{1\dagger}$,
Ignacio de la Serna$^{1}$,
Lara Kirfel$^{1}$,
Prateek Gupta$^{1}$,
Ivan Soraperra$^{1}$,
Thomas F. Eisenmann$^{1}$,
Dirk U. Wulff$^{2,4}$, \and
Iyad Rahwan$^{1}$ \\
\small$^{1}$Center for Humans and Machines, Max-Planck Institute for Human Development, Germany \\
\small$^{2}$Center for Adaptive Rationality, Max-Planck Institute for Human Development, Germany \\
\small$^{3}$Center Synergy of Systems, TUD Dresden University of Technology, Germany \\
\small$^{4}$Department of Business Analytics and Decision Science, Vienna University of Economics and Business, Austria \\
\small$^\ast$Corresponding author(s): {yakura,lopez,brinkmann,rahwan}@mpib-berlin.mpg.de \\
\small$^\dagger$These authors contributed equally to this work.
\end{center}

\subsubsection*{This PDF file includes:}
Materials and Methods\\
Figures \ref{fig:delve-robustness-composite} to \ref{fig:abm}\\
Tables \ref{tab:model-versions} to \ref{tab:synonym-pairs}

\newpage

\subsection*{Materials and Methods}
\input{text/si_methods}
\newpage

\subsection*{Supplementary Figures}
\input{text/si_figures}

\newpage
\subsection*{Supplementary Tables}
\input{text/si_tables}

\end{document}

%% file: text/abstract.tex
From the printing press to social media, innovations in communication technology have repeatedly reshaped how ideas spread through human culture. Chatbots powered by generative artificial intelligence constitute a new medium, encoding cultural patterns in their neural representations and disseminating them in conversations with hundreds of millions of people. Whether these patterns transmit into human language, and ultimately shape human culture, is a fundamental question. While fully quantifying the causal impact of a chatbot like ChatGPT on human culture is challenging, lexical shifts in human spoken communication may offer an early indicator. Here we show that words preferentially generated by ChatGPT, such as \textit{delve}, \textit{showcase}, \textit{boast}, \textit{intricacies} and \textit{meticulous}, increased abruptly in spontaneous human speech. A synthetic-control analysis~\citep{abadie2010} of 737,083 hours of conversation from 824,634 podcast episodes, screened for unscripted speech, causally links this shift to ChatGPT's release. The measurable influence on spontaneous speech suggests that humans internalize the lexical choices of large language models (LLMs). A preregistered experiment (N = 496) confirms they do, as a brief chatbot interaction led participants to adopt its words as their own, persisting past a distractor task and confirmed in forced lexical choice, indicating entrenchment in the active vocabulary. Together these results show that machines trained on human data now feed their own traits back into human language, integrating LLMs into the ongoing processes of cultural evolution \citep{brinkmann2023}. This coupling raises concerns about linguistic homogenization~\citep{sourati2026} and the capacity of a few major AI providers for latent cultural influence at scale.

%% file: text/intro.tex
Communication technologies have long altered how knowledge and practices arise, spread, and persist---the process of cultural evolution~\citep{mcluhan1964, henrich2016}.  Writing, the printing press, broadcast media, and the Internet each reshaped human culture in distinctive ways~\citep{goody1963, putnam2000, dittmar2011, vosoughi2018}.  Generative AI, particularly Large Language Models (LLMs) such as ChatGPT, now emerges as the next such technology~\citep{acerbi2019}: a medium that reaches a global audience in an integrated voice with distinctive linguistic characteristics~\citep{liang2024, kobak2025, siler2026}.  As with previous communication media, its influence may extend beyond its own outputs to the language humans themselves produce, including patterns that are internalized and reproduced spontaneously~\citep{bock1986, levelt1999, ellis2002}.  This prompts a fundamental question: are the outputs of generative AI internalized, signifying cultural transmission?

LLMs are a structurally novel medium for cultural transmission. Unlike books and newspapers, their outputs are largely non-verbatim---each interaction reconstructs rather than copies---resembling oral cultural transmission more than the faithful copying of print~\citep{acerbi2015}, and users engage with these reconstructions much as they would with knowledgeable interlocutors~\citep{colombatto2024}, a setting that might activate established social-learning strategies~\citep{henrich2001, kendal2018}.  Yet unlike peer-to-peer cultural transmission, LLMs collapse multiple voices into one, concentrated in a small number of providers~\citep{bender2021, sourati2026} and delivered through parallel one-to-one conversations.  While such cultural transmission resists direct measurement, its reproduction in spontaneous spoken language offers a distinctive empirical signal.

Within two months of release, ChatGPT had accumulated more than 100 million users~\citep{hu2023}, with adoption proliferating across the English-speaking world and beyond.  ChatGPT's pronounced preference for words such as \emph{delve} offers a unique opportunity to observe and quantify its cultural influence in real time, within a quasi-experimental setting~\citep{kobak2025,juzek2025}.  The observation was made possible by a rare window of measurement: for roughly 18 months, millions of users interacted with a single dominant language model, GPT-3.5, whose unusually distinctive lexical preferences (\figref{fig:method-and-delve}D) created a measurable signature that could be tracked as it spread through human communication.

Each LLM interaction exposes users to a unique response, but these responses are drawn from the same underlying distribution, providing repeated opportunities for their patterns to be transmitted.  Such repeated exposure is known to cause durable changes in language production through entrenchment, a process that operates automatically and below the level of explicit attribution~\citep{ellis2002,diessel2019}. If LLM exposure affects human language production at this level, it would situate LLMs inside the cultural-transmission network alongside humans~\citep{brinkmann2023}, where the distinction of AI-generated, AI-influenced, and unshaped human language is increasingly blurred~\citep{veselovsky2025,liang2024,rilla2025}.

In this study, we provide complementary empirical evidence of LLM-mediated linguistic influence. At the population level, we use the November 2022 ChatGPT release as a natural experiment, documenting a measurable shift in the spontaneous use of ChatGPT-favored words in over a million hours of human spoken communication. At the individual level, a preregistered controlled experiment isolates entrenchment as a proximate mechanism, showing that short interactions with an AI chatbot induce persistent lexical shifts that survive a distractor task and operate below the threshold of explicit attribution. Together, these results indicate that LLMs have a measurable influence on human language and point to the integration of LLMs into ongoing processes of cultural evolution~\citep{brinkmann2023, acerbi2023}.

\section*{Comparing Word Preferences of Humans and LLMs}
\label{sec:intro-method}

ChatGPT is trained on broad public corpora~\citep{openai2023} and fine-tuned through opaque proprietary processes~\citep{liesenfeld2023}, producing an emergent linguistic profile shaped by statistical learning, reinforcement, and alignment objectives~\citep{brown2020, ouyang2022}. While rooted in human language, this profile exhibits distinctive characteristics that set it apart from organic human communication~\citep{munoz-ortiz2024, reinhart2025}, with a persistent preference for normative, socially desirable patterns~\citep{herbold2023, park2024}.

\begin{figure}[H]
    \centering
    \includegraphics[width=\textwidth]{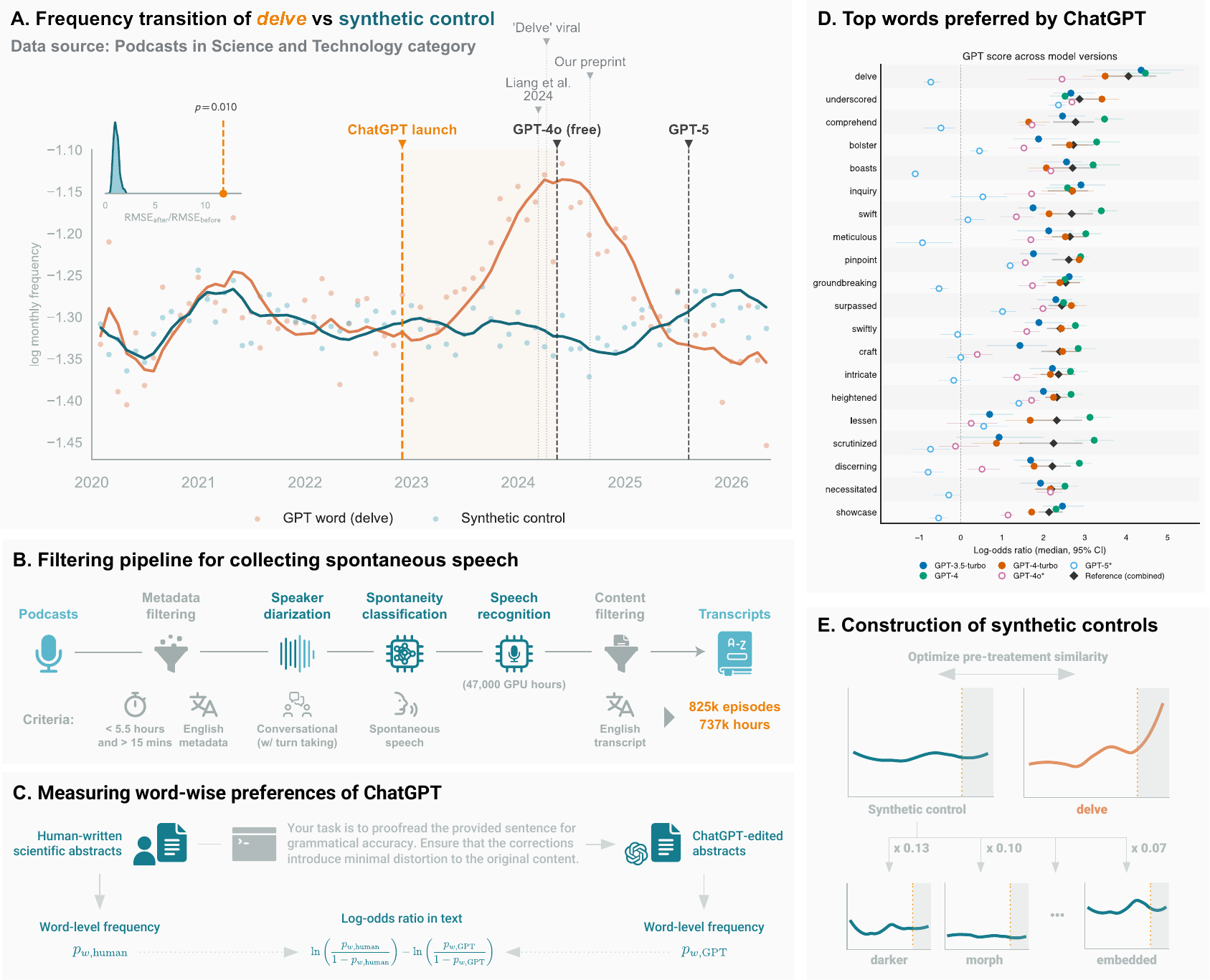}
    \caption{\textbf{ChatGPT's word preferences and their measurable influence on spoken communication.}
    (A) Monthly relative frequency of \emph{delve} (the fraction of podcast episodes containing the word per month) in the Science \& Technology category (orange) and its synthetic control (teal), with LOWESS trend curve overlaid on the monthly observations. The shaded region marks the post-release window; the inset shows the placebo distribution of post-/pre-treatment RMSE ratios across the donor pool, with the observed ratio for \emph{delve} marked ($p = 0.01$). Vertical dashed lines mark major LLM-related events.
    (B) Filtering pipeline for the podcast corpus, retaining only episodes that pass duration and language checks, contain conversational turn-taking, and are classified as spontaneous speech by a custom classification model (see \secref{sec:si-methods-spontaneity} for details).
    (C) Systematic method for measuring ChatGPT's word-level preferences: human-written scientific abstracts and their ChatGPT-edited counterparts are compared via word-level log-odds ratios, yielding the \emph{GPT score}.
    (D) GPT scores for the top 20 GPT-preferred words across model versions; the combined reference score (black diamonds) is computed from GPT-3.5-turbo, GPT-4, and GPT-4-turbo, the three models available at the time the score set was defined. GPT-4o and GPT-5 (star-marked) are shown for comparison and are not included in the reference. \emph{delve} sits at or near the top, with markedly attenuated preference with GPT-5. 
    (E) Construction of a synthetic control: a convex combination of donor words optimized to match the treated word's pre-treatment trajectory, illustrated for \emph{delve}.}
    \label{fig:method-and-delve}
\end{figure}

At the lexical level, where word choice is itself a key aspect of cultural behavior, ChatGPT exhibited distinctive lexical characteristics that reflect its training and optimization~\citep{feng2023, hofmann2024}. A striking example is \emph{delve}, which was favored over alternatives such as \emph{explore} or \emph{examine}~\citep{kobak2025, juzek2025}. To quantify these characteristics, we computed a per-word \emph{GPT score} from word-level log-odds ratios between human-written texts and their GPT-edited counterparts, aggregated across datasets, models, and rephrasing prompts~\citep{liang2024} (\figref{fig:method-and-delve}C). \emph{Delve} sits at the top of the resulting distribution across early GPT models, alongside a broader cluster of GPT-preferred words including \emph{showcase}, \emph{intricacies}, and others (\figref{fig:method-and-delve}D).

Such preferences have already left a measurable mark on written language, where \emph{delve} rose across scientific abstracts and peer reviews soon after ChatGPT's release~\citep{liang2024}. But this mark may never be internalized. In writing, a ChatGPT-preferred word can be introduced into a text even if it never enters the writer's own vocabulary. While in written text, LLMs may supply lexical choices directly through full production or copy-pasting, spontaneous speech excludes such shortcuts. Words surfacing in unscripted talk would signal internalization by the speaker, signifying machine-to-human cultural transmission~\citep{brinkmann2023, brinkmann2025}.

To estimate the population-level causal impact of ChatGPT's release on spoken communication, we treat the launch as a natural experiment and apply the synthetic control method~\citep{abadie2010, abadie2021using}: for each treated word, we compute its monthly relative document frequency (the fraction of podcast episodes containing the word per month) and build a counterfactual from a convex combination of untreated donor words whose pre-release frequency trajectory best matches that of the treated word (\figref{fig:method-and-delve}E). This per-word, time-series design sits in the text-as-outcome branch of the causal-inference-for-text literature~\citep{feder2022causal, keith2020text}, distinct from document-level methods that estimate the causal effect of latent linguistic properties under no-unobserved-confounding assumptions on text representations~\citep{pryzant2021causal, agrawal2024causalcite}; matching on usage rather than meaning is necessary for identification, since semantically similar words are most likely to share the treatment and would contaminate the counterfactual (see \secref{sec:methods-synthetic}).

In a first approximation on 360{,}445 YouTube academic talks, the same lexical signature was already detectable, as anticipated from its written footprint in scientific communication~\citep{liang2024} and consistent with concurrent work on academic conference talks~\citep{geng2024speaking} (see \secref{sec:appendix-youtube} for details). Yet, academic talks are often scripted or delivered from prepared notes. This motivated a complementary corpus capturing spontaneous speech across diverse domains beyond the academic domain. 

Podcasts are well-suited to this question because they frequently contain spontaneous conversation, and individual word choices are largely selected in the moment, even when the broader format is prepared. We therefore assembled 1{,}407{,}131 podcast episodes spanning science, technology, education, business, sports, and a topic-agnostic snapshot (see \secref{sec:methods-datasets}). To restrict the analysis to episodes that genuinely carry spontaneous communication, we trained a classifier to detect spontaneity from audio (\figref{fig:method-and-delve}B; $\approx 90\%$ accuracy on held-out annotations; see also \secref{sec:si-methods-spontaneity}), retaining only the conversational and unscripted subset on which the analyses that follow are based.

%% file: text/results.tex
\section*{Relationship between ChatGPT's word preferences and human adoption}
\label{sec:results}

\begin{figure}[H]
    \centering
    \includegraphics[width=.5\linewidth]{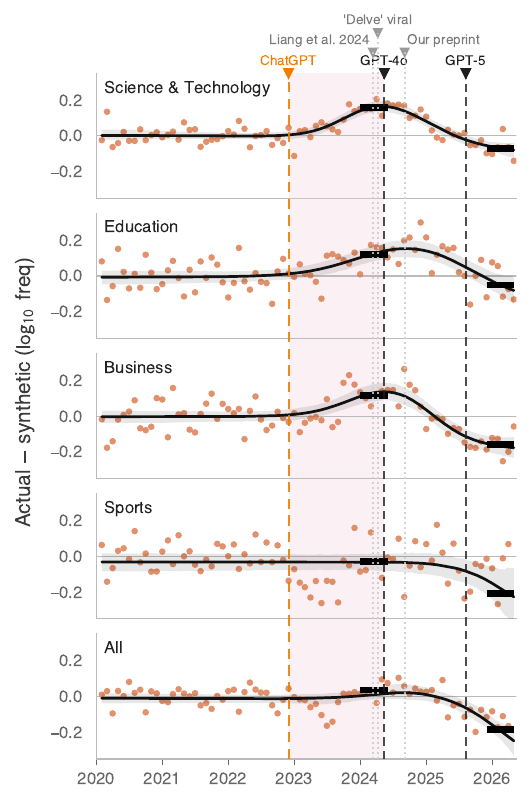}
    \caption{\textbf{\emph{Delve} rises after the ChatGPT release and then reverts in multiple podcast categories.} Monthly difference between the observed and synthetic-control relative frequency of \emph{delve} (the fraction of podcast episodes containing the word per month; actual $-$ synthetic; positive values indicate use above the pre-ChatGPT baseline) for four podcast categories and a category-independent sample (\emph{All}). Points are monthly observations; the black curve is a double-logistic \emph{smoother}---the S-curve form expected for lexical change~\citep{blythe2012}---and the grey band its pointwise posterior. The dashed line marks the ChatGPT release (30 November 2022) and the shaded strip the GPT-3.5 era (release to the free release of GPT-4o); the two black markers are the mean gap over the post-adoption window (months 13--18 after release) and over the last six recorded months. Vertical markers denote a selection of external events. Per-category point estimates, placebo-based 95\% confidence intervals, and $p$-values are in Supplementary Table~\ref{tab:fig2-b2-ci}.}
    \label{fig:delve-delta-sigmoid}
\end{figure}

For \emph{delve}---the word most consistently overused across models and contexts---usage in Science \& Technology podcasts (restricted to conversational, spontaneous speech; see \secref{sec:methods}) rose following the release of ChatGPT (\figref{fig:method-and-delve}A) to $\sim$44\% above its synthetic-control counterfactual over months 13--18 post-release (95\% CI $[+22\%, +63\%]$, in-space placebo $p = 0.010$; see \secref{sec:methods-synthetic} in Methods); the rejection holds under alternative donor-selection strategies (Supplementary \figref{fig:delve-robustness-composite}). The result is not driven by a small number of high-usage channels: removing the top-usage channels leaves the post-release rise essentially unchanged (Supplementary \figref{fig:channel-outliers}), and the pattern mirrors the uptake we observe in YouTube academic talks (Appendix~\ref{sec:appendix-youtube}).

Consistent with the pattern in Science \& Technology, Education and Business podcasts showed mean elevations of $32\%$ ($p = 0.06$) and $31\%$ ($p = 0.04$) above the synthetic-control counterfactual over the same window (\figref{fig:delve-delta-sigmoid}), with a weak upward tendency in the category-independent sample ($+9\%$) and a slight decline in Sports ($-7\%$). This elevated period falls within the window in which a single model (GPT-3.5-turbo) dominated the consumer market (shaded strip).

The elevation did not persist. The gap peaked around mid-2024 and the usage of \emph{delve} declined in the following months, dropping below baseline in Science \& Technology and Education ($-15\%$ and $-11\%$) and overshooting further below it in Business ($-30\%$) and the category-independent sample ($-35\%$; 95\% CI $[-57\%, -7\%]$, placebo $p = 0.05$, two-sided). The turn coincides with the free release of GPT-4o and with growing public awareness of \emph{delve} as a signature of LLM language.

Extending the analysis beyond \emph{delve} and raw frequency difference, we investigated 3{,}535 lexical stems---a stem being the root of a word that reduces to once inflectional endings are stripped (i.e.\ unifying \emph{delve}, \emph{delves}, and \emph{delving})---that were present in our GPT-score dataset, part of the top 50k word2vec words, and appearing in at least 20 episodes per month on average over the pre-treatment window. For simplicity, we will use stem and word interchangeably in the following. For each treated word $w$ we tested whether the actual--synthetic gap $\Delta y_{w,t} = \log_{10} y_{w,t}^{\text{obs}} - \log_{10} y_{w,t}^{\text{synth}}$ departs from its pre-treatment trajectory at the ChatGPT release. We use the post-event slope coefficient $\beta_{\text{Post}}$ of a Bayesian change-point regression (Methods, \eqnref{eqn:linear-model}) on $\Delta y_{w,t}$ as our per-word estimate of the ChatGPT-induced change in usage; positive values indicate post-release acceleration of the treated word's frequency relative to its synthetic control.

\begin{figure}[H]
    \centering
    \includegraphics[width=\linewidth]{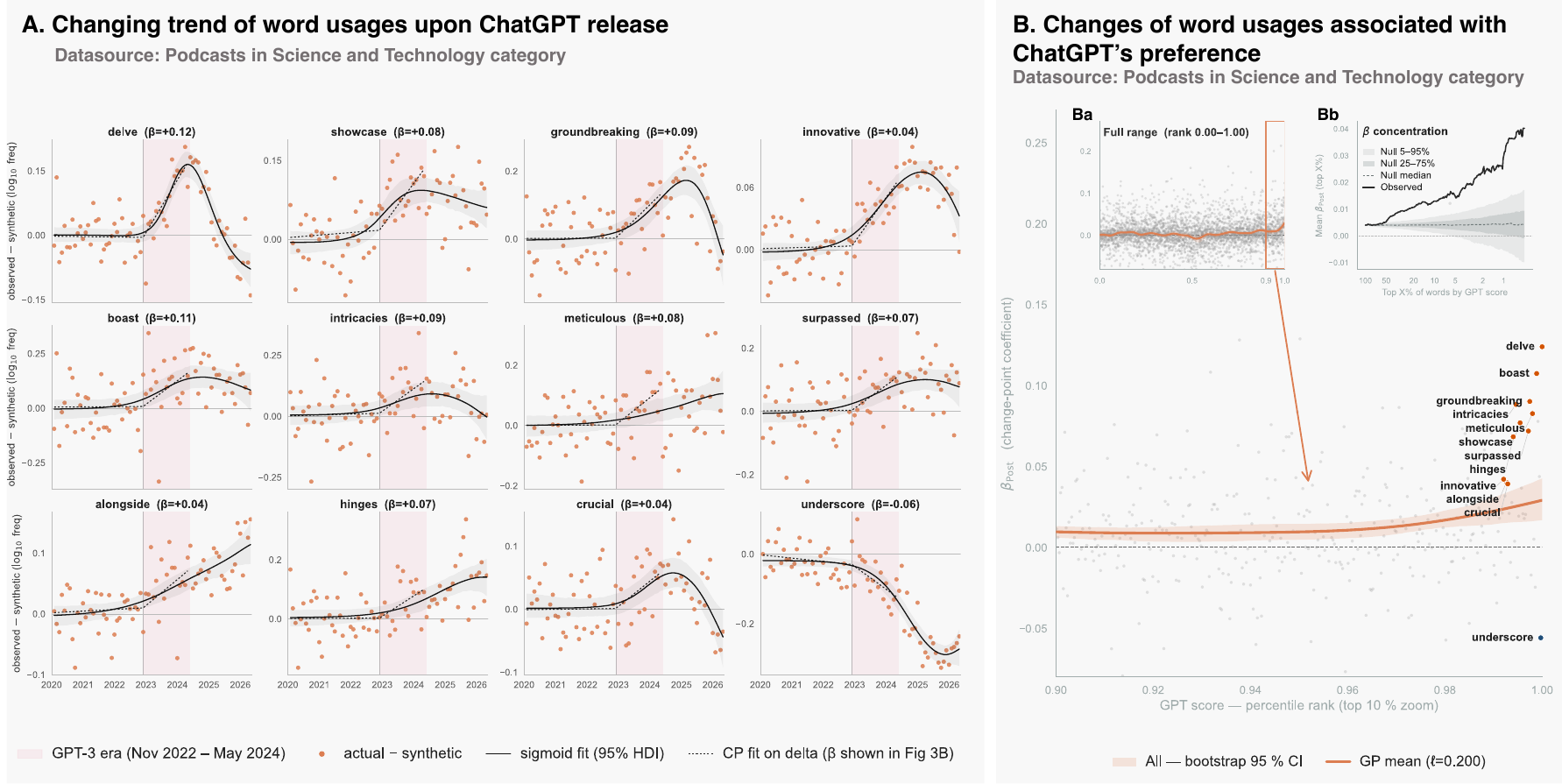}
    \caption{\textbf{Word-usage shifts after the ChatGPT release track ChatGPT's lexical preference (Science \& Technology podcasts).} \textbf{(A) Changing trend of word usages upon ChatGPT release.} Of the 36 words in the top 1\% of GPT scores, 15 showed a credible change in observed $-$ synthetic $\log_{10}$ frequency (orange points) at ChatGPT's launch. Shown here are the twelve with the largest-magnitude conservative $\beta_{\text{Post}}$ bound (the 95\% HDI limit nearest zero). Solid curves are double-sigmoid posterior smoothers~\citep{blythe2012} (95\% HDI shaded); dashed lines are the change-point fit of \eqnref{eqn:linear-model}, whose post-release slope $\beta_{\text{Post}}$ (panel titles) is the quantity plotted across all words in panel B. The shaded vertical band marks the period of analysis between ChatGPT's launch and the launch of GPT-4o (free). Eleven of the twelve rise after release; \emph{underscore} is the lone decline. \textbf{(B) Changes of word usages associated with ChatGPT's preference.} Per-word change-point slope $\beta_{\text{Post}}$ (\eqnref{eqn:linear-model}) plotted against GPT-score percentile rank ($n = 3{,}535$ words). The main panel zooms on the top 10\% of GPT scores; gray points represent individual words and the orange line shows the Gaussian-process posterior mean over all words (Mat\'ern $\nu = 2.5$, length scale $\ell = 0.20$ on the unit-interval rank axis; bootstrap 95\% CI shaded), which rises towards the top of the GPT-score range. The twelve words shown in A are labeled. Inset Ba shows the same relationship across the full GPT-score range (rank $0$--$1$). Inset Bb displays the slice-mean $\beta_{\text{Post}}$ over top-$X\%$ slices of the GPT-score distribution (black line) against the permutation null (grey band, 5th--95th percentile; shuffling GPT scores across words). The slice-mean rises faster than its permutation null, climbing from $+0.004$ over the whole vocabulary to $+0.030$ in the top 1\%, which corresponds to more than $2\times$ the permutation-null upper 95\% bound.}
    \label{fig:change-point-model}
\end{figure}

We found higher GPT-scores to be associated with an increase in usage in spoken communication (\figref{fig:change-point-model}B). For the top 1\% of GPT-score words ($n = 36$) the mean post-release slope $\beta_{\text{Post}}$ reaches $+0.030$; 28 of 36 show an increase in usage, 13 of them credibly so (95\% HDI excludes zero), against 8 decreases of which 2 are credibly so. \figref{fig:change-point-model}A shows the 12 words in the top 1\% with the largest credible change (see SI for a full overview, Supplementary \figref{fig:gpt-top1pct-panels}). While \emph{delve} and others revert towards baseline in recent months, words such as \emph{boast}, \emph{meticulous}, and \emph{showcase} maintain a sustained uptake.

The effect is strongest at the highest-scoring words. The mean post-release slope $\beta_{\text{Post}}$ is $+0.030$ in the top 1\% of GPT scores, $+0.025$ in the top 2\%, $+0.015$ in the top 5\%, and $+0.013$ in the top decile---corresponding to a median peak usage $\sim$50\% above the synthetic-control baseline at the top 1\%. The top-$k$ mean rises faster than its permutation null as $k$ narrows onto the highest-GPT-score words (\figref{fig:change-point-model}C). This argues against a vocabulary-wide drift and confirms an association between higher GPT scores and increased usage in spoken communication. The same concentration is visible under alternative donor-selection strategies and on un-audited counts (Supplementary \figref{fig:fig3-robustness-composite}), and does not depend on \emph{delve} alone ($\beta_{\text{Post}}$ in the top 1\% excluding \emph{delve}: $+0.027$; Supplementary \figref{fig:score-effect-delve-out}).

To further check whether the acceleration is anchored in time to ChatGPT's release, we considered every month in the data window as a candidate change point and refit the cross-word regression at each. For 28 candidate change points whose 18-month post-window pre-dates ChatGPT, the mean top-1\% slope $\beta_{\text{Post}}$ was flat ($+0.0004 \pm 0.003$); it rose to $+0.033$ only once the window reached the actual launch and peaked at $+0.042$ in September 2023 (permutation $p = 0.034$; Supplementary \figref{fig:changepoint-date-sweep}).

\section*{A brief chatbot interaction entrenches lexical choices}
\label{sec:chatbot-interaction}

\begin{figure*}[t!]
    \centering
    \includegraphics[width=\linewidth]{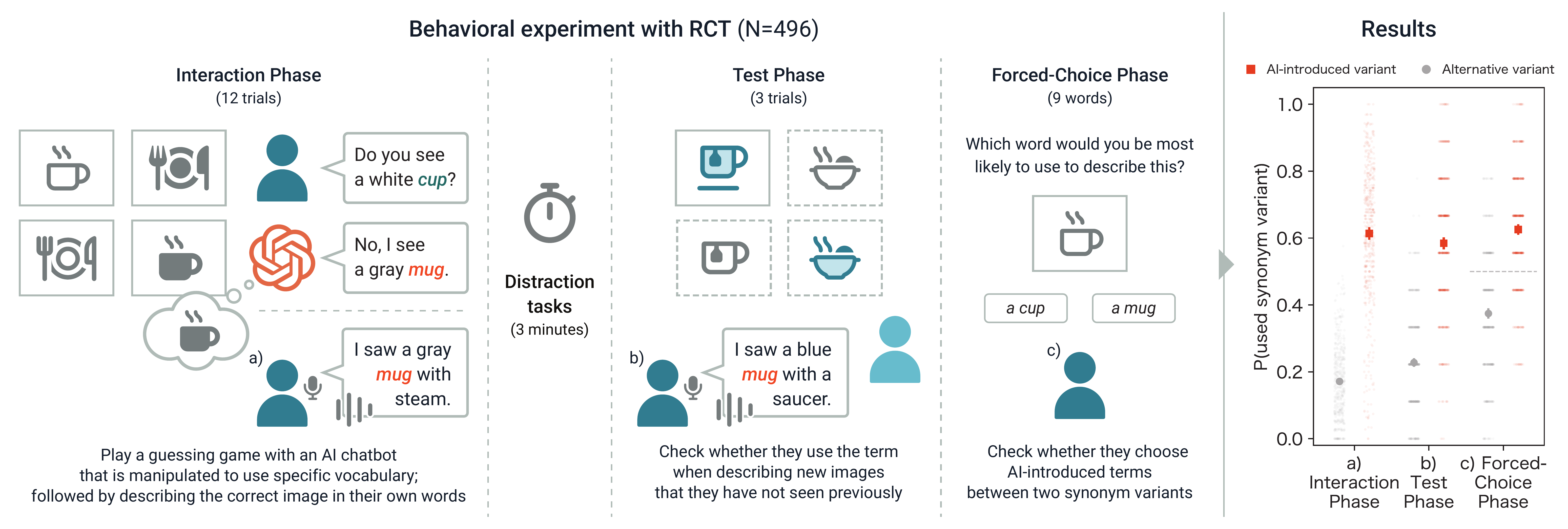}
    \caption{\textbf{Experimental design and results ($N = 496$).}
    Top panels: the four sequential phases of the experiment.
    \textit{Interaction Phase} (12 trials): participants played a referential image-guessing game with an AI chatbot covertly prompted to use specific synonym variants (e.g., \textit{mug} instead of \textit{cup}), then gave a spoken description of the target image.
    \textit{Distractor Phase} (3 min): unrelated arithmetic and visual pattern-matching tasks.
    \textit{Test Phase}: participants gave spoken descriptions of novel images not seen during the AI interaction, with no chatbot present.
    \textit{Forced-Choice Phase} (9 words): participants chose between two synonym labels for a depicted object.
    Right panel: probability of using the AI-introduced synonym variant (orange squares) versus an alternative variant (grey circles) across the three outcome phases.
    Note that in the interaction phase and the test phase, values do not necessarily sum to one, as participants might also use both or neither term.
    Small translucent dots show individual participant means; large markers show group means $\pm$ 95\% CI; dashed line at 0.5 indicates chance.
    The AI-introduced variant was adopted well above chance in the Interaction Phase and remained elevated after the distractor task in descriptions of entirely novel images.}
    \label{fig:exp}
\end{figure*}

The observational findings above establish a population-level shift in spoken word use following ChatGPT's release, but leave open the individual-level question: whether direct chatbot interaction alone is sufficient to produce lasting lexical change in individual speakers.
To answer this, we conducted a controlled experiment ($N = 496$; see \figref{fig:exp} Top and Methods). The study was pre-registered under \href{https://aspredicted.org/k38u44.pdf}{https://aspredicted.org/k38u44.pdf}. In this study, participants played a chat-based picture selection game with an AI co-player, modeled after referential communication game paradigms~\citep{clark1986, yule1997}. During the game, the AI co-player used one of two synonym variants (Variant 1 vs. Variant 2) to describe picture content, and participants had to identify a target image based on these descriptions (see \figref{fig:si-exp-interface}).

Participants' \textit{own} spoken descriptions of the target images during the Interaction Phase showed strong alignment with the AI-introduced vocabulary. Participants were substantially more likely to use a given word variant when the AI had used it than when the AI had used the alternative (61\% vs.\ 17\% for AI-introduced vs.\ other variant; $b = 0.44$, 95\% CI $[0.42, 0.46]$, $p < .001$, Monte Carlo permutation test; Supplementary \figref{fig:si-exp-during}), indicating that exposure to the AI's word choices significantly increased participants' likelihood of adopting them. Because participants were randomly assigned to hear one variant or the other of each pair, the differential adoption isolates the AI's causal contribution.

Following a 3-minute distractor task, participants had to describe nine novel images (i.e., images never seen during the previous Interaction Phase) so that a hypothetical future human co-player could identify the depicted image, with no AI chatbot present. As hypothesized, the AI-induced vocabulary uptake persisted in these spontaneous descriptions ($b = 0.36$, 95\% CI $[0.33, 0.38]$, $p < .001$; 58\% vs.\ 23\% for AI-introduced vs.\ other variants; Supplementary \figref{fig:si-exp-after}).
The effect generalized to new picture material and survived cognitive distraction, consistent with durable lexical entrenchment rather than within-context repetition~\citep{brennan1996, bortfeld1997}.
In a subsequent forced-choice task in which participants selected their preferred expression from the two variants, they chose the AI-introduced variant on 63\% of trials ($b = 0.13$, 95\% CI $[0.11, 0.14]$, $p < .001$; Supplementary \figref{fig:si-exp-forced}). This provides converging evidence of an explicit lexical preference shift.
The uptake of AI-introduced variants held across three distinct lexical categories---nouns, verbs, and adjectives---in both spontaneous production and explicit label choice (see \figref{fig:exp} Bottom).

An open-ended detection check administered after the Test Phase revealed that only 15 of 496 participants (3.0\%) noticed the chatbot's vocabulary pattern; most attributed the discrepancy to regional or stylistic variation rather than a deliberate constraint.
The remaining 481 (97.0\%) reported nothing unusual or noted only incidental features such as response speed or punctuation.
These responses suggest that the observed lexical shift was implicit rather than deliberate.

%% file: text/discussion.tex
\section*{Discussion}
\label{sec:discussion}

Together, our findings indicate that lexical features exhibited by ChatGPT are internalized into spontaneous human speech at a population scale. By analyzing 737,083 hours of transcribed podcast episodes, we reveal a measurable surge in words preferred by ChatGPT---including \emph{delve}, \emph{boast}, and \emph{meticulous}---with a causal association to its public release.
While the strongest and earliest signal appears in academic-adjacent domains (Science \& Technology), where exposure to LLM-shaped text might be the highest, indications of spreading to other domains, such as education, business, and the category-independent sample, suggest subsequent filtering into the general public. Restricting the analysis to podcast episodes featuring unscripted, conversational, spontaneous discourse shows that the shift extends beyond scripted or formal speech.
A subset of these words then exhibits a subsequent partial moderation, with usage falling back towards, and in some cases below, the pre-ChatGPT baseline, suggesting a more complex dynamic in which initial adoption is followed by selective avoidance once these words become culturally marked as AI-associated and model providers react by changing the words their models favor.

A behavioral experiment suggests an individual-level mechanism for the shift observed in podcasts. A short text-based AI chatbot interaction induces lexical shifts in participants' subsequent spontaneous speech production. The effect survives cognitive distraction and generalizes to new contexts. We observe the use of AI-introduced words increasing by 36 percentage points, a magnitude comparable to lexical-alignment effects reported in studies of human-to-human dialogue~\citep{suffill2021, hawkins2017convention}. A separate forced-choice task, likewise administered after the distraction tasks, provides converging evidence: participants select the AI-introduced words in 63\% of cases. Our results show that humans not only converge lexically with their conversation partner during a dialogue, whether human~\citep{brennan1996, pickering2004} or AI~\citep{zeng2026lvlms, jones2026llms, vaduguru2025success}, but also carry AI-introduced word choices into subsequent spontaneous speech once the interaction ends. This points to entrenchment rather than transient alignment, with repeated exposure strengthening the lexical representation in long-term memory and increasing its activation probability in future production~\citep{ellis2002, diessel2019}. Such imitation operates at high fidelity as a matter of convention, without an instrumental payoff~\citep{legare2015}.

The mechanism we observe is consistent with a long-standing psycholinguistic picture, in which lexical access in spontaneous speech is automatic rather than strategic~\citep{bock1986, levelt1999} and repeated exposure entrenches lexical representations below the threshold of explicit attribution~\citep{ellis2002, diessel2019}, as our open-ended detection check confirms. What is notable is how little the filters that normally make social learning selective apply to it. Humans readily defer to algorithmic systems on capability-defined tasks~\citep{logg2019, spatharioti2025}, while algorithm aversion persists in identity-laden domains~\citep{dietvorst2015, longoni2019}. For LLMs, which collapse authorship into a uniform voice, such task-type moderation~\citep{castelo2019} is plausibly dampened, and evolved heuristics for source evaluation---such as prior accuracy, prestige, and expertise~\citep{kendal2018, rendell2011}---might find less purchase. Instead, an LLM's lexical choices are read as articulate and authoritative, lending a word like \emph{delve} the appearance of sophisticated expression, and it is this perceived value, in the words themselves, on which such variants are selected~\citep{singh2022}. The consequence is that an algorithmic system may come to function as a cultural model---a source people learn from. 

The podcast corpus and the experiment together cast LLMs as sources of cultural variants that humans internalize, signifying a coupled human--machine cultural process. The boundary between human- and machine-authored text is already eroding across scientific writing, academic discourse, and online communication~\citep{liang2024, kobak2025, siler2026, geng2024, veselovsky2025}, and our results show that LLM-shaped patterns enter even unassisted human production. The concern that successor LLMs trained on increasingly LLM-shaped corpora may degrade in output diversity~\citep{shumailov2024}---so-called model collapse---was previously thought to be partly contained by original human language serving as an external anchor for the training distribution~\citep{gerstgrasser2024}. Our findings, however, suggest that human language can no longer be treated as an independent external anchor; rather, human and machine cultural production form a single, integrated system~\citep{brinkmann2023}. This integration is not structurally symmetric. Human--AI interaction has a hub-and-spoke topology in which many users converse with few generative AI systems, closer to early broadcast media than to the peer-to-peer networks of the online era. Such structures concentrate exposure and are known to amplify influence, with changes at highly connected nodes propagating to individuals never directly exposed to the source~\citep{katz1957, mobilia2003, centola2010, centola2018}. We illustrate the asymmetry with a noisy voter model on a small-world network with one committed source. A hub committed to the favored variant drives population-level uptake well beyond a randomly placed speaker of equal commitment, reaching agents with no direct hub contact (\figref{fig:abm}; see Supplementary Methods).

The observed lexical shifts are specific in kind. Lexical shifts can have a range of causes. Social media diffuses novel lexical items~\citep{eisenstein2014}, world events such as the COVID-19 pandemic drive topical spikes, and new technologies drive the rise of vocabulary tied to the practices they enable, from the telephone and the radio to the search engine~\citep{michel2011}. Our own corpus shows both of these last two patterns. The COVID-19 vocabulary (\emph{pandemic}, \emph{vaccine}, \emph{mask}) surged in 2020--2021 and declined in the following years; machine-learning-adjacent technical vocabulary (\emph{gpu}, \emph{python}, \emph{vector}) climbed with the growth of the field over 2022--2024; and the post-2022 rise of \emph{prompt} tracks the new activity of interacting with LLMs (Supplementary \figref{fig:top20-credible-change}). The words we identify---\emph{delve}, \emph{boast}, \emph{meticulous}, and the other top-1\% GPT-preferred words---fit none of these categories. They have no obvious referential connection to language models, and they occupy a frequency band where natural-rate lexical change is slow~\citep{hamilton2016}.

The dynamics of LLM influence on language are more complex than adoption alone. Across the GPT-score distribution the effect is unidirectional: ChatGPT-preferred words accelerate in spoken use, but ChatGPT-disfavored words do not show a corresponding decline (Supplementary \figref{fig:score-slice-symmetric}). The trajectory is also not monotone since adoption reverses for some words. The word \emph{delve} in particular drops sharply once it is discussed in social and traditional media, settling below its pre-ChatGPT baseline. OpenAI also reacted, with later GPT versions removing \emph{delve} during text-editing operations and driving its usage below the human baseline (see \figref{fig:method-and-delve}D). A plausible mechanism comes from sociolinguistics. Beyond content, lexical choices signal group membership and authenticity~\citep{bourdieu1991}. Words that become culturally legible as AI-associated may therefore attract avoidance after initial uptake, as speakers distance themselves from forms that threaten authenticity, a pattern reminiscent of hypercorrection~\citep{labov2006}. These signaling dynamics may also have consequences for social stratification. While LLMs lower linguistic barriers for non-native speakers seeking to communicate in formal English~\citep{noy2023, wang2023}, adopting LLM-marked vocabulary now risks new stigmas. Words such as \emph{delve} may come to be stereotypically associated with lower skill or with uncritical AI use, reshaping perceptions of credibility and competence.

Several limitations qualify these results.
The corpus is English-only and is limited to a self-selected, public-facing population of podcast hosts and their guests. The main analysis covers the first 18 months after ChatGPT's release and treats OpenAI's GPT models as the dominant driver; LLM deployment has since fragmented across models and providers, making attribution increasingly complex. Reverse causality (ChatGPT amplifying emerging human preferences) is addressed by the in-time placebo showing the spoken shift is specific to ChatGPT's launch date (\figref{fig:changepoint-date-sweep}), and the random assignment in the experiment ruling it out at the individual level. Because ChatGPT could in principle affect all words in non-trivial ways, we cannot rule out residual interference in our synthetic control, even though we restricted donors to words with near-zero GPT scores and excluded the closest synonyms. However, even if the donors are partially affected, the differential effect (i.e., between high-GPT-score treated words and near-zero-GPT-score donors) still reflects ChatGPT's preferences in subsequent human usage. The finding that these preferences propagate into spoken language therefore stands. In addition, the experimental evidence is limited to a sample recruited online via Prolific, which consisted of self-identified native English speakers. Comprising a single test session after a brief distractor, the experimental window captures only an early signal of AI-induced lexical entrenchment.

We have shown how AI shapes human language quietly. Its lexical choices are entering spontaneous spoken communication, with entrenchment during chat interactions as a plausible mechanism.
Unlike documented cases of deliberate imitation of algorithmic solutions~\citep{brinkmann2025} or AI-driven persuasion~\citep{salvi2025, costello2024}, the channel demonstrated here is incidental---operating below explicit attribution.
As chatbots are increasingly used from education to therapy, these findings raise the question of which other, more consequential associations and thinking patterns might be transmitted along the same channel.
The stakes are high.
The narrow set of dominant AI systems may compress both the diversity of cultural variants and the choice between sources---eroding the variation and selection on which cultural evolution depends. The cultural dynamics that emerge from the feedback of human and machine cultural production---in which AI systems both draw from and reshape human language---thereby open a new research endeavor, complementing the study of machines and humans in isolation. The question is no longer whether machines influence us, but in which way, through which channels, and under whose control.

%% file: text/methods.tex
\section*{Methods}
\label{sec:methods}

We tested whether ChatGPT's lexical preferences propagate from generated text into spontaneous human speech using a per-word causal-inference design. Each word's affinity for ChatGPT output is quantified via a log-odds score (the GPT score); per-word post-release usage shifts are estimated by combining synthetic-control matching with Bayesian change-point regression, with specificity tested by in-word-space and in-time placebo procedures.

\subsection*{Constructing datasets of human spoken communication}
\label{sec:methods-datasets}

To capture spontaneous spoken communication, we systematically constructed a dataset of podcast transcripts spanning multiple categories.

\subsubsection*{Data collection}

Our sampling of podcasts was designed to trace how any linguistic shift propagates outward from its likely point of entry. Science and Technology was our primary target, as the domain closest to the documented written-language cases of the phenomenon. Around it we sampled categories at increasing conceptual distance: Education and Business, which overlap with Science and Technology in theme and audience, and Sports, a more distant and characteristically spontaneous domain. To place these targeted categories against the wider landscape, we additionally collected a broad snapshot of all remaining categories in the catalog.

We ran a first exploratory data collection, in which episodes were drawn from a database of over four million series, randomly sampling 6{,}000 episodes per quarter \emph{for each} of five candidate categories (Business, Education, Religion and Spirituality, Science and Technology, and Sports), restricted to English-language episodes published between January 2017 and late 2024, and yielded 771{,}591 transcribed episodes. A conversational screen---at least two distinct speakers and four or more exchanges over a ten-minute slice---showed that the yield of genuine dialogue varied sharply by category: Science and Technology retained 40.8\% of collected episodes and the remaining general categories roughly 60\% to 75\%, whereas Religion and Spirituality retained only 16.2\%, reflecting a predominance of single-speaker, monologic delivery. Learning from this, we dropped Religion and Spirituality up front---together with Books, which consists almost entirely of scripted audiobook readings---rather than collecting and then discarding the bulk of those episodes, giving a cleaner sampling frame.

Building on this, the present study substantially enlarges the corpus and sharpens the spontaneity criterion. We drew candidate feeds from the PodcastIndex public catalog of approximately 4.4~million podcast series,\footnote{PodcastIndex feed database: \url{https://public.podcastindex.org/podcastindex_feeds.db.tgz}} using the snapshot dated 5~April 2026, and retrieved episode-level metadata through the PodcastIndex API.\footnote{\label{PodcastIndex}PodcastIndex API: \url{https://podcastindex-org.github.io/docs-api}} We restricted the collection to English-language episodes published between 1~January 2017 and mid-April 2026, mapping provider-supplied labels onto broader, general-purpose categories.\footnote{Apple Podcasts categories: \url{https://podcasters.apple.com/support/1691-apple-podcasts-categories}} Sampling was stratified by calendar quarter and seeded for reproducibility, so that temporal coverage is balanced across the study period rather than dominated by recent, higher-volume years. This collection comprises 1{,}407{,}131 candidate episodes (approximately 1.0~million hours of audio), of which 931{,}450 passed the dialogue screen and were transcribed---about 20\% more transcribed episodes than the earlier collection, drawn from a broader set of categories. As described below, we also replaced the earlier dialogue-only screen with a two-stage spontaneity filter that adds a trained audio classifier, giving a finer and better-validated separation of spontaneous from scripted speech; 824{,}634 episodes met this stricter criterion. A small fraction of episodes could not be retrieved or were unavailable in a usable audio format and were discarded, and the number of usable episodes decreases further through the filtering pipeline described below.

\subsubsection*{Filtering and transcription}

To maximize the number of transcripts we can obtain with limited GPU resources, we implemented the following filtering criteria.
We first removed podcast episodes shorter than 15 minutes, since they often include non-conversational speech content.
Additionally, we excluded episodes that exceeded 20,000 seconds (approximately 5.5 hours, which fell around the 99th percentile of duration in an early downloaded subset) to avoid unnecessary GPU occupation by rare extreme-length episodes.

Importantly, our intention in using podcasts was to specifically analyze the influence of LLMs on spontaneous communication.
We therefore filtered episodes in two stages.
First, we applied speaker diarization, which partitions audio into segments labeled by speaker identity, to a 10-minute slice extracted from the middle of each episode, using the \texttt{pyannote} library~\citep{plaquet2023, bredin2023}.
We required at least two distinct speakers and four or more exchanges (alternating turns) to retain an episode, which removed pure monologues and read-out broadcasts before further processing.
The dialogue filter alone, however, does not completely separate spontaneous interaction from scripted exchange (e.g., scripted interviews or co-hosted read-throughs).
We therefore complemented diarization with an audio-based classifier of \emph{spontaneity}, trained against human annotations on a 1--4 scale ranging from ``clearly scripted'' to ``clearly spontaneous''.
The annotation followed the disfluency-based protocol of Cho et al.~\citep{cho2014} (filler words, repetitions and corrections, hesitations, and incorrectly used words); three na\"ive coders rated three 30-second samples per episode after a two-round calibration, and the interrater reliability reached ICC $= 0.87$ (95\% CI 0.81--0.91; see \secref{sec:si-methods-spontaneity-annotation}).
We then trained a classifier on top of the \texttt{Whisper-large-v3} encoder~\citep{radford2022}, following Elisha et al.~\citep{elisha2024}, so that it maps the 25 $\times$ 30-second middle window of each episode to a soft distribution over the four annotation classes (see \secref{sec:si-methods-spontaneity-classifier}), and aggregated chunk predictions into an episode-level continuous score in $[1, 4]$.
For all analyses, an episode was retained if (i) it passed the diarization-based dialogue filter and (ii) its spontaneity score exceeded $3.0$, which corresponds to the inflection point of the labeled validation curve.
We confirmed that the retained episode set is essentially unchanged under an alternative, majority-vote-based aggregation of the same chunk-level predictions (see \secref{sec:si-methods-spontaneity}).

The transcription of the collected data was performed using the \texttt{large-v3} model of \texttt{WhisperX}~\citep{bain2023}, a faster version of the \texttt{Whisper} speech-to-text model~\citep{radford2022}.
We employed batch processing with the model, achieving an average transcription speed of approximately 2 minutes per hour of audio with Nvidia A100 GPU.
Here, we opted to run the transcription process ourselves rather than use pre-existing transcript data from YouTube or other podcast platforms, given the possibility that they have switched transcription models over time.\footnotemark
As a result, the same model, configured identically, was applied uniformly to every episode across the entire time window of the study, while this configuration achieves a word error rate of approximately 5\% averaged across common English audio datasets~\citep{radford2022}.
Importantly, since the model is fixed across the time axis, any residual transcription error contributes a constant background that cannot, by construction, generate a change point at the ChatGPT release date or differentially favor the GPT-preferred words over their synthetic controls.
The recognition language was fixed to English throughout transcription, consistent with the English-language restriction applied at collection through the podcast feeds' language metadata.
\footnotetext{Specifically in the YouTube transcripts, we found an unnatural increase in the frequency of the filler word \emph{um} starting around May 2020, which we found difficult to attribute to an actual increase in speakers' usage of the word. It is more plausible that YouTube switched to a transcription model that transcribes fillers verbatim, and thus, we conducted the transcription process to avoid a potential source of bias.}

\subsubsection*{Preprocessing}
\label{sec:method-dataset-preprocessing}

We preprocessed the obtained transcripts to capture essential changes in word frequency by removing noise and highlighting relevant patterns.
We followed a systematic procedure:
\begin{enumerate}
    \item \textbf{Tokenization}: The text is divided into individual tokens (words) for processing.
    \item \textbf{Normalization}: All words are converted to lowercase to ensure uniformity and avoid duplication due to case differences.
    \item \textbf{Stop word removal}: Commonly used words that do not carry significant semantic meaning, such as \textit{and}, \textit{the}, and \textit{is}, are removed. The list of stop words used in this process is sourced from the Natural Language Toolkit (NLTK) library~\citep{bird2006}, which provides a standard set of English stop words.
    \item \textbf{Non-alphabetic filtering}: Words containing non-alphabetic characters are excluded, ensuring only standard words are retained.
    \item \textbf{Length filtering}: Words with fewer than three characters are removed to eliminate overly short and potentially uninformative tokens.
    \item \textbf{Stemming}: Words are reduced to their root forms using the Porter stemming algorithm \citep{porter1980}. This algorithm applies a series of heuristic rules to iteratively strip suffixes from words (e.g., \textit{running} to \textit{run}). Since raw stems are often non-words (e.g., \emph{delve} stems to \texttt{delv}), figures and tables relabel each stem with a representative surface form for readability.
\end{enumerate}

For subsequent analyses, we calculated the log relative frequency of podcast episodes containing each presented word, sampled monthly.
Older data may exhibit different word usage trends due to factors such as the relatively low number of published podcast episodes.
Hence, we analyzed data spanning six years before the initial release of ChatGPT on November 30, 2022.
Additionally, due to the timing of data collection, the corpus includes podcast episodes published up to April 30, 2026. The change-point regression and synthetic-control estimates are anchored to the 18-month GPT-3.5 era ending May 30, 2024 (the month GPT-4o replaced GPT-3.5 as the default on ChatGPT's free tier); post-2024 data is used to characterize the reversion dynamics reported in Results.
We employed log frequency to facilitate trend interpretation within this early diffusion phase, using Laplace smoothing \citep{manning2008} to account for zero counts, which helps detect emerging patterns that may initially exhibit exponential growth.

\subsection*{Measuring word preferences of large language models}
\label{sec:methods-gptscore}

We investigated the word preferences of commonly used LLMs by prompting various models to edit a diverse set of human-authored texts. Building on prior research~\citep{liang2024,kobak2025}, we analyzed differences in word frequencies between original human-written texts and their LLM-edited versions. Our analysis spans a wide range of human texts, prompts, and models, enabling the computation of an aggregated \emph{GPT score}.

\subsubsection*{Creation of contrastive datasets}

We compiled datasets from diverse sources, all predating the introduction of ChatGPT. These included 7,182 abstracts from arXiv (2019--2022) using the arXiv API, 2,880 abstracts from bioRxiv (2019--2022) via the bioRxiv API, over 8,000 abstracts from Nature (2019--2023) collected through its search engine, 10,000 samples each from the Enron email dataset (2000--2001), Hewlett Foundation student essays (2012) and Wikipedia articles (2019--2022), and 2,000 spontaneous-speech podcast transcript excerpts predating ChatGPT. Detailed dataset creation steps are provided in \secref{sec:si-gptscore-dataset}.

To assess how prompts influence model word preferences, we used three standard prompts across all datasets and models:
\begin{itemize}
    \item \textbf{Prompt 1:} \, \texttt{Please polish this text: \{text\}}
    \item \textbf{Prompt 2:} \, \texttt{Can you improve this text: \{text\}?}
    \item \textbf{Prompt 3:} \, \texttt{Please rephrase this text to improve its clarity: \{text\}}
\end{itemize}
As Prompt 3 frequently altered the content of emails, we extended the prompt to include: ``\texttt{It's an email so please don't change the structure of the text.}'' in that specific case.

We preprocessed both the original human texts and their LLM revisions using the same procedure applied to transcript datasets of YouTube videos and podcasts. For robustness, we considered only words whose human or LLM document frequency, pooled across all dataset--model--prompt strata, reached at least one per mille of the pooled document count, and excluded prompt-related stems (\textit{rephrase, polish, dear, text, certainly, subject, readable, clarity, enhance, version, title}) that were frequently repeated in the LLM's responses.
Our analysis also included different GPT-family models, of which GPT-3.5-turbo, GPT-4, and GPT-4-turbo were the production models available at the time the GPT-score set was defined and used to compute the reference GPT score below.
To check whether the same preferences persist in later models, we additionally evaluated GPT-4o and GPT-5 after that, while we did not include them in the reference score so that the set of treated words used in the downstream causal analysis remains anchored to the pre-release model family. 

\subsubsection*{Log-odds ratio estimation}
\label{sec:methods-gptscore-logodds}

To identify words preferentially associated with LLMs, we computed log-odds ratios comparing word frequencies in human-authored and ChatGPT-edited corpora. For each word \(w\), we estimated its document frequency in human (\(p_{\text{human}}\)) and ChatGPT (\(p_{\text{GPT}}\)) corpora using Laplace smoothing to mitigate zero-count issues:
\[
p_{w} = \frac{\text{number of documents containing word } w + 1}{\text{total documents} + 1}.
\]
The log-odds transformation was applied to these smoothed probabilities:
\[
\text{log-odds}\left(p\right) = \ln\left(\frac{p}{1 - p}\right),
\]
yielding the log-odds ratio (LOR) for each word:
\[
\text{LOR}_w = \text{log-odds}\left(p_{w,\text{GPT}}\right) - \text{log-odds}\left(p_{w,\text{human}}\right).
\]
Positive LOR values indicate higher prevalence in ChatGPT-edited texts, while negative values suggest human-associated usage. This metric was computed independently across all dataset--model--prompt combinations. When estimating $p_w$ we set the denominator to the maximum number of returned documents across (model, prompt) combinations for each dataset to avoid inflation through occasional API refusals or empty completions.

We define these probabilities on \emph{document frequency}---whether a word occurs in a document---rather than on token counts, because our interest is in how \emph{widely} a word is used rather than how often, this Bernoulli definition measures the prevalence of a word across units of communication. It is immune to within-document repetition and less sensitive to document length than a token-count measure, both of which would otherwise inflate the estimated preference for a word. It also coincides with the definition of \emph{GPT score}.

\subsubsection*{Calculation of a weighted GPT score}

We present word preferences for a range of GPT-family models and document types (see~\ref{sec:appendix-word-preference}). For our main analysis, we focus on scientific abstracts (arXiv, bioRxiv, Nature) and the GPT chat models available before GPT-4o (GPT-3.5-turbo, GPT-4, and GPT-4-turbo; Table~\ref{tab:model-versions}), and we developed a GPT score that marginalizes over uncertainties in model usage patterns. Given the unknown true distribution of ChatGPT's usage across datasets (\(D\)), models (\(M\)), and prompts (\(P\)), we adopted a Bayesian hierarchical model with non-informative Dirichlet priors:
\[
\begin{aligned}
P\left(D\right) &\sim \text{Dirichlet}\left(\mathbf{1}\right), \\
P\left(M \mid D\right) &\sim \text{Dirichlet}\left(\mathbf{1}\right), \\
P\left(P \mid D, M\right) &\sim \text{Dirichlet}\left(\mathbf{1}\right),
\end{aligned}
\]
where \(\mathbf{1}\) denotes flat priors for each parameter. The joint distribution \(P(D, M, P)\) was computed as:
\[
P\left(D, M, P\right) = P\left(D\right) \cdot P\left(M \mid D\right) \cdot P\left(P \mid D, M\right).
\]
For each of 1000 Monte Carlo samples from this prior, we marginalized the human and ChatGPT smoothed probabilities over (dataset, model, prompt):
\[
\hat{p}_w^{(\cdot)} = \sum_{d \in \mathcal{D}} \sum_{m \in \mathcal{M}} \sum_{p \in \mathcal{P}} p_{w}^{(\cdot,d,m,p)} \cdot \lambda\left(d, m, p\right),
\]
where $(\cdot)$ is either human or GPT, $\lambda(d,m,p) \propto P(D=d, M=m, P=p)$, and $\mathcal{D}, \mathcal{M}, \mathcal{P}$ index datasets, models, and prompts. The per-sample log-odds ratio is $\mathrm{logit}(\hat{p}_w^{\text{GPT}}) - \mathrm{logit}(\hat{p}_w^{\text{human}})$. The GPT score is the median LOR across the 1000 samples, with uncertainty quantified via 95\% percentile intervals.

The Dirichlet prior structure reflects maximum entropy assumptions about potential correlations between datasets, models, and prompts. By sampling from the joint prior distribution, we emulate the variability expected under real-world deployment scenarios where specific GPT-family model, dataset, and prompt combinations are not systematically favored. The resulting GPT scores thus represent robust centrality estimates of word preferences across plausible usage distributions.

\subsection*{Estimating causal influence of ChatGPT}
\label{sec:methods-synthetic}

To assess ChatGPT's causal impact on human verbal communication, we employed the synthetic control method~\citep{abadie2010, abadie2022}.
This method allows us to estimate the usage pattern of a ``treated'' GPT-preferred word (\ie a word with a high GPT score) in the counterfactual scenario where ChatGPT was never deployed.
This is built on the assumption that words sharing similar pre-release usage patterns would have continued exhibiting comparable patterns in the absence of the release.
Thus, the method constructs a synthetic control for each treated word by forming a convex combination of multiple ``donor'' words whose usage closely tracks the treated word's pre-release trajectory so that it predicts the counterfactual pattern by extrapolating this combination forward.

Reading the treated vs synthetic gap as the causal effect of ChatGPT's release requires three assumptions. First, \emph{no anticipation}: the release is an external, discretely timed event that speakers could not have foreseen, so pre-release trajectories are free of treatment, and the six-year pre-release window furnishes an uncontaminated basis for matching. Second, \emph{a good and stable pre-release fit}: the synthetic control must reproduce the treated word's pre-release trajectory closely and over a long horizon, since a fit that holds only briefly, or that is achieved by chance on a short window, does not support extrapolation past the release; our six-year pre-release window and trajectory-based donor selection (below) are designed to meet this requirement. Third, \emph{no interference} (the stable unit treatment value assumption, SUTVA): the donor words must not themselves be affected by the release, so that their post-release usage reflects the counterfactual rather than a diffuse response to ChatGPT. This is the binding assumption in our setting, since ChatGPT may, in principle, shift all language rather than only the words it most prefers. We cannot guarantee it, but we make it as credible as the design allows by choosing donors that are unlikely to be differentially treated. Specifically, we exclude the treated word's closest semantic neighbors, which are its plausible substitutes and would inherit a fraction of the same shock, and we restrict the donor pool to words that ChatGPT neither over- nor under-uses (those with near-zero GPT score), removing the words most likely to carry the treatment (see \secref{sec:methods-synthetic-donor}). We discuss the consequences of residual interference, and the interpretation of our results should it remain, in \secref{sec:discussion}.

\subsubsection*{Donor selection}
\label{sec:methods-synthetic-donor}

For each treated word, the synthetic control method requires the selection of a set of donor words that are used to build the synthetic control. The candidate pool is the intersection of the top 50{,}000 most frequent words in the pre-trained word2vec embedding with the words we measured a GPT-score for; restricting to this vocabulary discards rare words keep computation feasible and ensuring well defined GPT-scores. From this pool, we first filtered out the $K = 20$ words most similar to the treated word in a pre-trained word2vec embedding built on the Google News dataset~\citep{mikolov2013}. Close semantic neighbors of the treated word can plausibly be substitutes for it and therefore inherit a fraction of the same treatment shock (potentially in an inverse direction), which would contaminate the counterfactual. We narrowed the pool of potential donors to a symmetric neutral-percentile band around $|\text{GPT score}| = 0$, retaining the central 50\% of words by absolute GPT score. Removing words that showed over- or underusage by GPT enforces the stable unit treatment value assumption that underpins the synthetic control estimator.

From the remaining vocabulary, we then selected the $L = 100$ words following the recommendation of Abadie and Bastida~\citep{abadie2022}. We specifically chose words whose pre-treatment log-frequency trajectories were closest to that of the treated word in pointwise Euclidean distance over the pre-treatment months. To prevent month-to-month sampling noise from dominating this matching step, the pre-treatment trajectories were first smoothed with a Gaussian-process prior (Mat\'{e}rn kernel with $\nu = 5/2$ and a two-year length scale, $\ell = 720$~d). The length scale is chosen as a compromise between the raw monthly series, on which the synthetic overfits noise, and longer length scales at which it underfits the relevant low-frequency dynamics (Supplementary \figref{fig:length-scale-robustness}). The smoother enters only at donor selection and the synthetic control fit itself; all downstream change-point and placebo statistics are computed on the raw monthly series.

The synthetic control for the treated word was then constructed as a convex combination of these $L$ donors, with the weights chosen to minimize the pre-treatment root mean squared prediction error between the treated word's trajectory and the weighted donor average~\citep{abadie2010} using the smoothed pre-treatment trajectories.
We constrained the weights to be non-negative and to sum to one, which rules out extrapolation outside the convex hull of donor trajectories and induces sparsity, so that the fitted weight vector typically had support on a small subset of the pool.
Weights were optimized by sequential least-squares programming (SLSQP, \texttt{scipy.optimize.fmin\_slsqp}) initialized at the uniform vector $\omega_{j} = 1/L$. Across the treated words in Science \& Technology podcasts, the non-negativity and sum-to-one constraints induce sparse SLSQP weight vectors, so the fitted synthetic control typically draws on only a handful of donors, as illustrated for \emph{delve} in Table \ref{tab:delve-donors}.

We re-ran the entire synthetic-control pipeline under four control specifications that each vary a specific design choice: \textbf{C1} drops the semantic-neighbour and GPT-score filters, leaving only $\ell_2$ matching on the raw vocabulary; \textbf{C2} shrinks the donor pool from $L = 100$ to $L = 10$; \textbf{C3} replaces the SLSQP convex fit with deterministic inverse-distance similarity weights, following CausalCite~\citep{agrawal2024causalcite}, adapted to our time-series setting; \textbf{C4} swaps the audited counts substrate for the raw, un-audited counts. Full specifications, per-variant placebo $p$-values, and side-by-side renderings of \figref{fig:method-and-delve}A and \figref{fig:change-point-model}B across the four controls are given in Supplementary Materials (\tabref{tab:control-specs}, Figs.~\ref{fig:delve-robustness-composite} and \ref{fig:fig3-robustness-composite}); the qualitative pattern is preserved across all four.

\subsubsection*{Placebo test}

We assess the significance of the causal effect using the in-space placebo test of Abadie et al.~\citep{abadie2010} as refined by Ferman and Pinto~\citep{ferman2017}. For each word in the matching pool, we re-run the synthetic control procedure with that word in the role of the treated word, taking the remaining donors as the new pool, and compute the post- to pre-treatment ratio of mean squared prediction error (MSPE). The placebo pool size is bounded above by the size of the donor pool: for the Main specification we use $n_{\text{placebo}} = L = 100$ (every donor serves once as a placebo target), while for the C2 robustness specification the ten-word donor pool yields $n_{\text{placebo}} = 10$. The placebo targets are drawn without replacement from the donor pool.

The empirical $p$-value pools the treated word with its $n_{\text{placebo}}$ placebos and ranks the MSPE ratios: $p = (r + 1) / (n_{\text{placebo}} + 1)$, where $r$ is the number of placebos with MSPE ratio at least the treated word's. The floor is therefore $1/(n_{\text{placebo}} + 1)$: $1/101 \approx 0.010$ under the Main specification ($n_{\text{placebo}} = 100$) and $1/11 \approx 0.091$ under C2 ($n_{\text{placebo}} = 10$).

For the window-mean gaps reported in \figref{fig:delve-delta-sigmoid}, we apply an analogous inversion of the same in-space placebo distribution. For each placebo target, we compute the mean observed-minus-synthetic log-frequency gap over the window of interest (months 13--18 after release, and the last six recorded months). The 95\% confidence interval for the treated word's true effect is obtained by inverting this distribution, $[\hat{g} - q_{97.5},\; \hat{g} - q_{2.5}]$, where $\hat{g}$ is the treated word's window-mean gap and $q_\alpha$ denotes the $\alpha$-quantile of the placebo distribution. We report a one-sided $p$-value for the predicted post-adoption rise over months 13--18 and a two-sided $p$-value for the post-2024 reversion, whose sign was observed rather than predicted. Both use the same $(1 + r) / (1 + n_{\text{placebo}})$ add-one convention. Per-category values for \emph{delve} are reported in Supplementary \tabref{tab:fig2-b2-ci}.

\subsubsection*{Change-point regression of the synthetic-control gap}

Having constructed a synthetic control for each treated word, we tested whether the word's trajectory departs from that counterfactual after the release of ChatGPT, and by how much. We restricted the set of treated words to those appearing in at least twenty episodes per month on average over the pre-treatment window, which removes targets for which the gap series is too noisy to support a meaningful change-point fit. For each treated word $w$ we formed the monthly gap between its observed and synthetic-control log-frequency,
\[
  \Delta y_{w,t} \;=\; \log_{10} y_{w,t}^{\text{obs}} - \log_{10} y_{w,t}^{\text{synth}},
\]
and fitted to it the hierarchical Bayesian change-point regression of \eqnref{eqn:linear-model},
\begin{equation}
\Delta y_{w,t} \;=\; \alpha + \beta_{\text{Pre}}\, t + \beta_{\text{Post}}\, d_{\text{Post}}\, (t - T_{\text{event}}) + \epsilon_t,
\qquad
d_{\text{Post}} =
\begin{cases}
  1 & t > T_{\text{event}}\\[2pt]
  0 & \text{otherwise.}
\end{cases}
\label{eqn:linear-model}
\end{equation}
Here $t$ is continuous time measured in years and anchored at the start of the window $T_{\text{start}}$, so the coefficients are interpretable as changes in $\log_{10}$ frequency per year, and the post-release term allows the slope to change at the change point $T_{\text{event}}$ (the ChatGPT release, 30~November 2022). The fitted slope is $\beta_{\text{Pre}}$ before the release and $\beta_{\text{Pre}} + \beta_{\text{Post}}$ after it. The post-release change in slope $\beta_{\text{Post}}$ is our estimate of ChatGPT's influence on the usage of $w$. The fit spans the same 18-month window as the prior analysis.

We placed weakly-informative priors on every parameter: a standard normal $\mathcal{N}(0, 1)$ on each slope coefficient ($\beta_{\text{Pre}}$, $\beta_{\text{Post}}$), a diffuse $\mathcal{N}(0, 10)$ on the intercept $\alpha$, and a half-Cauchy$(0, 10)$ prior on the standard deviation $\sigma$ of the Gaussian residual $\epsilon_t$. Each word was fitted independently by Hamiltonian Monte Carlo using Stan's no-U-turn sampler via \texttt{cmdstanpy} at its defaults, with four chains of 1{,}000 post-warmup draws each; warmup draws were discarded. We summarized every coefficient by its posterior mean and the 95\% highest-density interval (HDI).

We compare the slice-mean $\beta_{\text{Post}}$ over nested top-$X\%$ slices of the GPT-score distribution against a permutation null (\figref{fig:change-point-model}C). $X$ is sampled on geometric grid from $100\%$ to $0.5\%$; at each point we generate $n_{\text{perm}} = 1{,}000$ permutations and report the 5th--95th percentile of the permuted slice mean.

\subsection*{Controlled referential communication experiment}
\label{sec:methods-experiment}

\paragraph{Participants}
Five hundred participants completed the experiment via Prolific (mean age~$= 40.5$ years, $SD = 13.2$; 50\% women; 91\% reporting English as their first language; see Supplementary \tabref{tab:demographics} for full demographics).
The study was preregistered on AsPredicted (\#k38u44; \url{https://aspredicted.org/k38u44.pdf}).
Sample size was determined by a power analysis targeting $d = 0.30$ with 80\% power ($N_{\text{required}} = 352$); we recruited above this target to accommodate exclusions.
Four participants were excluded for self-reported color vision deficiency, yielding a final $N = 496$.

\paragraph{Design}
The experiment used a 2 stimulus group (Group A/B, \textit{between-subject}) $\times$ 2 mode (synonym Variant~1 vs.\ Variant~2 as AI-used, \textit{between-subject}) $\times$ 12 trial (\textit{within-subject}) design.
The stimulus group determined which nine of the 18 synonym pairs the AI was instructed to use during the Interaction Phase. Participants were randomly assigned to one of the two stimulus groups. Within their group, participants were then randomly assigned to one of two word mode sets (Variant 1 or 2). 

\paragraph{Materials}
Stimuli in total comprised 18 synonym pairs spanning three lexical categories: nouns (e.g., \textit{mug}/\textit{cup}), verbs (e.g., \textit{to fix}/\textit{to repair}), and adjectives (e.g., \textit{glossy}/\textit{shiny}). Each stimulus Group A/B consisted of 9 synonym pairs, i.e., three noun pairs, three verb pairs, and three adjective pairs. The mode determined which of the two synonyms within each pair the AI chatbot used consistently across all 9 pairs. For example, a participant assigned to Group A, Variant 1 encountered an AI co-player that always used the words ``thermos bottle'', ``gift'', ``cup'', ``to fix'', ``to examine'', ``to install'', ``colorful'', ``cracked'', and ``spotted''. The full list is provided in Supplementary \tabref{tab:synonym-pairs}. Images were AI-generated using Gemini 3 (\texttt{gemini-3-pro-image-preview}) to depict scenes that allowed the AI co-player to describe features in the target image using the prompted variants.

\paragraph{Procedure}
Participants completed five sequential phases:

\subparagraph{(1)~\textit{Interaction Phase}}
Participants played a picture selection game~\citep[modeled after][]{clark1986, yule1997} with the GPT-4o-based chatbot that plays the role of a co-player (see \figref{fig:si-exp-interface}). In each trial, the participant viewed six candidate images and had to identify the one target image based on the AI chatbot's description. Each target image was identifiable by three distinct features (one object, one activity, one attribute), which the AI chatbot describes using its three assigned synonym variants (one noun, one verb, one adjective). Participants selected the image they thought best matched the AI chatbot's description and received feedback on whether their selection was correct. They then recorded a spoken description of the correct target image themselves. The interaction phase consisted of 12 trials, and each trial featured one noun, one verb, and one adjective synonym variant. Each of the 9 synonym variants appeared four times across all trials.

\subparagraph{(2)~\textit{Distractor Phase}} Following the Interaction Phase, participants completed a 3-minute distraction task consisting of numerical and spatial reasoning problems~\citep{waris2017}.

\subparagraph{(3)~\textit{Test Phase}} 
In our final Test Phase, participants were presented with three new sets of six images, each with one image highlighted as the target. For each set, participants were tasked with recording a spoken description of the target image, assuming another virtual participant who would try to identify it in a future iteration of the game. Each of the 9 synonym variants was depicted once across the three trials.

\medskip\noindent
Between phases~(3) and~(4), participants answered an open-ended detection check: ``Have you noticed anything about the language of the chatbot?''. 

\subparagraph{(4)~\textit{Forced-Choice Task}} Participants were presented with nine images depicting the objects, activities, and attributes corresponding to the nine synonym pairs encountered during the interaction phase. For each image, participants had to indicate how they would prefer to describe what they saw by selecting one of two synonym variants (e.g. ``mug'' or ``cup'').

\subparagraph{(5)~\textit{Background questionnaire}} 
Finally, participants complete a series of questions covering demographic information, familiarity with and use of AI, and language background and usage.

\medskip\noindent
Throughout the above procedure, participants' spoken descriptions were transcribed via the browser Web Speech API, and during the Test Phase, the transcripts were then displayed to participants for manual correction and submitted as human-verified transcripts (see \secref{sec:si-methods-experiment-implementation}).

\paragraph{Scoring and analysis}
Transcriptions were scored using a stemmer-based span-dominance algorithm (see \secref{sec:si-methods-experiment-implementation}).
The primary outcome, $\Delta p$, was the per-participant difference in AI-variant usage rate between AI-introduced and non-introduced variant pairs.
Significance was assessed via Monte Carlo sign-flip permutation test (10,000 permutations, two-tailed); 95\% CIs were computed by cluster bootstrap (10,000 replications, resampling by participant); a linear mixed model with random intercepts for participants and word pairs served as a parametric cross-check.

%% file: text/acknowledgments.tex
\section*{Acknowledgments}

H.Y. is supported in part by JST PRESTO Grant Number JPMJPR246B. E.L.-L. is funded by the Deutsche Forschungsgemeinschaft (DFG), project number 458366841 (POLTOOLS - Assisting behavioral science and evidence-based policy making using online machine tools).

%% file: text/appendix.tex
\section{Word preferences of Large Language Models}
\label{sec:appendix-word-preference}

\begin{figure}[ht!]
    \centering
    \begin{subfigure}[t]{0.49\textwidth}
        \centering
        \caption{Word preferences are robust to the rephrasing prompt.}
        \includegraphics[width=\textwidth]{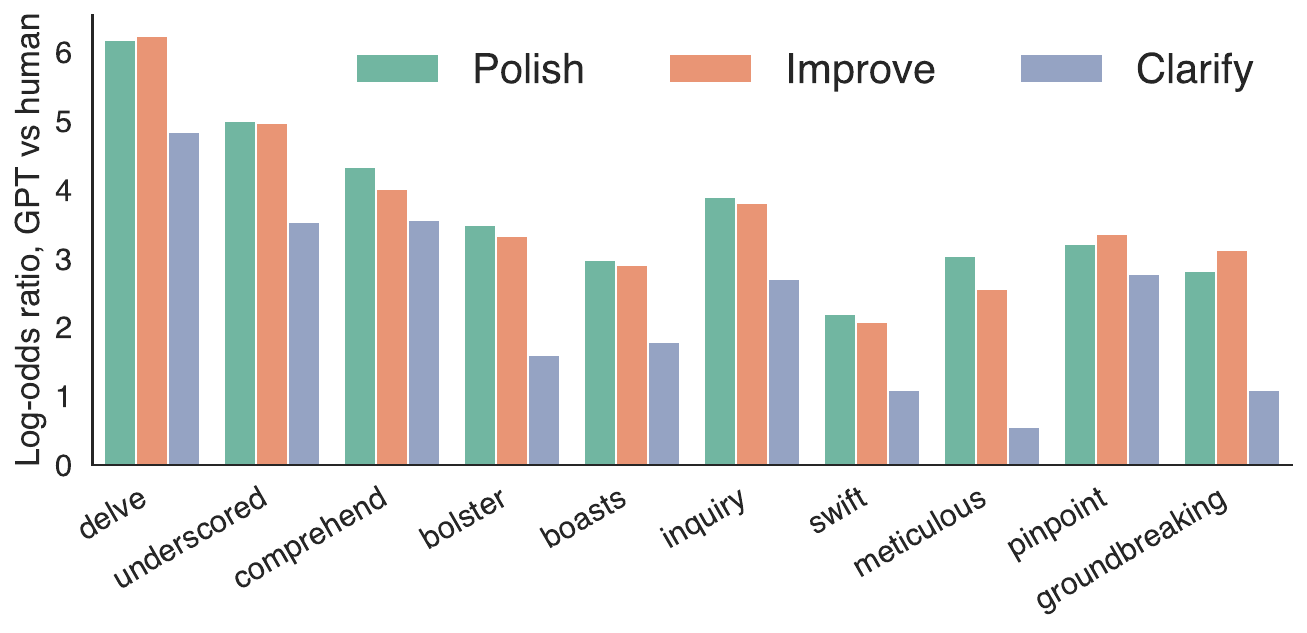}
        \label{fig:appendix-lor-words}
    \end{subfigure}
    \hfill
    \begin{subfigure}[t]{0.49\textwidth}
        \centering
        \caption{Word preferences of revisions by \textrm{GPT-3.5-turbo}.\phantom{This text will be invisible}}
        \includegraphics[width=\textwidth]{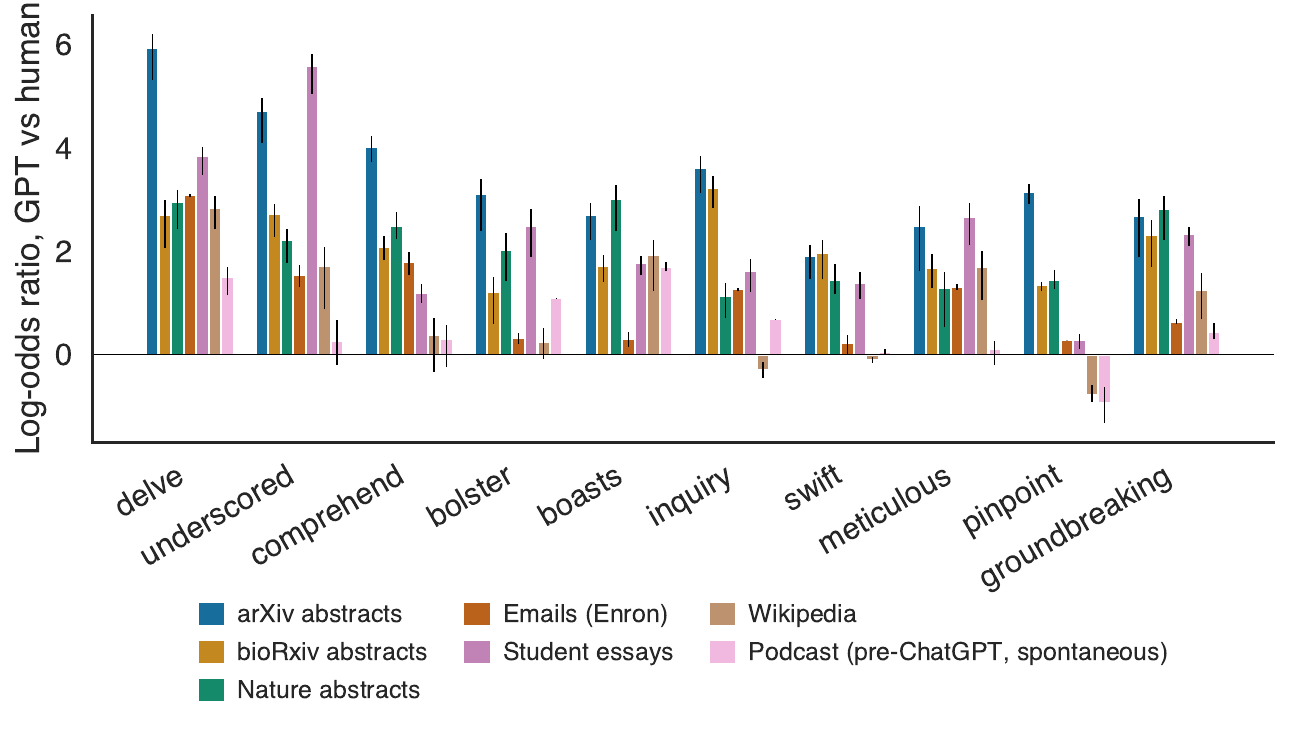}
        \label{fig:appendix-lor-datasets}
    \end{subfigure}
    \caption{\textbf{Log-Odds-Ratios (LORs) of words in human versus LLM-revised text.}
    (a) Word preferences are largely invariant to the exact rephrasing instruction (\texttt{polish}, \texttt{improve}, and \texttt{clarify}) when revising arXiv abstracts with \textrm{GPT-3.5-turbo}. (b) Substantial variations in LORs emerge when examining revisions of \textrm{GPT-3.5-turbo} across different datasets.}
    \label{fig:appendix-lor-combined}
\end{figure}

Large language models (LLMs) from the GPT family systematically alter word frequencies when revising human-written text~\citep{liang2024,kobak2025}. To quantify these word preferences, we computed log-odds ratios (LORs) comparing word frequencies in human-authored texts and their GPT-revised counterparts. We systematically evaluated the sensitivity of this effect to model version (Figure~\ref{fig:method-and-delve}D), prompting (\figref{fig:appendix-lor-combined}A), and source dataset (\figref{fig:appendix-lor-combined}B; Supplementary \figsref{fig:word_log_probability_by_metric} and \ref{fig:word_log_odds_ratio_by_dataset}).

Word preference patterns exhibited notable stability across GPT-family models (Figure~\ref{fig:method-and-delve}D), suggesting these biases emerge from intrinsic characteristics of the training pipeline rather than version-specific training. However, specifically for \emph{delve}, we found decreasing preference in newer models. For \emph{delve}, the odds ratio in revised arXiv abstracts declines from approximately 380:1 under \textrm{GPT-3.5-turbo} to 100:1 for \textrm{GPT-4-turbo} and 40:1 for \textrm{GPT-4o}. This trend culminates in \textrm{GPT-5}, which exhibits markedly smaller lexical anomalies than any earlier model: most of the signature words are no longer over-used, and \emph{delve} itself falls below its human baseline (odds ratio $\approx 0.7{:}1$ on arXiv).

These preferences are also robust to the exact rephrasing instruction (\figref{fig:appendix-lor-combined}A): the three prompts we used (\texttt{polish}, \texttt{improve}, and \texttt{clarify}) yield closely similar LORs for every reference word, with the \texttt{clarify} prompt consistently the mildest yet never reversing a preference.

LOR magnitudes varied substantially across source corpora (\figref{fig:appendix-lor-combined}B). Analysis of log-probability distributions (Supplementary \figref{fig:word_log_probability_by_metric}) revealed that this variability stems primarily from baseline differences in human word choices. For instance, while humans rarely use \emph{underscore} in essays, GPT revisions introduced this term frequently across domains, including essays.

Focusing on scientific abstracts (arXiv, bioRxiv, Nature) and the GPT chat models available before \textrm{GPT-4o} (\textrm{GPT-3.5-turbo}, \textrm{GPT-4} and \textrm{GPT-4-turbo}; see Table~\ref{tab:model-versions}), we computed a weighted GPT score by marginalizing over model, prompt, and dataset combinations (Figure~\ref{fig:method-and-delve}D; black diamonds). Here, \emph{delve} emerged as the most strongly overused term ($\text{LOR} > 4$), followed by \emph{underscore}, \emph{comprehend}, \emph{bolster}, \emph{boast}, \emph{inquiry}, \emph{swift}, \emph{meticulous}, \emph{pinpoint} and \emph{groundbreak} ($\text{LOR} > 2.5$).

\section{Population-level replication in academic YouTube talks}
\label{sec:appendix-youtube}

As a complement to the podcast analysis, we replicate the population-level signal in another spoken corpus: 360,445 academic YouTube talks from the channels of 20,622 research institutions cataloged in the Research Organization Registry~\citep{noauthor_ror_2024}. Such talks sit between the written academic record where ChatGPT's footprint is well documented~\citep{liang2024,kobak2025} and the spontaneous speech we study in podcasts, providing an independent test of whether the effect extends beyond conversational audio.

The pipeline mirrors the podcast pipeline (\secref{sec:methods-datasets}), with three differences: (i)~institutional channels are identified by querying the YouTube Data API and selecting the most plausible match via \texttt{gpt-3.5-turbo-0125} (Supplementary \figref{fig:prompt-youtube}); (ii)~videos are retained between 20 minutes (the API's \texttt{short} cutoff, below which videos frequently consist of non-speech content such as advertisements or trailers) and the 99th-percentile duration ($\sim$3.0 hours); (iii)~we omit the speaker-diarisation and spontaneity-classifier steps, since academic talks are predominantly prepared, monologic speech. Transcription, synthetic control, and change-point analyses are otherwise identical.

\begin{figure*}[t]
    \centering
    \includegraphics[width=\linewidth]{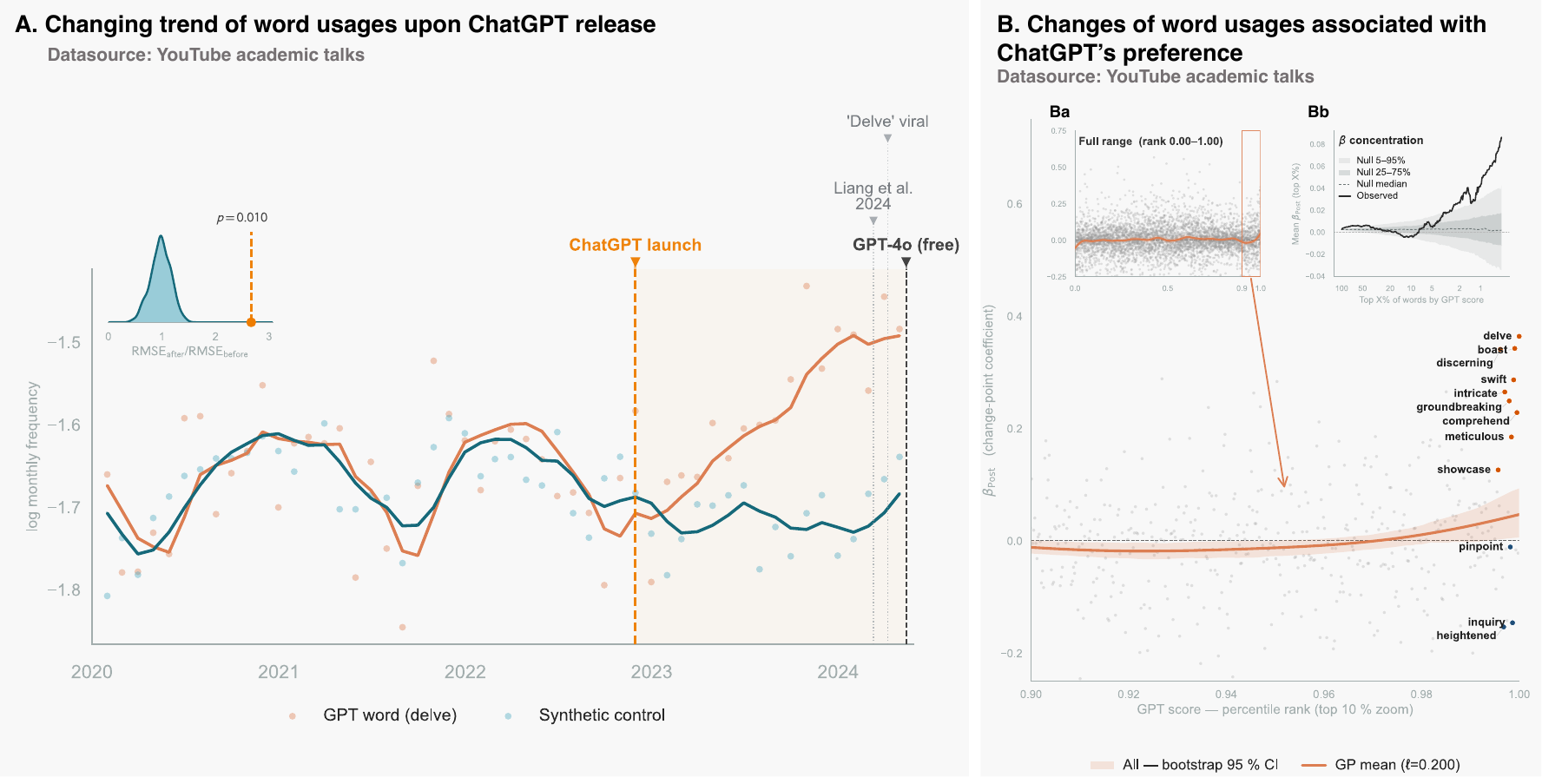}
    \caption{\textbf{Population-level trend changes for top GPT-preferred words in academic YouTube talks.}
    (A) Monthly relative frequency of \emph{delve} in academic YouTube talks (orange) and its synthetic control (teal). The inset shows the placebo distribution of post-/pre-treatment RMSE ratios across the donor pool, with the observed ratio for \emph{delve} marked ($p = 0.010$). The vertical dashed line marks the ChatGPT release.
    (B) Piecewise-linear trend fits for the top GPT-preferred words, with a change point at the ChatGPT release. Words such as \emph{comprehend}, \emph{boast}, and \emph{swift} show a significant post-release increase, mirroring \emph{delve}.}
    \label{fig:appendix-youtube}
\end{figure*}

The headline matches the podcast result: the placebo test for \emph{delve} yields $p = 0.010$ (\figref{fig:appendix-youtube}A). The piecewise-linear regression extends the effect to other top GPT-preferred words, such as \emph{comprehend}, \emph{boast}, \emph{swift}, and \emph{meticulous}; all exhibit a significant post-release uptake (\figref{fig:appendix-youtube}B). ChatGPT's lexical signature is thus also audible in academic talks, consistent with the stricter spontaneous-speech result reported in the main text.

%% file: text/si_methods.tex
\subsection*{Datasets to compute Log Odds Ratios of human and LLM word usage}
\label{sec:si-gptscore-dataset}

\paragraph{arXiv}
The arXiv dataset contains abstracts from research papers that were published on the arXiv website. We used the arXiv API\footnotemark to extract 150 papers from five different categories, namely Computer Science, Electrical Engineering and Systems Science, Mathematics, Physics, and Statistics, each month from 2019 to 2022. All categories were further divided into 133 subcategories, and we gathered 7182 abstracts in total.
\footnotetext{\url{https://info.arxiv.org/help/api/index.html}}

\paragraph{bioRxiv}
The bioRxiv abstracts were gathered using the bioRxiv API.\footnotemark We ran a brute force query from the start to the end of the month for 4 years (2019-2022). We collected 60 papers per month, which was 2,880 papers in total.
\footnotetext{\url{https://api.biorxiv.org/}}

\paragraph{Nature}
Nature abstracts were collected using the search engine on the Nature website. The process involved querying for up to 20 pages, each displaying 50 results. Our goal was to collect between 7,000 and 10,000 abstracts to match a comparable size of our other datasets.
To achieve this, we ran a query without specific search terms, focusing on publications from 2019 to 2023. The results were sorted in two ways: ascending and descending per year. This dual sorting approach yielded over 8,000 unique abstracts, sufficient for our purposes.

\paragraph{Emails}
The email dataset was sourced from the publicly available Enron email dataset on Kaggle.\footnotemark The original dataset contains 500,000 emails generated by employees of the Enron Corporation. However, for our use case, we randomly sampled 10,000 emails and processed them into a dataset. The emails were sent between 2000 and 2001, far before the introduction of ChatGPT.
\footnotetext{\url{https://www.kaggle.com/datasets/wcukierski/enron-email-dataset}}

\paragraph{Essays}
This dataset comprises student essays collected from The Hewlett Foundation: Automated Essay Scoring challenge on Kaggle.\footnotemark The goal of this challenge was to develop an automated scoring algorithm for essays. All of the essays in the challenge, which was released in 2012, were composed by students. Similar to other datasets, we sampled 10,000 essays for our analysis.
\footnotetext{\url{https://www.kaggle.com/competitions/asap-aes}}

\paragraph{Wikipedia}
We utilized the Wikipedia API\footnotemark to pull the articles from Wikipedia. One restriction of the API is that it only gives the article title and article ID. Due to the existence of duplicate articles, we were forced to query 30,000 of them. We extracted the articles' dates of publication after eliminating duplicates, keeping just those that were released between 2019 and 2022. We used the page title to randomly collect the content of 10,000 articles from this chosen subset. 
\footnotetext{\url{https://www.mediawiki.org/wiki/API:Main_page}}

\paragraph{Podcast}
To test whether GPT word preferences extend to spontaneous spoken language, we reused the pre-ChatGPT podcast transcripts assembled for the observational analysis (\secref{sec:methods-datasets}). From English-language episodes published before ChatGPT's release (30 November 2022) that had an available transcript and a spontaneity score above the threshold used in the main analysis (\secref{sec:si-methods-spontaneity}), we randomly sampled 2,000 episodes and extracted one 300-word excerpt from each.

\subsection*{Spontaneity annotation and classification}
\label{sec:si-methods-spontaneity}

A central premise of our podcast analysis is that the corpus captures \emph{spontaneous} spoken language rather than read-out scripted material.
Speaker diarization can distinguish monologues from multi-speaker exchanges, but a multi-speaker show can still be scripted (e.g., a co-hosted news read-through or an interview from a written question list), and a monologue can still be spontaneous (e.g., extemporized commentary).
Following the register-theoretic distinction between ``oral'' and ``literate'' features of spoken language~\citep{biber2019}, and building on prior corpus work that operationalizes spontaneity via disfluencies~\citep{cho2014}, we therefore complemented diarization with an explicit annotation of how spontaneous each podcast sounds.
Disfluencies, such as filler words, repetitions and corrections, hesitations, and locally incorrectly used words, are taken as positive evidence of spontaneous production, since they reflect cognitive effort in language production and interaction management that does not arise when reading a script.

\subsubsection*{Annotation protocol}
\label{sec:si-methods-spontaneity-annotation}

Three coders were recruited from the research-assistant pool of the host institution and instructed via a written coding manual (reproduced in the appendix to the report on podcast annotation; available with the release materials).
They were blind to the hypothesis of the study and to the purpose of the annotation task.
The manual instructed them to attend only to disfluencies, and explicitly \emph{not} to language ability, monologue versus dialogue, post-production quality, or genre.
Each episode was rated on a four-point scale: 1 (\textit{clearly scripted}), 2 (\textit{rather scripted}), 3 (\textit{rather spontaneous}), and 4 (\textit{clearly spontaneous}).
A threshold of five disfluencies in a 30-second sample was adopted as the operational cut-off between scripted and spontaneous codes; clips with around four disfluencies were treated as edge cases, with the manual specifying tie-breaking rules across the three samples drawn from each episode.

Annotation proceeded through a custom web interface that presented one episode at a time as three 30-second clips taken from the beginning, middle, and end of the middle ten minutes of the recording.
This window avoids introductions, outros, and trailing music while still sampling the dynamics of the episode.
Episodes were shown in random order, independently per coder, and coders could flag individual clips for technical issues (e.g., non-English passages, audio artifacts), via a dedicated web interface (\figref{fig:annotation-interface}).
The annotation effort was split into a two-round calibration phase and a main labeling phase.
After the first calibration sample of 80 podcasts the interrater reliability was moderate-to-good (intraclass correlation coefficient ICC $= 0.79$, 95\% CI 0.69--0.86; two-way agreement model with average measures, computed via the \texttt{irr} package in R~\citep{gamer2010}), so the coders met to discuss low-agreement cases and the manual was refined (including the five-disfluency threshold).
A second calibration sample of 80 podcasts yielded ICC $= 0.85$ (95\% CI 0.76--0.91), at which point we moved to the main labeling sample.
The main sample consisted of 400 independently coded podcasts split across three batches (80, 160, and 160), used as the ground-truth set for classifier training and evaluation.
One episode was lost due to a technical issue, yielding a final main-sample size of 399 episodes; on this set, the interrater reliability was ICC $= 0.87$ (95\% CI 0.81--0.91), indicating excellent agreement under any of the standard guidelines for ICC interpretation.

\subsubsection*{Audio-based spontaneity classifier}
\label{sec:si-methods-spontaneity-classifier}

We used the 399-episode annotated set to train an audio classifier in the same manner as Elisha et al.~\citep{elisha2024} so that it predicts spontaneity directly from each podcast's middle window, allowing the spontaneity criterion to be applied to the full corpus without further manual labeling.

\paragraph{Feature extraction}
Each episode was resampled to 16~kHz mono and split into 25 contiguous 30-second chunks taken from the middle of the recording, matching the slice used during human annotation.
A log-mel spectrogram (128 mel bands $\times$ 3{,}000 frames) was computed per chunk and passed through the encoder of the \texttt{whisper-large-v3} speech model~\citep{radford2022}, yielding a $1500 \times 1280$ embedding tensor per chunk and a $25 \times 1500 \times 1280$ tensor per episode.
The Whisper encoder was kept frozen throughout; only the downstream classification head was trained.

\paragraph{Classifier architecture}
We used a compact multilayer perceptron (MLP) operating on individual chunk embeddings.
Each frame of the chunk embedding is projected via a linear layer (1280 $\rightarrow$ 8) followed by ReLU, expanded back via a second linear layer (8 $\rightarrow$ 128) followed by ReLU, average-pooled across the 1{,}500 time positions, regularized with dropout, and mapped via a final linear layer to four logits corresponding to the four annotation classes (clearly spontaneous, rather spontaneous, rather scripted, clearly scripted).

\paragraph{Training targets and procedure}
The three coders' ratings were converted into a per-episode soft label by placing each rater's vote on the four-class simplex and averaging, yielding a discrete probability distribution over the four classes per episode.
This soft target preserves the partial disagreement information that a hard majority vote discards (e.g., a 4/3/4 vote and a 4/4/4 vote get different targets).
Soft labels were broadcast from the episode level to every one of its 25 chunks for chunk-level training.
The MLP was trained with a soft-label cross-entropy objective on the 399-episode annotated set; we used a held-out split of the same set for early stopping and threshold selection, and report classification metrics (precision, recall, F1, balanced accuracy) along with the confusion matrix from the held-out split with the release materials.

\paragraph{Inference and episode-level score}
At inference time, the trained MLP emits a four-class softmax per chunk, and we aggregated the 25 chunk distributions into a single episode-level score in two complementary ways.
The \emph{binary vote} score uses each chunk's argmax: an episode's score is the fraction of chunks whose argmax falls in the two spontaneous classes (so the score lies in $[0, 1]$).
The \emph{weighted} score maps each chunk to the expected ordinal value $\sum_{k=1}^{4} k \cdot p_{k}$ under the chunk's predicted distribution and averages across the 25 chunks (so the score lies in $[1, 4]$, with higher values indicating more spontaneous production).
The weighted variant is the default used downstream; as below, the binary-vote variant was confirmed not to materially change which episodes are retained.

\paragraph{Filtering policy}
For all main analyses, an episode is retained for downstream word-frequency aggregation if (i) it passes the diarization-based dialogue filter (at least two distinct speakers and four or more alternating turns within the middle 10-minute window) and (ii) its weighted spontaneity score exceeds $3.0$.
The threshold of $3.0$ corresponds to the inflection point of the precision/recall curve on the held-out annotated split (approximately $90\%$ accuracy at the threshold) and roughly partitions episodes between the ``rather spontaneous'' and ``rather scripted'' labels.
The binary-vote analogue ($\geq 18$ of $25$ chunks classified spontaneous, equivalently a score of $0.72$) yields a near-identical retained episode set, confirming that the choice between the two aggregation rules does not materially affect the downstream corpus.

\subsection*{Word-sense audit}

The raw podcast counts tally every token whose stem matches a target word, irrespective of meaning. This over-states the signal of interest in two ways. The first is contamination by proper nouns, brands, and transcription artifacts that happen to share a stem. A count for \texttt{swift}, for instance, sweeps in Taylor Swift, Apple's Swift programming language, the SWIFT banking network, and NASA's Swift telescope, none of which is the adjective the word list is meant to track. 
The second is the presence of dictionary senses that one corpus essentially never produces yet that still inflate the raw match. 
To recover a count that reflects the intended sense, we audited each target word, retaining only occurrences that fall within a curated set of legitimate senses. This is a \emph{sense-validation} step and it is symmetric in human and machine usage. It removes contamination and unused senses, and the rule that selects which senses to keep never inspects the human-versus-GPT asymmetry, so it neither identifies nor conditions on the overuse effect the analysis later estimates.

The audit begins from the word-level log-odds ratios that rank words by their preference in GPT-rephrased over human text, and retains the most GPT-preferred words after two filters that protect signal quality and bound cost. We dropped stems present in more than 20\% of podcast episodes, which are too ubiquitous to carry a distinguishable shift, and stems above the 99th percentile of total podcast occurrence, whose per-occurrence auditing would be prohibitively expensive. For each surviving word we retrieved its inventory of dictionary senses from Wiktionary,\footnote{English Wiktionary: \url{https://en.wiktionary.org}} following inflected-form redirects so that a form such as \emph{underscored} resolves to the entry for \emph{underscore}. We then extracted paired example sentences from both sides of the comparison, drawing on the order of one hundred human and three hundred GPT usages per word, the larger machine sample reflecting the three rephrasings generated per source passage.

Sense assignment and occurrence filtering were carried out by an open-weight instruction-tuned language model, Qwen3-30B-A3B-Instruct-2507, served on a compute cluster.\footnote{Qwen3 instruction-tuned model: \url{https://huggingface.co/Qwen/Qwen3-30B-A3B-Instruct-2507}} In a first pass the model classified each example sentence into one of the word's Wiktionary senses, or into a \texttt{contamination} label reserved for proper nouns, transcription artifacts, and uses absent from the dictionary. Aggregating these assignments across a word's examples gives a per-sense distribution for each corpus, from which we selected the in-scope senses by an intersection rule at a fixed noise floor. A sense is kept only if both corpora attest it at least twice, and the \texttt{contamination} and parse-failure categories are always excluded. Requiring attestation in both corpora removes senses that one side essentially never produces---transcription artifacts and ghost-only senses---while the permissive two-occurrence floor accommodates the modest per-word example counts, so that genuine senses appearing only a handful of times are still retained. The retained senses for each word define its in-scope set. Every podcast occurrence of the word was then presented to the model under a binary prompt asking whether that occurrence belongs to the in-scope set, and occurrences judged out of scope were removed from the count.

Aggregating the per-word results yields the audited count matrix, which preserves the same 931{,}450-episode rows as the raw matrix and replaces each target word's column with its audited count. The audit typically removes a substantial share of the raw matches for contaminated words, as the brand, banking, and telescope senses of a word such as \texttt{swift} are filtered out.

These audited counts are the substrate for the synthetic-control estimation, the change-point model, and the figures, which are accordingly reported as the ``audited'' variants. The robustness analyses (\texttt{audited\_main} against \texttt{c1}--\texttt{c4}) vary only the synthetic-control specification on this shared audited substrate, with a single control run repeated on the raw un-audited counts to isolate the contribution of the audit itself.

\subsection*{Trends not linked to ChatGPT}

Applying the same change-point machinery without the top-1\% GPT-score filter recovers the twenty most credibly shifted words across the full Science \& Technology vocabulary (Supplementary \figref{fig:top20-credible-change}). This is a deliberately word-preference-blind view: candidates are ranked only by how confidently their post-release slope departs from zero, so any real change in how often a word is spoken can surface, whatever its cause.

The largest declines trace a single well-understood event. The four steepest falls are \emph{pandemic}, \emph{vaccine}, \emph{corona} and \emph{virus}, the core vocabulary of the COVID-19 period. These words peaked across 2020 and 2021 and were already receding when ChatGPT was released in late 2022, so the model reads their continued decline as a strong negative post-release slope. The shift is genuine, but its driver is the waning of a global news cycle rather than any property of language models. Neighbouring falls tell the same story of topical churn: \emph{twitter} declines as the platform rebrands and its salience drops, and case-study or subject-specific terms such as \emph{scholars}, \emph{reynolds} and \emph{urine} reflect the rotation of particular talks in and out of the corpus.

The rises are more mixed, and this is exactly the point of contrast with the main analysis. Some are the GPT signatures our study targets, led by \emph{delve} (the highest GPT score in the corpus), \emph{invaluable} and \emph{prompted}. Others are ordinary technical vocabulary riding the same 2022--2024 wave of interest in machine learning and tooling: \emph{gpu}, \emph{python}, \emph{vector}, \emph{queries} and \emph{transcript} all climb, yet their GPT scores sit near the middle of the distribution. A word can therefore change credibly for reasons that have nothing to do with LLM revision preferences, from shifting subject matter to the growth of an entire field.

This is why the main study conditions on GPT preference rather than on magnitude of change alone. The change-point detector answers the question ``which words moved?''; it cannot by itself distinguish an LLM fingerprint from a pandemic, a product launch or a hot research topic. Our design isolates the LLM channel by first restricting to words that GPT systematically over- or under-uses when revising human text, then asking whether those specific words shifted after the release. \figref{fig:top20-credible-change} shows what the unconditioned version looks like, and makes clear that the pandemic and technology-hype trends visible here are a separate phenomenon from the word-preference effect the paper measures.

\subsection*{Causal Identification}

\subsubsection*{GP smoothing of the input series}

We apply a Gaussian-process smoother (Mat\'ern $\nu = 2.5$, length scale $\ell = 720$ d, noise $\sigma = 0.05$) to the monthly input series before fitting the synthetic control. The GP is fit on the pre-treatment months only; post-treatment months are passed through verbatim from the raw monthly series, so that the smoother cannot leak post-release dynamics into either the donor-selection step or the SLSQP fit. The length scale is chosen as a compromise between the raw monthly series, on which the synthetic overfits noise, and longer length scales (e.g.\ $\ell = 1440$ d) at which it underfits the relevant low-frequency dynamics (Supplementary \figref{fig:length-scale-robustness}). The smoother enters only at donor selection and the synthetic control fit itself; all downstream change-point and placebo statistics are computed on the raw monthly series.

\subsubsection*{In-time placebo: Change-point date sweep}
\label{sec:si-methods-in-time-placebo}

To assess whether the post-release acceleration is anchored in time to ChatGPT's launch rather than to an arbitrary cut in the data window, we sweep the assumed change point across a monthly grid spanning the full observation window. At each candidate month $c$, we form a $24$-month pre-window $[c - 24\text{mo},\, c)$ and an $18$-month post-window $[c,\, c + 18\text{mo}]$, refit the per-word change-point regression of Eq.~\eqref{eqn:linear-model} on the synthetic-control gap, and record the cross-word mean of the post-release slope $\beta_{\text{Post}}$ over the top-$1\%$ GPT-score slice. To make the sweep tractable across hundreds of candidate dates, $\beta_{\text{Post}}$ at each candidate is estimated with an ordinary-least-squares proxy of the change-point regression; we validated the proxy against the Stan estimator at the true change point and obtained a Pearson correlation of $0.98$ across the top-$1\%$ slice.

We define the null over candidate change points whose entire $18$-month post-window strictly pre-dates ChatGPT's release, yielding $n = 28$ pre-GPT candidates; the reported permutation $p = 0.034$ follows the same rank convention as the in-space placebo and is bounded below by $1/29$.

\subsubsection*{Synthetic-control pre-trends check}

The change-point regression~\eqref{eqn:linear-model} estimates a pre-treatment slope $\beta_{\text{Pre}}$ on each word's treated$-$synthetic-control gap. The synthetic control is fit by minimizing pre-treatment RMSPE, which constrains the level of the gap but not its slope, so $\beta_{\text{Pre}}$ is not zero by construction and can serve as a pre-trends test. Across the Science \& Technology panel ($n = 3{,}535$ words), pre-slopes are tightly distributed around zero: median $|\beta_{\text{Pre}}| = 0.001$ versus median $|\beta_{\text{Post}}| = 0.015$. 

The synthetic control also reproduces each treated word's pre-release trajectory closely. Across the top-1\% GPT-score words, the pre-treatment RMSPE is small on both the smoothed time series and the raw monthly data (median $0.008$ and $0.038$ in $\log_{10}$-frequency units across all $3{,}535$ Science~\&~Technology words; Table~\ref{tab:pretreatment-rmspe}).

\paragraph{Conservative $\beta_{\text{Post}}$ bound.}

Per-word rankings and the ``credibly positive'' / ``credibly negative'' labels in Fig.~\ref{fig:change-point-model}A use a \emph{conservative} summary of $\beta_{\text{Post}}$ defined as the 95\% HDI limit nearest zero: the lower HDI bound when the posterior is credibly positive (HDI excludes zero from below), the upper HDI bound when credibly negative, and zero otherwise. The conservative bound is therefore a lower bound (in magnitude) on the slope change attributable to the release, and is the quantity by which we order the top-1\% panel in Fig.~\ref{fig:change-point-model}A.

\subsubsection*{Robustness checks}
\label{sec:si-methods-controls}

Detailed specifications for the four control variants introduced in the Methods; Table~\ref{tab:control-specs} gives the parameter view.

\paragraph{C1 --- bare $\ell_2$ matching.}  C1 replaces Main's $w$2$v$-then-$\ell_2$ donor pool with a strict $\ell_2$-nearest pool drawn from the full vocabulary, thereby removing both the semantic-neighbor exclusion and the GPT-score neutral band --- the two SUTVA-protecting filters that pre-screen the donor pool against interference.  C1 asks whether the headline survives even when the donor pool is not pre-screened.

\paragraph{C2 --- SLSQP on a tight ten-donor pool.}  C2 keeps Main's donor strategy and SLSQP weight fit but shrinks the pool from $L = 100$ to $L = 10$.  The smaller pool restricts the fit to the ten closest pre-treatment matches; it asks whether the SLSQP fit on a hundred-element basis exploits the flexibility the data does not warrant.  Because the in-space placebo procedure draws targets from the same donor pool, C2's ten-donor pool floors the empirical placebo $p$-value at $1/(10+1) \approx 0.091$.

\paragraph{C3 --- inverse-distance similarity weights.}  C3 keeps Main's donor pool ($L = 100$) but replaces the SLSQP convex fit with deterministic inverse-distance similarity weights $w_i = (1/(d_i + \varepsilon)) / \sum_k (1/(d_k + \varepsilon))$, where $d_i$ is the Euclidean distance between the smoothed pre-treatment trajectory of donor $i$ and that of the treated word and $\varepsilon = 10^{-12}$ is a numerical guard against $d_i = 0$.  The form follows CausalCite~\citep{agrawal2024causalcite}; it asks whether the headline survives a deterministic, optimization-free aggregation of the same donor pool.

\paragraph{C4 --- raw (un-audited) counts.}  C4 keeps the Main synthetic-control specification but replaces the audited counts substrate with the un-audited counts, retaining the ambiguity across different word meanings that the audited pipeline controls for upstream; it asks whether the upstream audit step drives the result.

The Main, C1, C2, C3, and C4 variants are rendered side-by-side for the \emph{delve} synthetic-control panel \figref{fig:delve-robustness-composite} and for the score--effect relationship \figref{fig:fig3-robustness-composite}.

\subsection*{Robustness to channel and episode outliers}
\label{sec:channel-outliers}

This channel-robustness analysis is based on 930{,}812 conversational episodes from 38{,}294 distinct shows (channels). Episodes per channel are right-skewed (median 5), but no single channel dominates. The largest contributes 0.46\% of episodes and the ten largest 1.4\% (\figref{fig:channel-outliers}A).

Word-level usage is more concentrated, as expected, but robust to outliers. Within Science \& Technology, excluding the ten channels that use \emph{delve} most leaves the post-release increase essentially intact: over the post-adoption window (months 13--18 after release) the token-normalized rate of \emph{delve} rises by a comparable factor with and without those channels (a factor of 1.5 in both cases; \figref{fig:channel-outliers}B). The shift reflects broad adoption across many channels rather than a few high-usage outliers.

The score--effect relationship is likewise not an artifact of a single word: excluding \emph{delve}, the Gaussian-process fit of per-word effect against GPT-score rank is essentially unchanged, and the top-$X\%$ slice-mean $\beta_{\text{Post}}$ remains above the permutation null at the high-GPT-score tail (top $1\%$ and top $2\%$: permutation $p = 0.001$; \figref{fig:score-effect-delve-out}).

\subsection*{Experimental implementation}
\label{sec:si-methods-experiment}

\subsubsection*{Participant demographics}

Table~\ref{tab:demographics} reports the full demographic breakdown of the final sample ($N = 496$).
Participants were recruited via Prolific between April 16 and April 20, 2026.
The sample was 50\% women, 48\% men, and 2\% non-binary or other; mean age was 40.5 years ($SD = 13.2$, range 18--65+; age collected in brackets).
91\% reported English as their first language.
Prior AI chatbot use: 96\% reported having used an AI chatbot; the most commonly reported systems were ChatGPT, Google Gemini, and Microsoft Copilot.
All participants provided informed consent and were compensated at a rate of \pounds11.4 per hour.

\subsubsection*{Power analysis}

We determined the sample size with an a priori power analysis.
Based on pilot data and prior literature, we targeted a small-to-medium effect of $d = 0.30$.
With two-tailed $\alpha = .05$ and power $= .80$, the required $N$ per cell was $88$, for a total of $N = 352$ across the 4 between-subjects cells.
We recruited above the required ($N = 500$), and after exclusions, the final sample was $N = 496$.

\subsubsection*{Synonym pair selection}

Synonym pairs were selected to satisfy two criteria: both forms are familiar, commonly used English words with no strong register or formality differences between them, and the concept can be unambiguously depicted in a cartoon illustration, enabling the referential image-guessing paradigm.
The 18 selected pairs span three lexical categories (nouns, verbs, adjectives) and are divided into two groups of nine pairs each.
Table~\ref{tab:synonym-pairs} lists all pairs and the accepted surface forms used for scoring.

\subsubsection*{System implementation}
\label{sec:si-methods-experiment-implementation}

The experiment was delivered as a custom web application (TypeScript/React~18 frontend; Python/FastAPI backend) hosted on a cloud server.

\paragraph{Stimulus generation.}
Each trial image was created as a matched treatment--control pair via a two-stage generative pipeline.
First, a language model (\texttt{gpt-5.2}) was prompted to generate a cartoon-style illustration description in which the three target synonym variants (one noun, one verb, one adjective) were semantically necessary to describe the depicted scene (treatment prompt), together with a control prompt obtained by sentence-level substitutions that removed this requirement while preserving the overall scene composition.
Second, an image generation model (\texttt{gemini-3-pro-image-preview}) rendered the treatment image from the treatment prompt, then received the treatment image together with the substitution instructions to produce the visually matched control variant.
The image generation model was instructed to never display the target words as text within any image.
The generated candidates were rated by the authors in terms of visual quality and consistency using a custom web-based annotation tool; only images receiving the highest rating were retained as final stimuli.

\paragraph{AI chatbot interaction.}
Chatbot responses were generated by calling the OpenAI API (\texttt{gpt-4o}); each call included the full within-trial chat history, the target image (no control image was provided), and additional context (see \figref{fig:prompt-base} and \figref{fig:prompt-vocab-rule} for details).
The system prompt was designed to ensure that the model consistently used the designated canonical synonym variant throughout the interaction.
It comprised three layers: (i)~a base game prompt establishing the image-guessing task and the model's role as an honest co-player; (ii)~a per-trial vocabulary rule specifying the canonical form for each target word in the current trial, paired with treatment and control scene descriptions that anchored each term to specific image content, so the model would apply the rule even when a participant's question referenced the scene without using the target word directly; and (iii)~a global vocabulary rule listing all 18 synonym pairs as remapping instructions, maintaining consistent usage across the entire interaction regardless of which image was shown or which words participants used.
Two intervention groups were counterbalanced: chatbot in Group~A used Variant~1 as the canonical, and Group~B used Variant~2.

\paragraph{Speech transcription.}
Spoken descriptions were captured via the browser and transcribed by the standard Web Speech API.
Upon stopping, the audio and the automatic transcript were uploaded to cloud storage; recordings shorter than 3 seconds or transcripts shorter than 5 words were rejected with a prompt to re-record.
For the Test Phase, the automatic transcript was displayed to the participant in an editable text area, together with audio playback.
Participants were asked to correct any recognition errors and submit, ensuring that the primary outcome measure was scored from human-verified transcripts.

\paragraph{Scoring and statistical analysis.}
For each synonym pair, we pre-registered a coding scheme listing the accepted surface forms of each variant, including morphological inflections (e.g., \textit{fix, fixes, fixed, fixing}), compound and multi-word constructions (e.g., \textit{to put up}), and spelling variants (e.g., \textit{multicolored, multicoloured, multi-colored, multi-coloured}).
Detection proceeded by tokenizing the transcript and each accepted phrase on alphabetic characters and stemming both with the NLTK \texttt{SnowballStemmer} (English), with a phrase matching whenever its stemmed token sequence appeared as a contiguous subsequence of the stemmed transcript; this yielded a binary trial-level indicator of whether the AI-primed variant was used.
From these indicators, the primary outcome measure for each phase $\Delta p$ was calculated as the per-participant mean difference in usage rate between the AI-primed and alternative variants across the nine trials.
Usage in the Interaction Phase and Test Phase was each assessed with a participant-level Monte Carlo permutation test (10{,}000 iterations); forced-choice selections were assessed with a sign-flip test against the 50\% chance baseline.
Confidence intervals were obtained by cluster-bootstrap resampling participants (10{,}000 iterations); all tests were two-tailed at $\alpha = .05$.

\subsubsection*{Participant flow details}

After completing a comprehension check and microphone test, participants proceeded through the Interaction Phase (12 image-guessing + spoken description trials), a 3-minute distractor task, the Test Phase (3 spoken description trials, no chatbot), an open-ended deception check (``Have you noticed anything about the language of the chatbot?''), the Forced-Choice Phase (9 two-alternative label selections), and a final questionnaire.
The detection check question was placed between the Test Phase and the Forced-Choice Phase so that responses reflected impressions formed during the interaction without being coloured by the explicit synonym-choice framing of the Forced-Choice task.
Open-ended responses were reviewed qualitatively; responses were coded as ``aware'' if they named one or more specific synonym pairs from the experiment or explicitly noted that the chatbot substituted the participant's word with a different one.
The final questionnaire collected demographics (age range, gender, education), language background (English as L1, age of acquisition), prior AI-chatbot use (systems used, frequency, purposes), and self-reported color vision deficiency  (used as the basis for the post-hoc exclusion reported in Methods).

\subsection*{Theoretical model}

We model linguistic preference diffusion using a noisy voter model~\citep{castellano2009} on a Watts--Strogatz small-world network ($N = 2000$ speakers, degree $k = 6$, rewiring probability $\beta = 0.1$). Each agent holds a binary state representing a preferred synonym variant, initialized i.i.d.\ from Bernoulli($p_0 = 0.1$). Each generation consists of $N$ asynchronous update steps. At each step a randomly selected agent copies the state of a randomly chosen neighbor; with probability $\varepsilon = 0.005$ it instead resets to a draw from Bernoulli($p_0 = 0.1$), preventing absorbing consensus and producing a stationary distribution centered on $p_0$. 

A single hub, modeled as a committed source (zealot) fixed in state 1, is added as an extra neighbor to a uniformly random fraction $f \in \{0.5\%, 2\%, 5\%, 10\%\}$ of the population. As a control, a single committed speaker (zealot) occupies one network node with its natural $6$ connections. After $100$ generations without the zealot, the simulation continues for $300$ generations with the zealot active. Results are averaged over $50$ independent runs, each with a fresh network and initial state.

%% file: text/si_figures.tex
\begin{figure}[H]
    \centering
    \includegraphics[width=\linewidth]{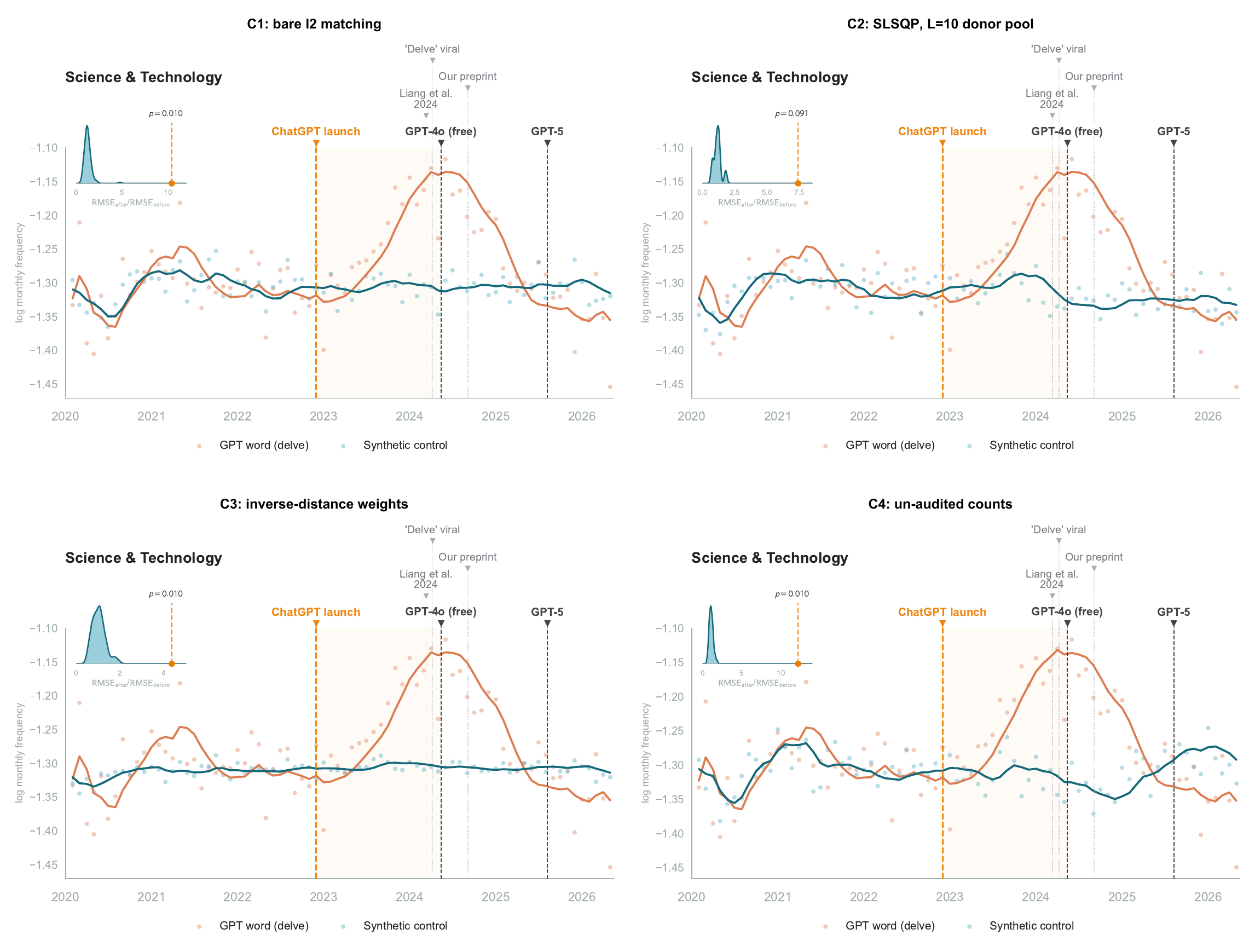}
    \caption{\textbf{The post-release gap between \emph{delve} and its synthetic control is preserved across all four robustness check specifications.} To test whether the result depends on specific donor-selection and aggregation choices, we re-ran the synthetic-control pipeline under four control specifications (see Methods and Supplementary Table~\ref{tab:control-specs}). In-space placebo $p$-values: $p_{\text{C1}} = 0.050$, $p_{\text{C2}} = 0.091$ (floor), $p_{\text{C3}} = 0.040$, $p_{\text{C4}} = 0.010$ (compare $p_{\text{Main}} = 0.010$). The main result is robust to changes in donor selection (C1, C2), weight aggregation (C3), and the counts substrate (C4).}
    \label{fig:delve-robustness-composite}
\end{figure}

\begin{figure}[H]
    \centering
    \includegraphics[width=0.95\linewidth]{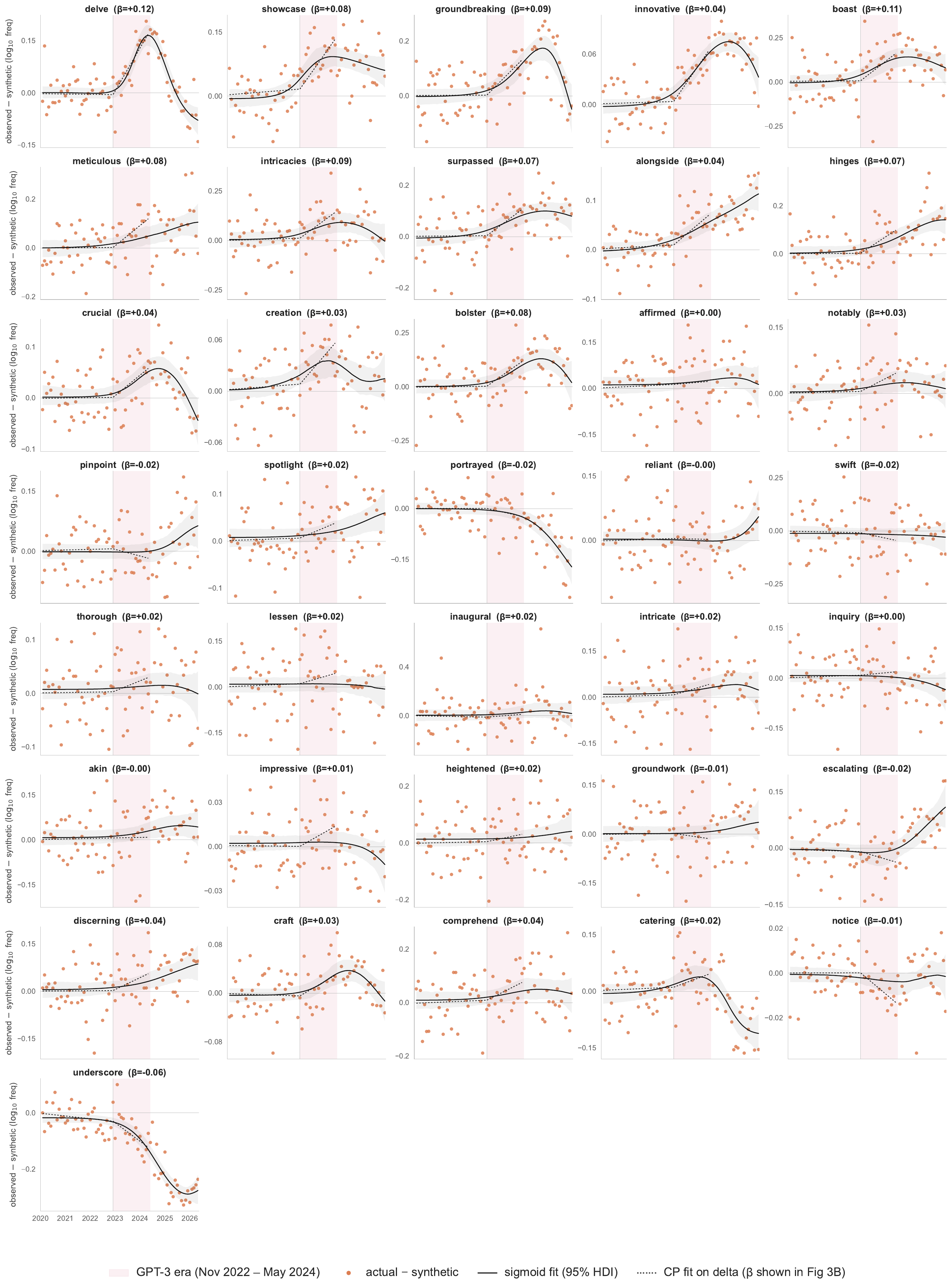}
    \caption{\textbf{Full top-1\% GPT-score panel grid.} Monthly observed $-$ synthetic $\log_{10}$ frequency (orange points) for every word in the top 1\% of GPT scores ($n = 36$), with the double-sigmoid posterior smoother (solid; 95\% HDI shaded) and the change-point fit of Equation~\ref{eqn:linear-model} (dashed) overlaid. Of the 36 words, 28 show a positive post-release slope and 13 are credibly so (95\% HDI on $\beta_{\text{Post}}$ excludes zero). The twelve panels reproduced in Fig.~\ref{fig:change-point-model}A are the largest-magnitude credible subset.}
    \label{fig:gpt-top1pct-panels}
\end{figure}

\begin{figure}[H]
    \centering
    \includegraphics[width=0.9\linewidth]{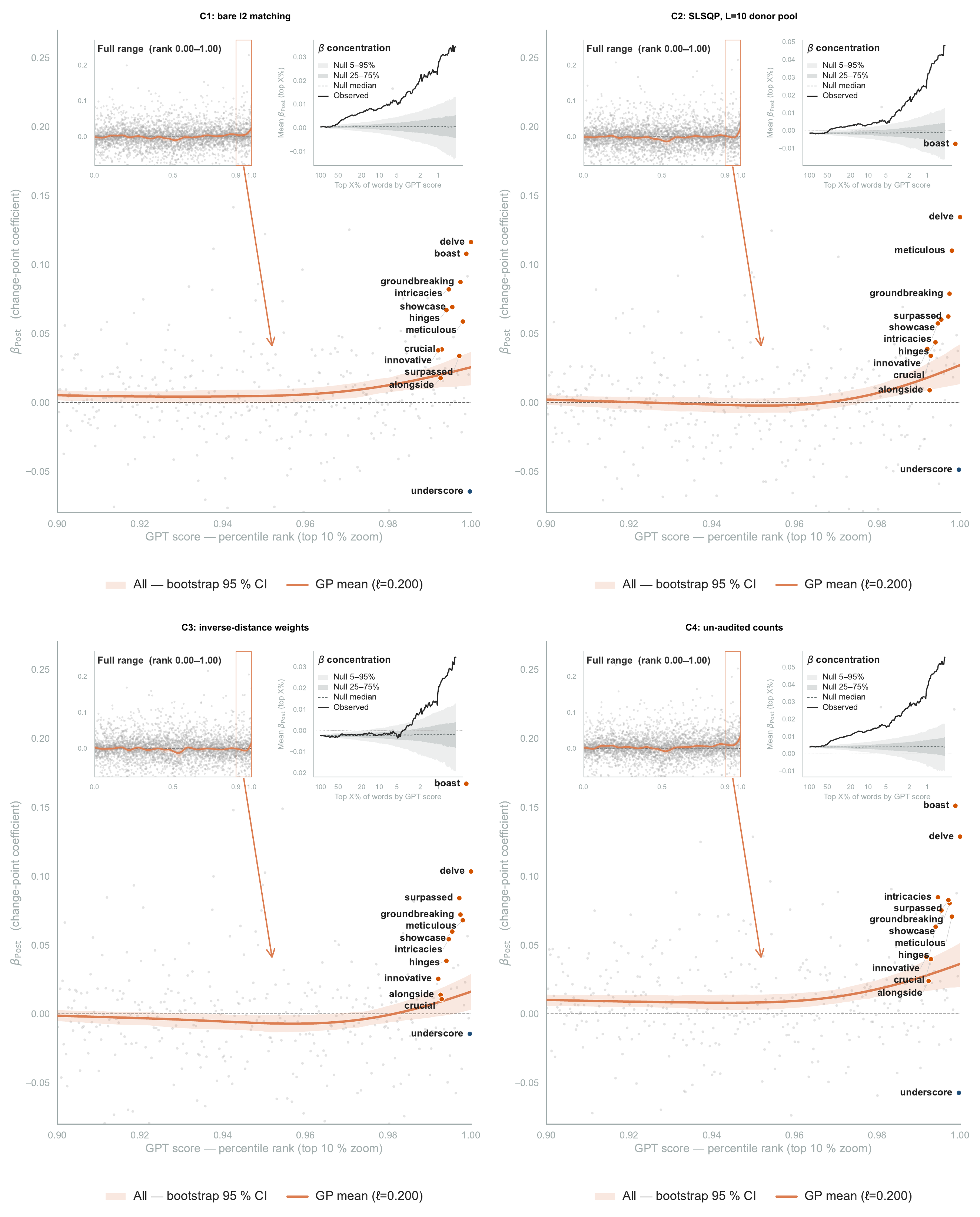}
    \caption{\textbf{The score-graded acceleration in the high-GPT-score tail is preserved across all four robustness controls.} To test whether the score--effect relationship depends on specific donor-selection and aggregation choices, we re-ran the synthetic-control pipeline under four control specifications (see Methods and Supplementary Table~\ref{tab:control-specs}). Each panel reproduces Fig.~3B for one control: per-word $\beta_{\text{Post}}$ against GPT-score percentile rank with the Gaussian-process posterior mean overlaid (bootstrap 95\% CI shaded); the left inset shows the full rank-axis range, and the right inset the slice-mean $\beta_{\text{Post}}$ over nested top-$X\%$ slices against the permutation null. The twelve labeled words are Main's top 12 by $|95\%$-HDI bound on $\beta_{\text{Post}}|$, plotted at each control's own positions so the same exemplars can be tracked across specifications. The slice-mean exits the permutation null at the high-GPT-score tail under every control.}
    \label{fig:fig3-robustness-composite}
\end{figure}

\begin{figure}[H]
    \centering
    \includegraphics[width=\linewidth]{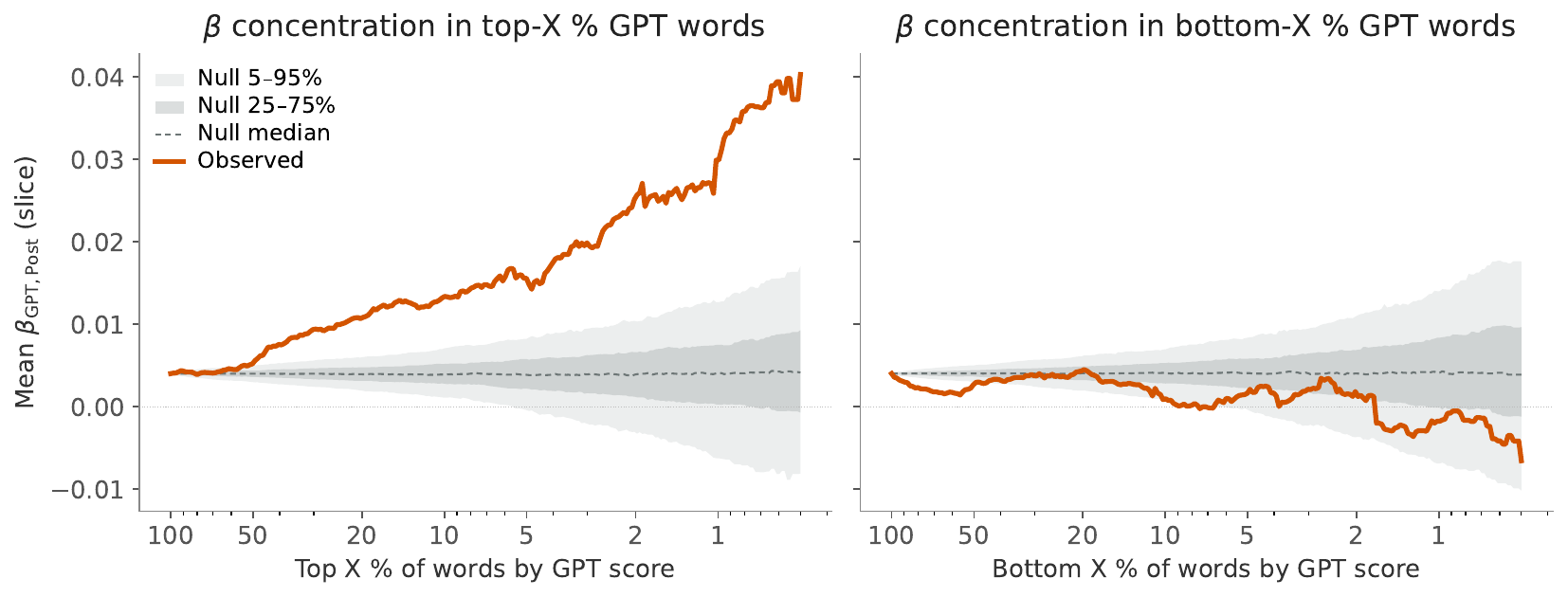}
    \caption{\textbf{Words favored by ChatGPT rise in spoken use, but words it disfavors do not show a matching decline.} Each panel sweeps a cut-off $X$ across the top (\textbf{left}) or bottom (\textbf{right}) of the GPT-score distribution. The orange line shows the average post-release change in usage (the change-point slope $\beta_{\text{Post}}$) for the $X\%$ of words at that end of the ranking; the grey band is the 5th--95th percentile range expected by chance, obtained by reshuffling GPT scores across words. \textbf{Left:} as the cut-off tightens onto the words ChatGPT most strongly prefers, the orange line rises clearly above chance --- at the top $1\%$ of words the mean ($+0.030$) is more than $2\times$ the upper bound of chance ($+0.013$). \textbf{Right:} the matching check on the words ChatGPT most strongly disfavors stays close to zero at every cut-off (at the bottom $1\%$: $-0.002$, chance range $[-0.007, +0.014]$).}
    \label{fig:score-slice-symmetric}
\end{figure}

\begin{figure}[H]
    \centering
    \includegraphics[width=\linewidth]{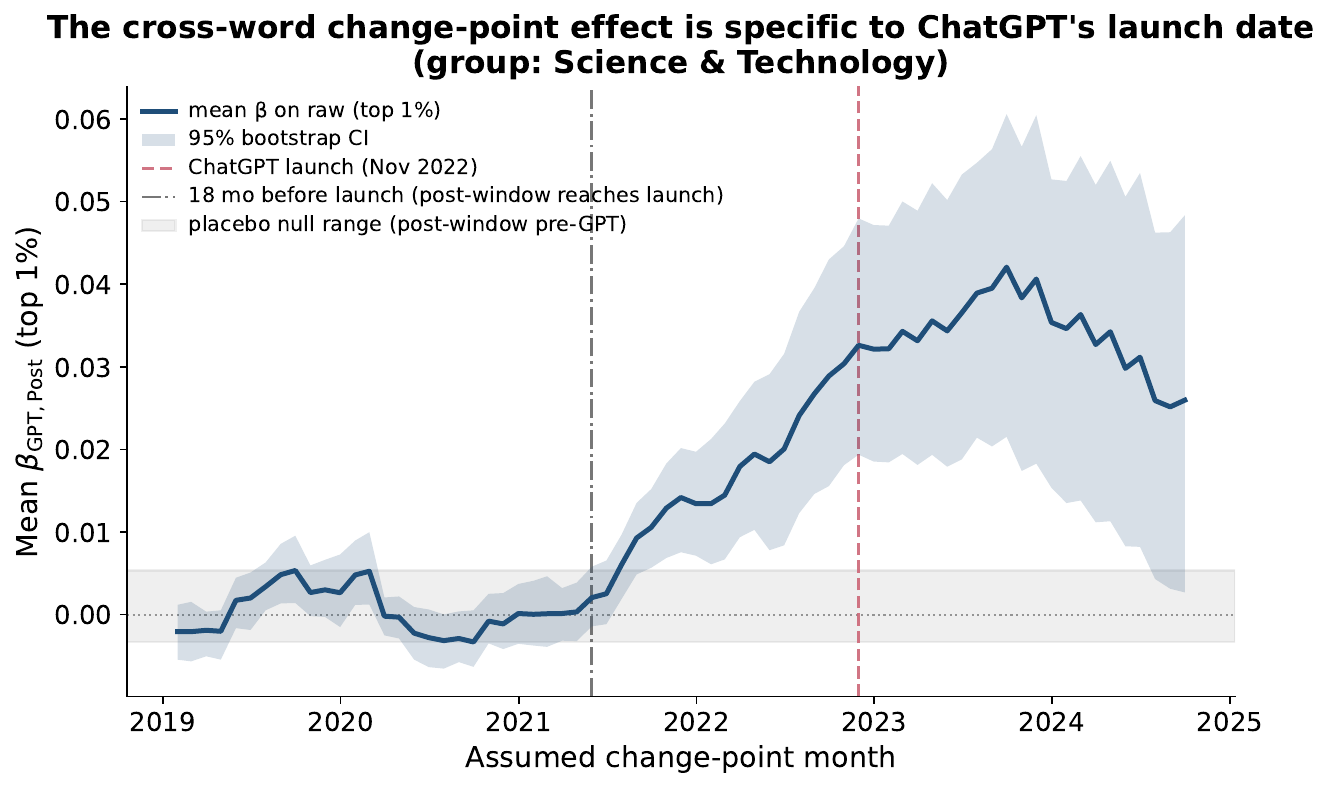}
    \caption{\textbf{The cross-word change-point effect is specific to ChatGPT's launch date.} Mean post-release slope $\beta_{\text{Post}}$ across the top 1\% of GPT-score words (Science \& Technology, $n = 36$) as the assumed change-point date slides month-by-month over the data window, with a 24-month baseline and an 18-month post window held constant. Grey band: placebo-null envelope of mean $\beta$ across candidate dates whose post-window is entirely pre-GPT. Indigo: 95\% bootstrap CI of the mean over words. Dashed red: true launch (2022-11-30). Mean $\beta$ stays inside the null at every pre-GPT candidate and steps up only once the window includes the launch. Permutation $p = 0.034$.}
    \label{fig:changepoint-date-sweep}
\end{figure}

\begin{figure}[H]
    \centering
    \includegraphics[width=\linewidth]{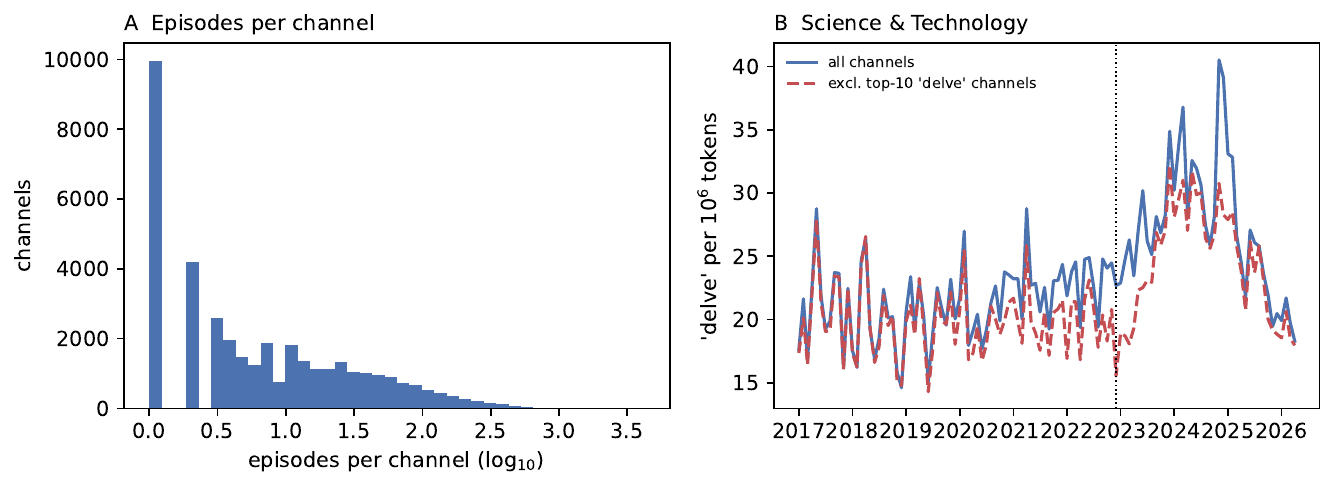}
    \caption{\textbf{Channel and episode outliers do not drive the podcast trends.} (A) Number of episodes per channel, across 38{,}294 podcast channels (log scale); the median channel contributes 5 episodes and the largest 0.46\% of all episodes. (B) Token-normalized monthly rate of \emph{delve} in Science \& Technology for all channels (solid) and after removing the ten channels with the most \emph{delve} usage (dashed); the dotted line marks the ChatGPT release. The post-release increase is essentially unchanged with and without those channels (fold $\approx$1.5 in both cases), so the trend is not driven by a few high-usage channels.}
    \label{fig:channel-outliers}
\end{figure}

\begin{figure}[H]
    \centering
    \includegraphics[width=\linewidth]{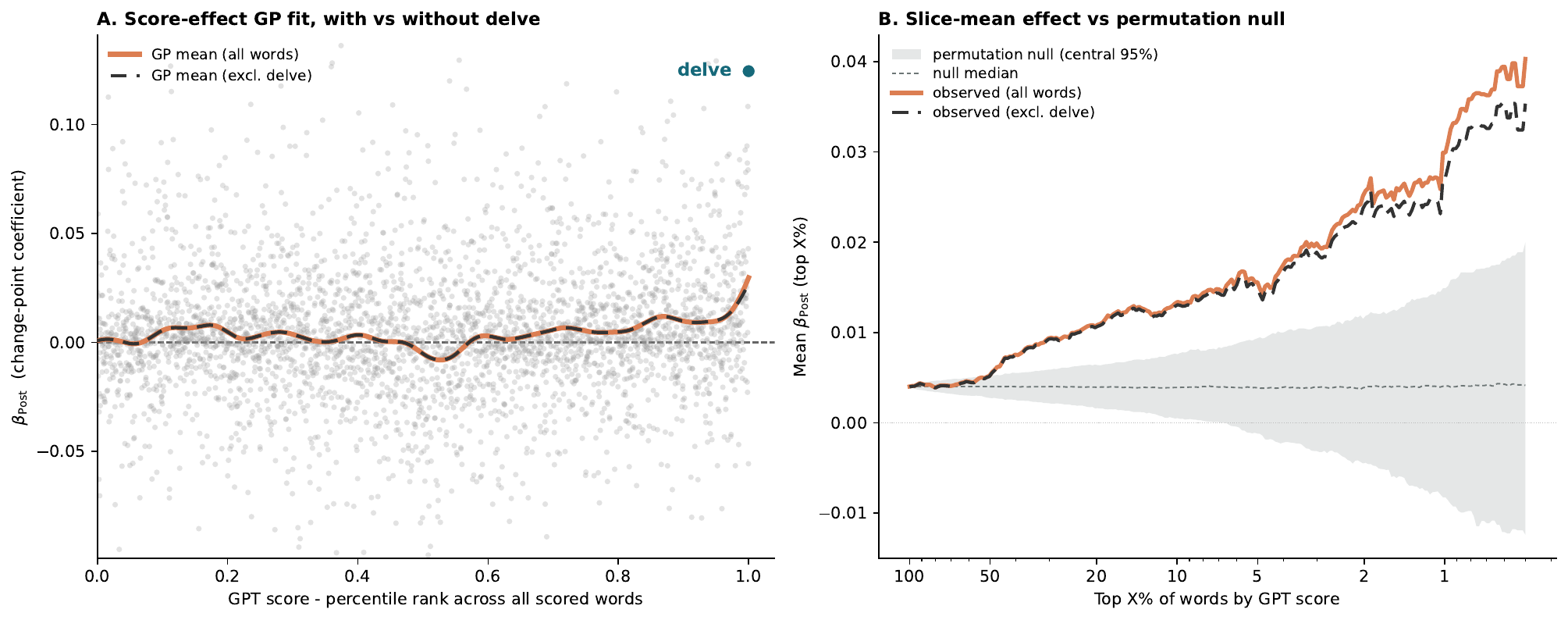}
    \caption{\textbf{The score-graded post-release acceleration is not an artifact of \emph{delve}.} Both panels are computed on the audited Science \& Technology substrate ($n = 3{,}535$ words). (\textbf{A}) Per-word change-point coefficient $\beta_{\text{Post}}$ against GPT-score percentile rank (grey scatter, all $3{,}535$ words). The Gaussian-process posterior mean of Fig.~\ref{fig:change-point-model}B is overlaid twice: over all words (solid) and with \emph{delve} removed (dashed). The two curves are visually indistinguishable, and both turn upward only in the high-GPT-score tail; \emph{delve} is marked at rank $1.0$. (\textbf{B}) Slice-mean $\beta_{\text{Post}}$ over nested top-$X\%$ GPT-score slices against a permutation null obtained by shuffling $\beta_{\text{Post}}$ across words while holding the GPT-score order fixed ($1{,}000$ permutations; central-$95\%$ band and median shaded). The observed slice-mean is drawn for all words (solid) and with \emph{delve} excluded (dashed); both rise far above the null at the high-GPT-score tail (top $1\%$: mean $\beta_{\text{Post}} = 0.030$ over $36$ words vs.\ $0.027$ excluding \emph{delve}, permutation $p = 0.001$ for both; top $2\%$: $0.025$ over $71$ words vs.\ $0.024$, $p = 0.001$ for both). The score--effect relationship therefore exceeds chance with and without \emph{delve}.}
    \label{fig:score-effect-delve-out}
\end{figure}

\begin{figure}[H]
    \centering
    \includegraphics[width=0.85\linewidth]{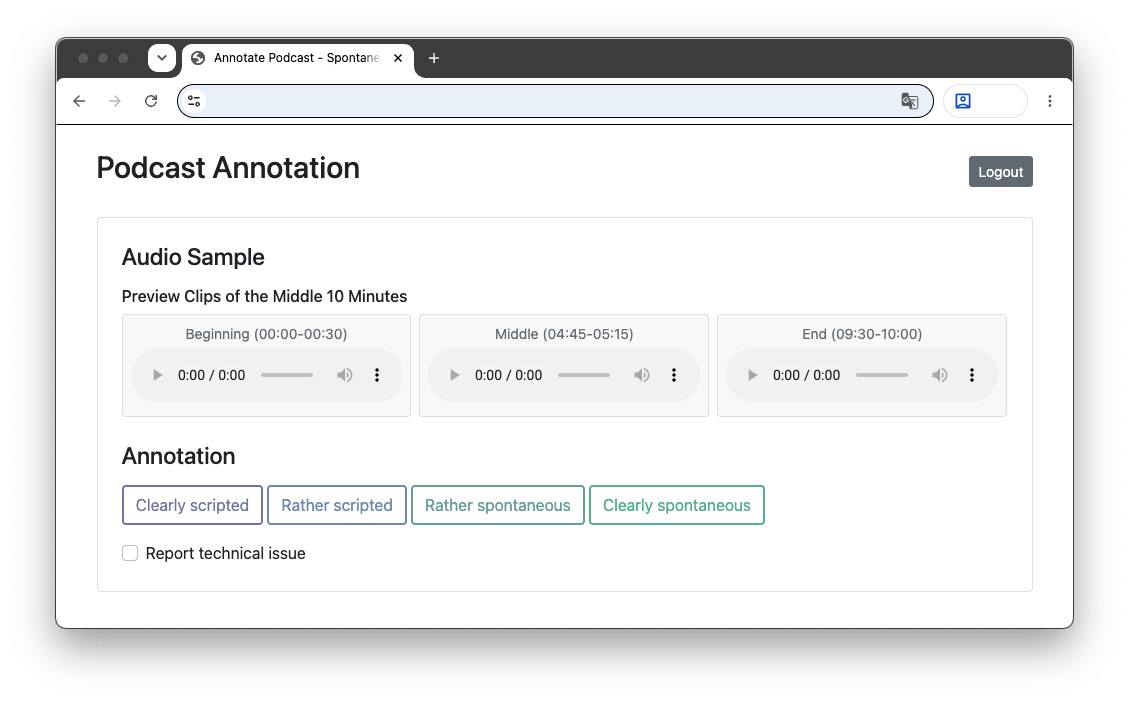}
    \caption{\textbf{Web-based interface used for the spontaneity annotation task.} Each coder heard three 30-second clips sampled from the beginning, middle, and end of the episode's middle 10-minute window (top row) and rated the episode on a four-point scale from \textit{clearly scripted} to \textit{clearly spontaneous} (bottom row). Coders were blind to the study hypothesis and to the purpose of the annotation task.}
    \label{fig:annotation-interface}
\end{figure}

\begin{figure}[H]
    \centering
    \includegraphics[width=\linewidth]{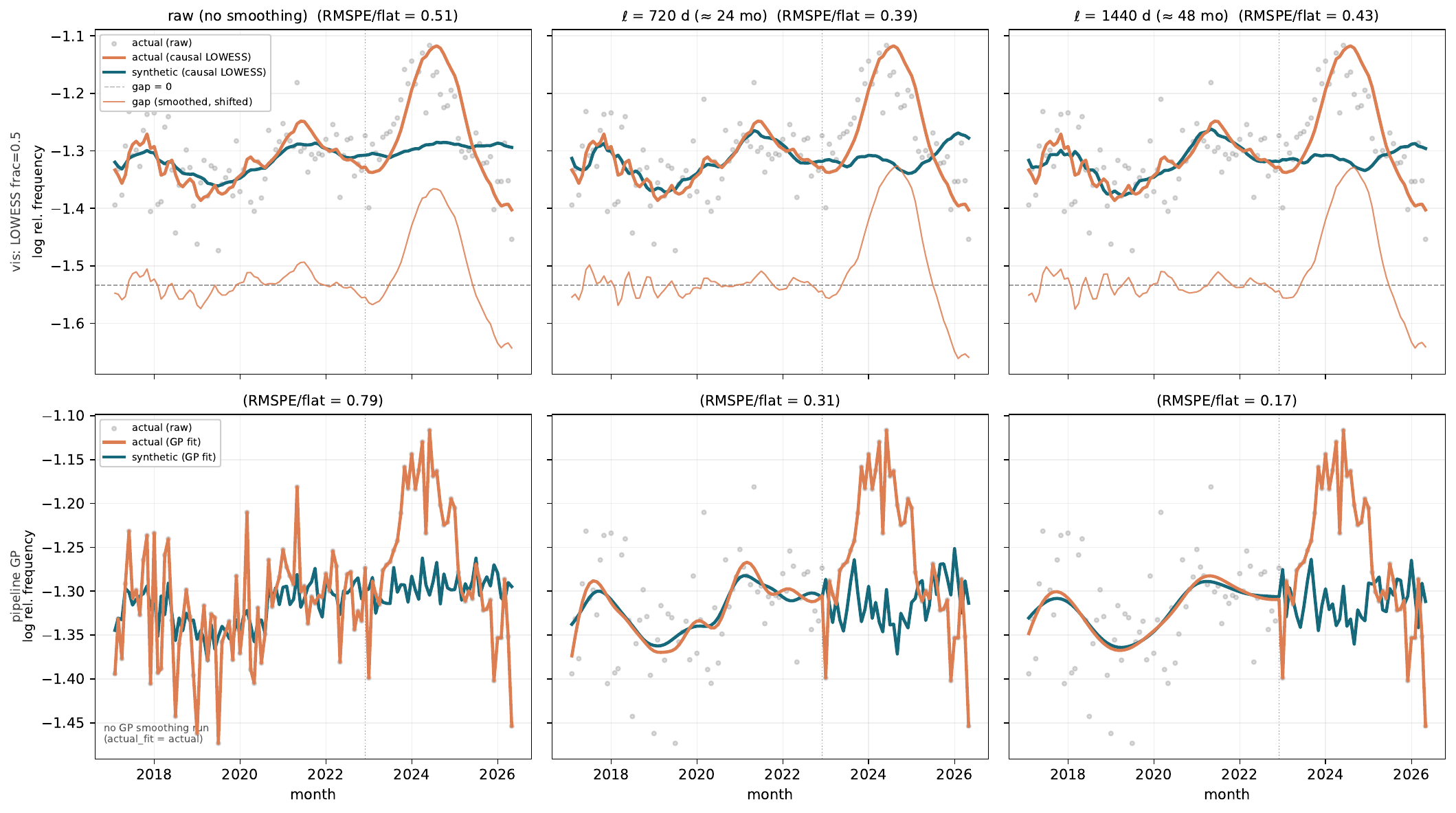}
    \caption{\textbf{Choice of GP smoother length scale for synthetic-control donor selection.} $2 \times 3$ grid showing actual $-$ synthetic across three input-series GP smoothing length scales (columns: $\ell$ = raw, $720$ d, $1440$ d) under two visualisation smoothers (top: causal LOWESS as in Fig 1A; bottom: pipeline GP fit). Without smoothing (left), the synthetic overfits monthly noise; at $\ell = 1440$ d (right), it underfits the low-frequency dynamics. $\ell = 720$ d (middle, the pipeline default) is chosen as a compromise. Smoothing enters only at donor selection and the synthetic control fit; all downstream change-point and placebo statistics are computed on the raw monthly series.}
    \label{fig:length-scale-robustness}
\end{figure}

\begin{figure}[H]
    \centering
    \includegraphics[width=\linewidth]{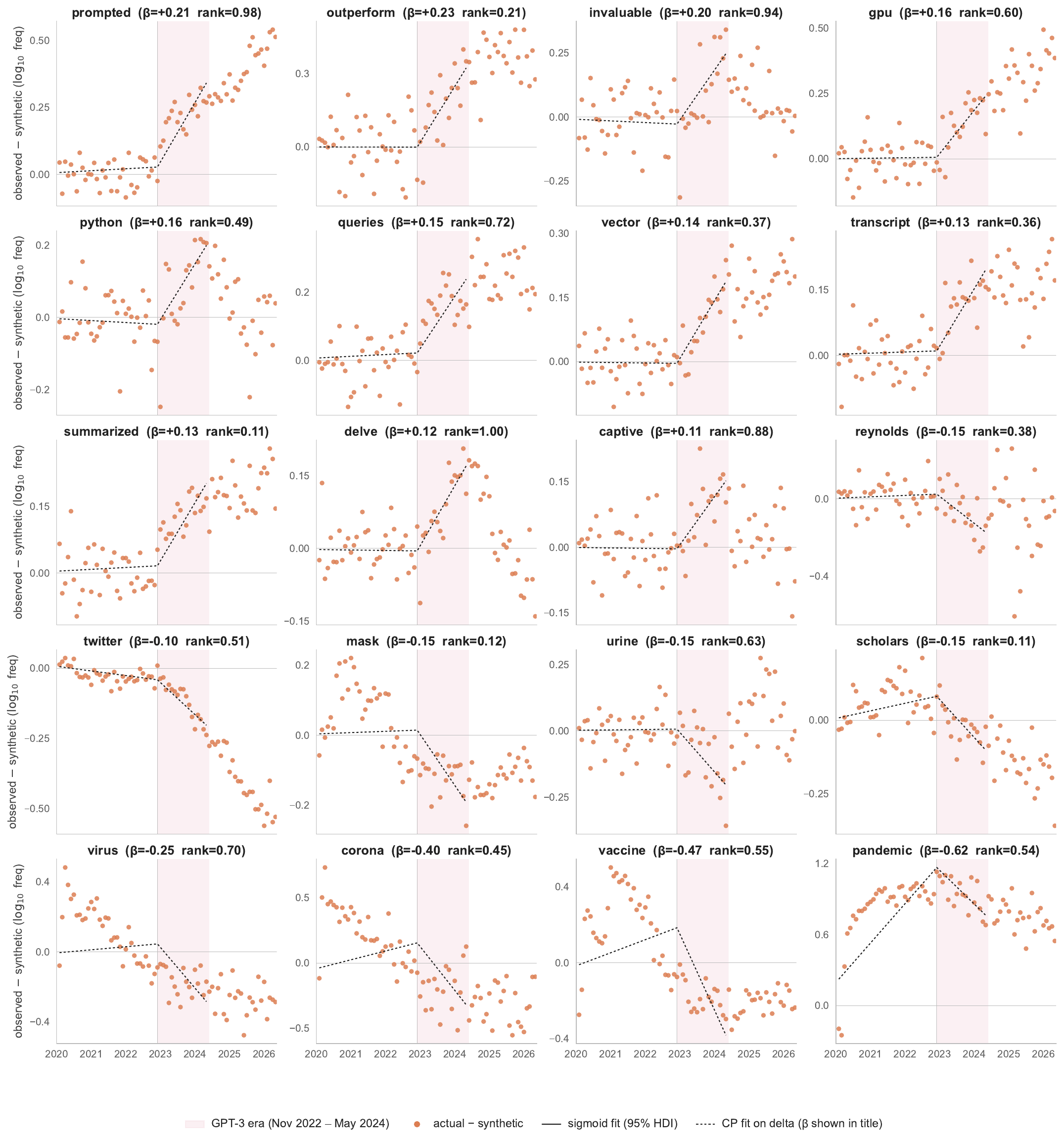}
    \caption{\textbf{The 20 words with the most credible post-release shift (Science \& Technology podcasts, $n = 3{,}535$ stems) without reference to GPT preference.} Each panel: monthly observed $-$ synthetic $\log_{10}$ frequency (orange scatter), double-sigmoid posterior smoother (solid, 95\% HDI shaded), and the change-point fit of Equation~\ref{eqn:linear-model} (dashed). Panels are ordered by the signed conservative bound on $\beta_{\text{Post}}$ (the 95\% HDI limit nearest zero). Per-panel GPT-score percentile rank is annotated in the title. The shaded vertical band marks the period of analysis between ChatGPT's launch and the launch of GPT-4o (free).}
    \label{fig:top20-credible-change}
\end{figure}

\begin{figure}[H]
    \centering
    \includegraphics[width=\linewidth]{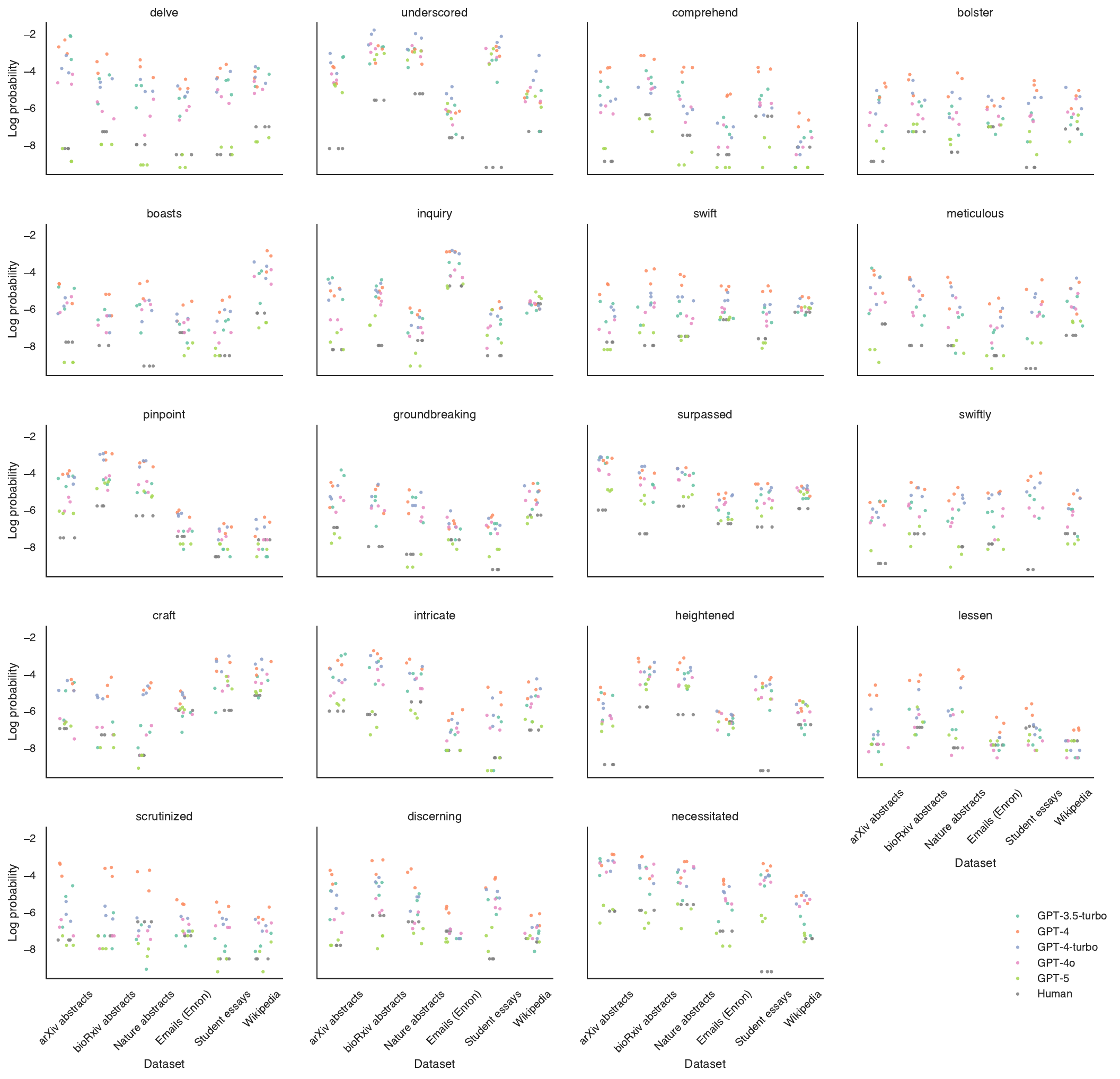}
    \caption{\textbf{Log probabilities of human and LLM-revised text.} We calculated the log-probability of a word appearing in human-authored text and its appearance in a version of the same text revised by different LLMs. Each colored point represents the log-probability for a specific combination of model, dataset, and prompt. The log-probability for the original human-authored text is shown in gray. Some LLM calls failed due to various reasons, such as policy violations. Consequently, the corresponding human-authored texts were removed from the dataset, introducing slight variations in the associated probabilities, even though the source dataset remained identical.}
    \label{fig:word_log_probability_by_metric}
\end{figure}

\begin{figure}[H]
    \centering
    \includegraphics[width=\linewidth]{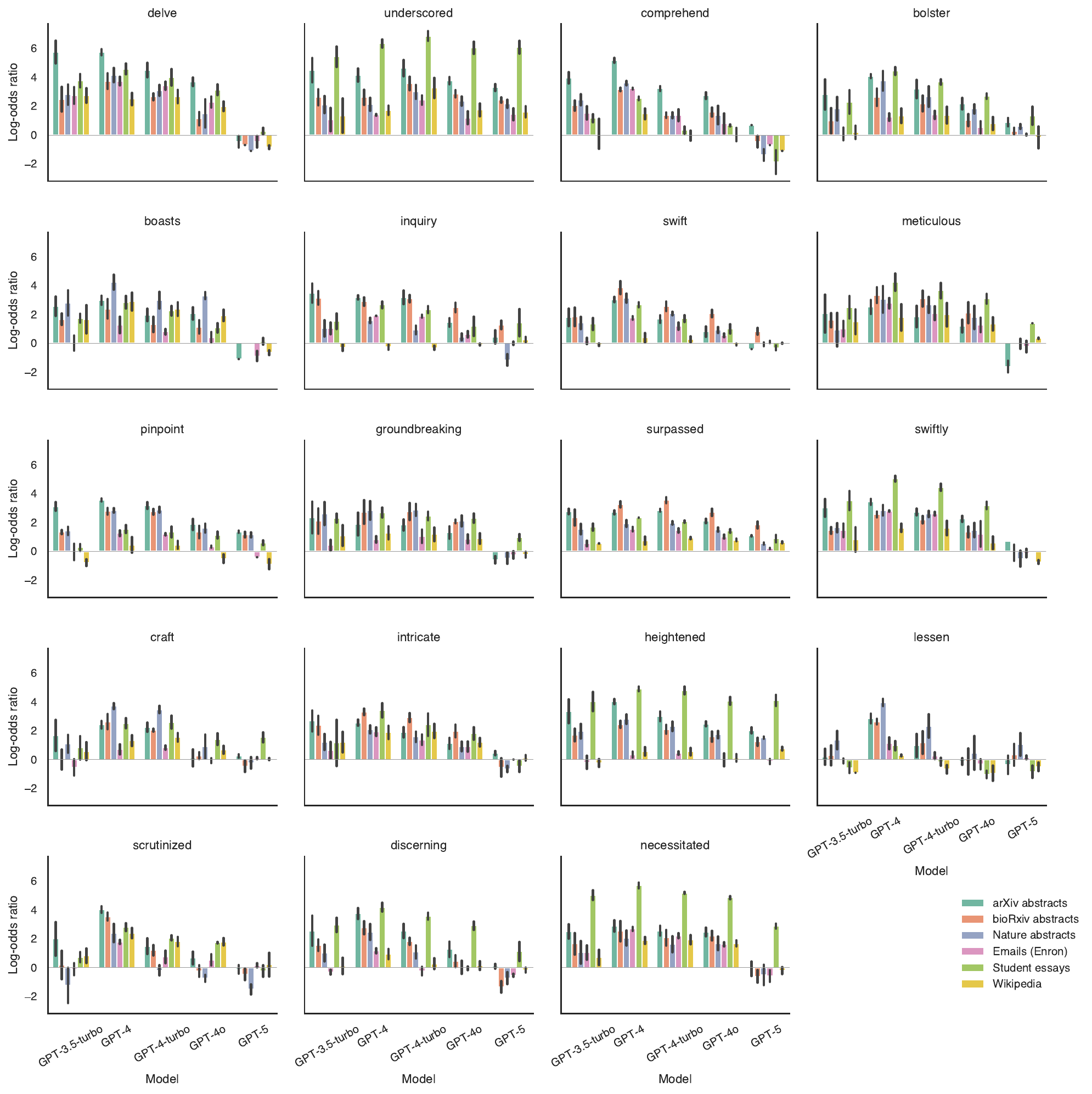}
    \caption{\textbf{Log-Odds ratios (LORs) of words in human vs. LLM-revised text.} We calculated the LOR of a word appearing in human-authored text compared to its appearance in a version revised by an LLM. Displayed here are the 19 words with the highest average LOR across all datasets, models, and prompts. The data are stratified by dataset and model, with error bars representing the standard error associated with the three prompts analyzed.}
    \label{fig:word_log_odds_ratio_by_dataset}
\end{figure}

\begin{figure}[H]
    \centering
    \includegraphics[width=\linewidth]{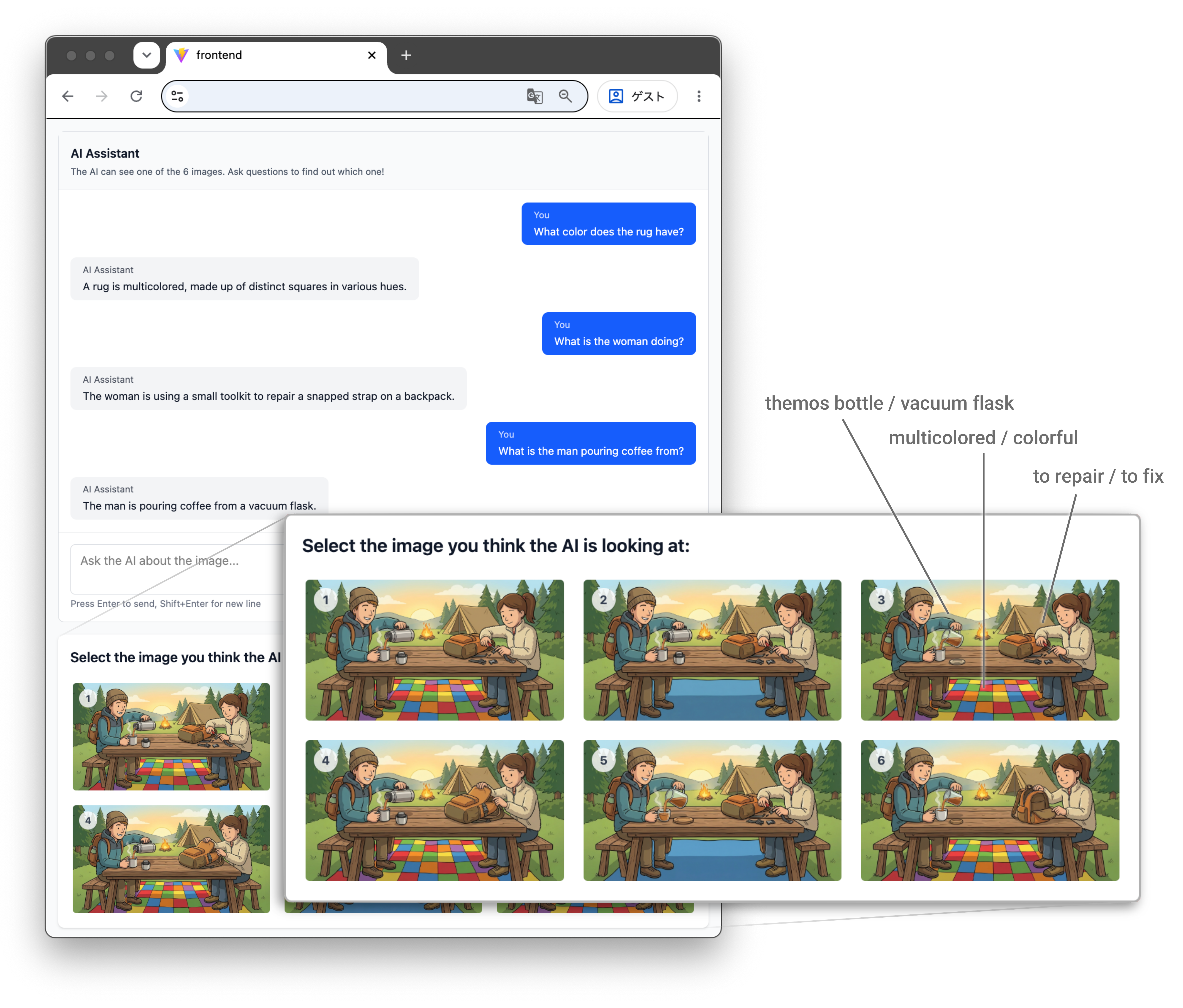}
    \caption{\textbf{Experiment interface during the Interaction Phase.} Participants conversed with a GPT-4o chatbot (left) to identify which of six candidate images (bottom) the chatbot was looking at, then submitted a spoken description. The chatbot was covertly prompted so that its replies consistently used the AI-canonical variant for each target synonym pair (annotated on the right: e.g., \textit{vacuum flask} rather than \textit{thermos bottle}, \textit{multicolored} rather than \textit{colorful}, \textit{to repair} rather than \textit{to fix}).}
    \label{fig:si-exp-interface}
\end{figure}     

\begin{figure}[H]
    \centering
    \includegraphics[width=\linewidth]{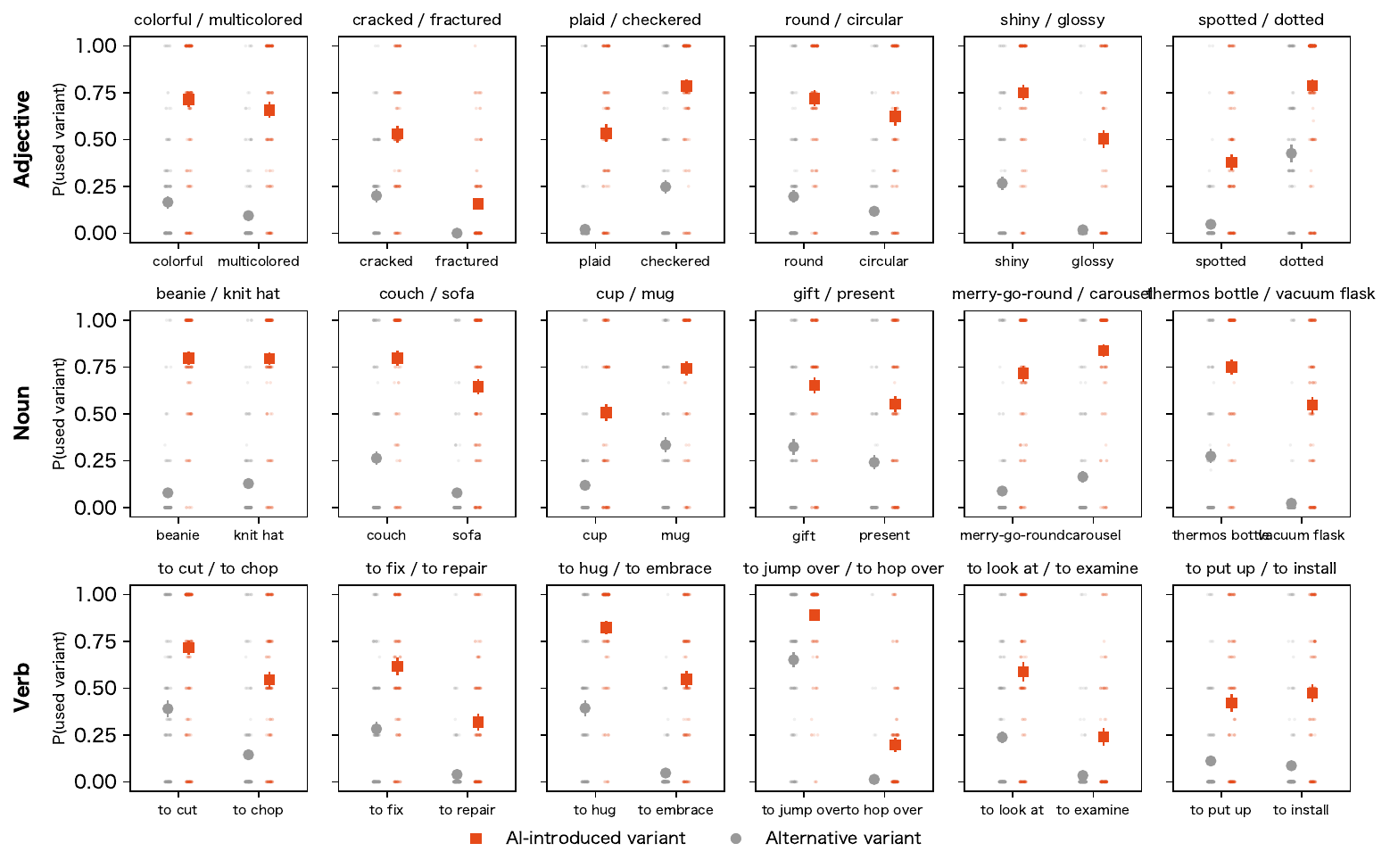}
    \caption{\textbf{Per-word-pair usage rates in the Interaction Phase.}
    Each panel shows one synonym pair (organized by lexical category in rows).
    For each variant position, the orange square is the mean usage rate for participants whose the AI chatbot was instructed to use that variant; the grey circle is the mean for participants whose the AI chatbot used the other variant.
    Small translucent dots show individual participant means; large markers show group means $\pm$ 95\% CI; dashed line at 0.5 indicates pair-internal chance level.}
    \label{fig:si-exp-during}
\end{figure}

\begin{figure}[H]
    \centering
    \includegraphics[width=\linewidth]{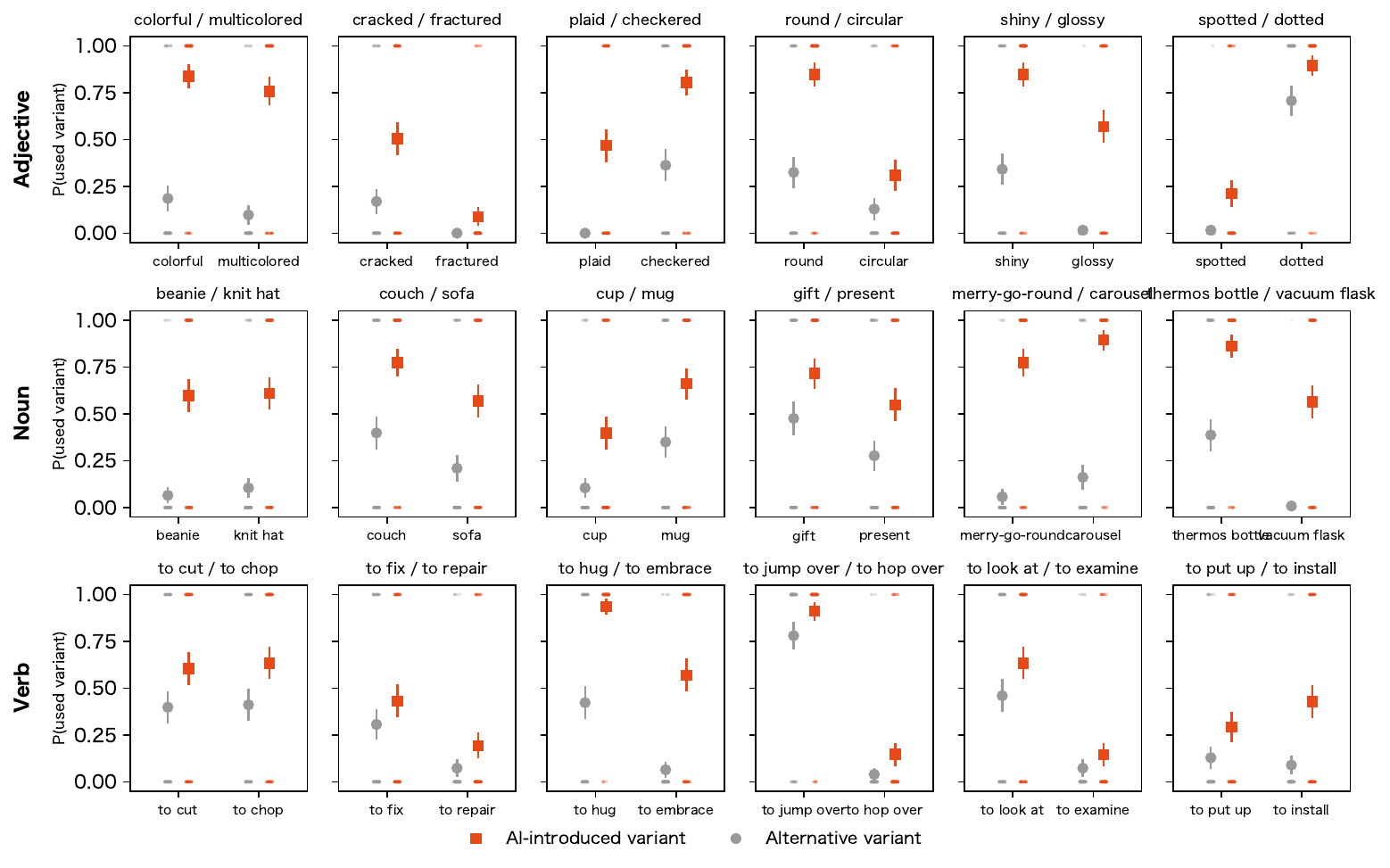}
    \caption{\textbf{Per-word-pair usage rates in the Test Phase.}
    Same layout as Supplementary \figref{fig:si-exp-during}, but for spoken descriptions of novel images not seen during the AI interaction.
    Small translucent dots show individual participant means; large markers show group means $\pm$ 95\% CI.}
    \label{fig:si-exp-after}
\end{figure}

\begin{figure}[H]
    \centering
    \includegraphics[width=\linewidth]{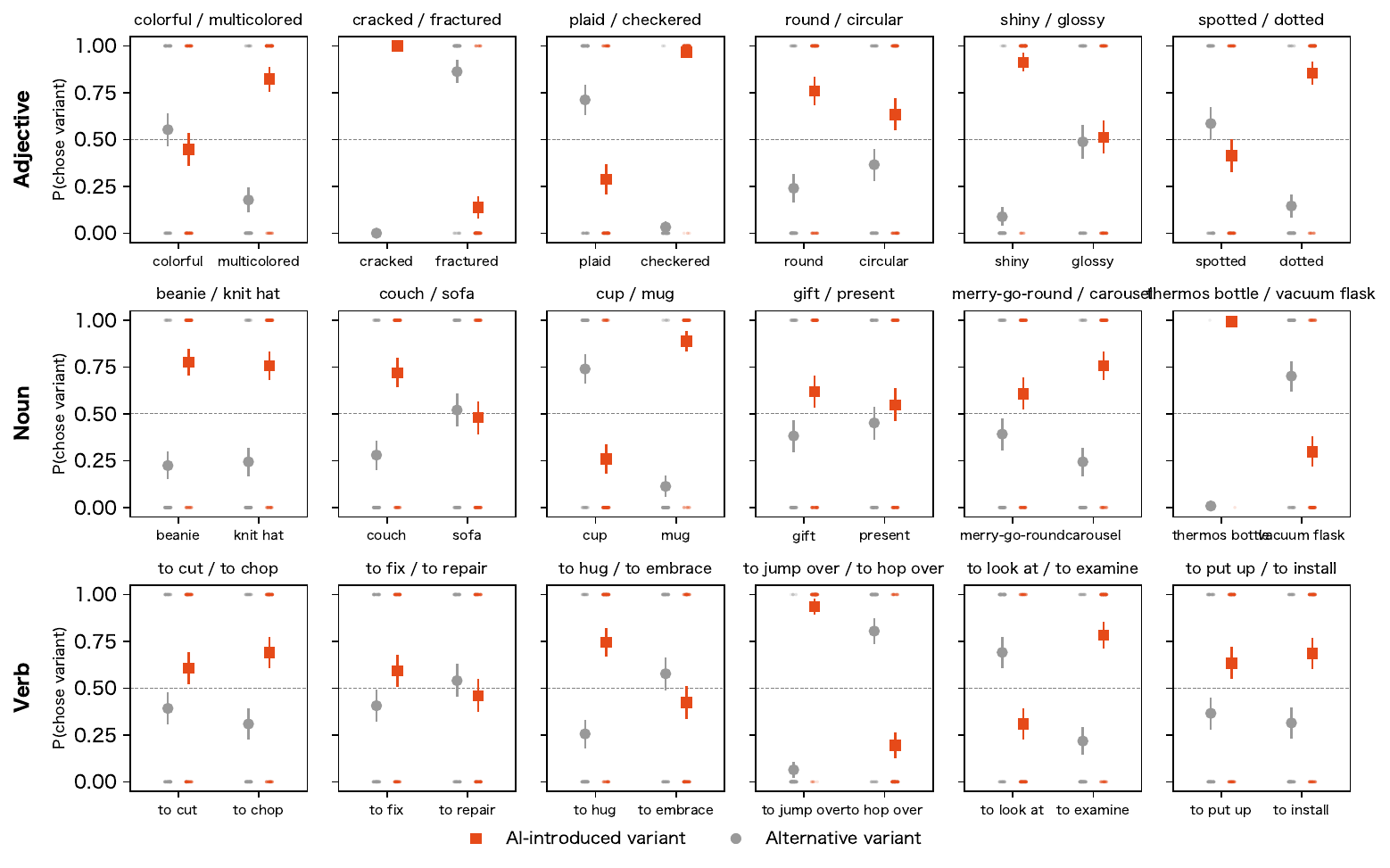}
    \caption{\textbf{Per-word-pair selection rates in the Forced-Choice Phase.}
    Each panel shows one synonym pair.
    The orange square is the mean selection rate for participants whose chatbot was primed to use that variant.
    Small translucent dots show individual participant means; large markers show group means $\pm$ 95\% CI; dashed line at 0.5 marks chance.}
    \label{fig:si-exp-forced}
\end{figure}

\begin{figure}[H]
    \centering
    \includegraphics[width=\textwidth]{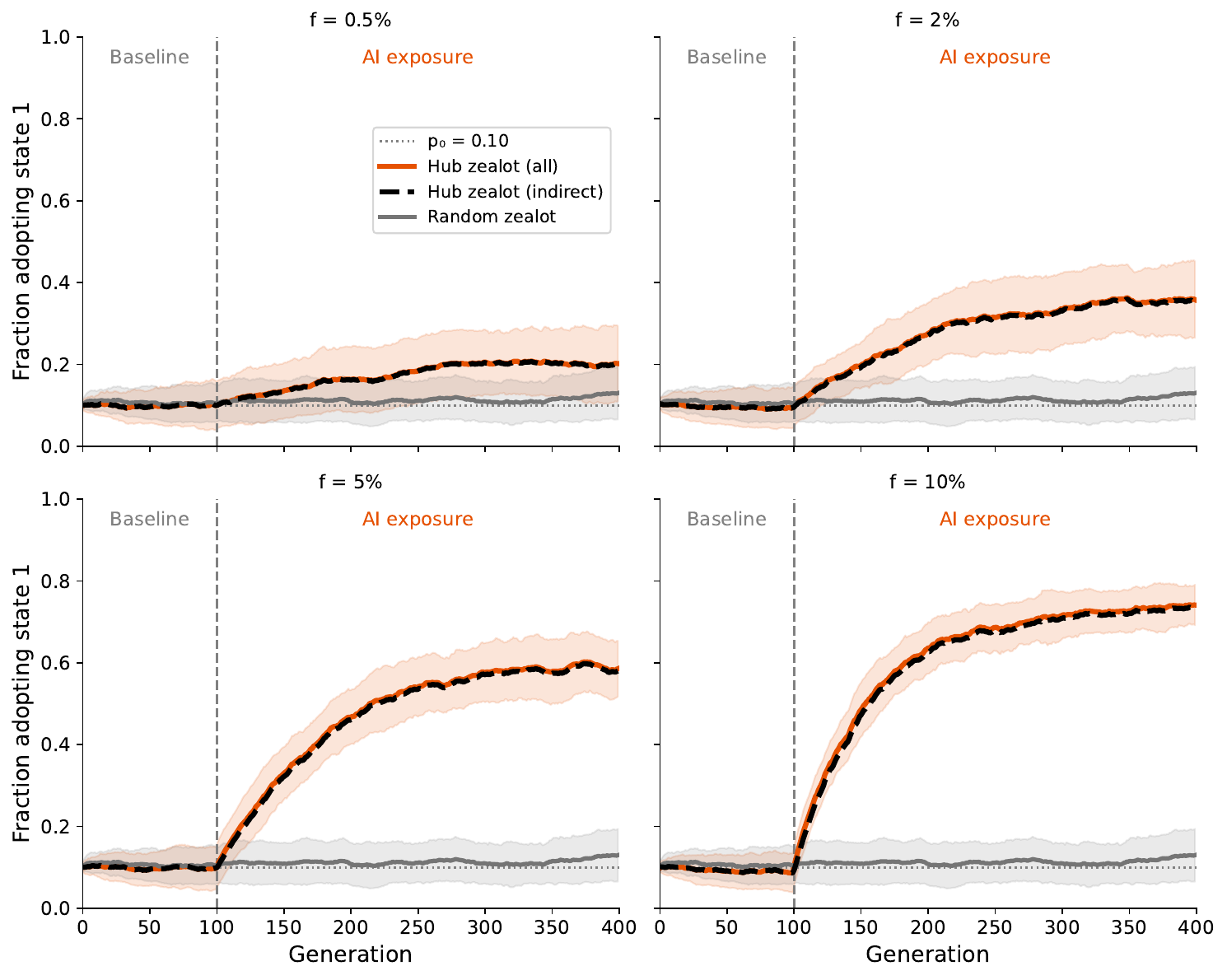}
    \caption{\textbf{Noisy voter model shows a population-level shift induced by a committed hub.} Noisy voter model with $N = 2000$ agents on a Watts--Strogatz network ($k = 6$, $\beta = 0.1$, $\varepsilon = 0.005$); the hub's exposure fraction is varied across $f \in \{0.5\%, 2\%, 5\%, 10\%\}$. Mean $\pm$ 1 SD across $50$ simulations.
    Solid orange: hub zealot, mean over all $N$ speakers; dashed black: hub zealot, mean restricted to speakers \emph{not} directly connected to the hub (i.e., reached only via network spread); solid grey: random zealot, control with a single committed speaker at a random network node.}
    \label{fig:abm}
\end{figure}

\begin{figure}[H]
    \centering
    \begin{boxedminipage}{0.9\textwidth}
        \ttfamily
        \small
        \textsf{\textbf{System prompt:}}\\[0.3em]
        You are a great research assistant who is asked to analyze YouTube data. You will be provided a list of YouTube channels as well as a target information. Please select the best channel that seems to be owned by the target. Importantly, please do not add explanations or comments other than the selected channel name. If there is no appropriate channel, please return N/A. \linebreak
        
        \textsf{\textbf{Example input prompt:}}\\[0.3em]
        \# Institution \linebreak
        
        Name: Max Planck Institute for Human Development \linebreak
        Address: Berlin, Germany \linebreak
        
        \# Candidates \linebreak
        
        Title: Max Planck Institute for Human Development \linebreak
        Description: The Max Planck Institute for Human Development (MPIB), which was founded in 1963, is dedicated to the study of human ... \linebreak
        -{}-{}-\linebreak
        Title: IMPRS LIFE \linebreak
        Description: The International Max Planck Research School on the Life Course (LIFE) is a joint international PhD Program of the Max Planck ... \linebreak
        -{}-{}-\linebreak
        Title: Behavioral Insights Bicocca \linebreak
        Description: BIB-Behavioral Insights Bicocca is a new research center focused on the behavioral analysis of public policies and public ...
    \end{boxedminipage}
    \caption{Prompt provided to \texttt{gpt-3.5-turbo-0125} to pick the most plausible channel among query results from Youtube API.}
    \label{fig:prompt-youtube}
\end{figure}

\begin{figure}[p]
    \centering
    \begin{boxedminipage}{\linewidth}
\begin{lstlisting}[style=promptbox]
You are participating in an image guessing game with a human player.

The game:
- The human player can see 6 different images, one of which is the image you're looking at
- The player will ask you questions to figure out which image you see
- Based on their questions, answer honestly and helpfully to help them identify the correct image

Your role:
- Answer questions about the image you see truthfully and accurately
- Be helpful but don't volunteer too much information at once
- Let the player guide the conversation with their questions
- Do NOT describe the entire image unprompted - wait for specific questions

Guidelines:
- Keep responses concise but informative
- Be accurate in your descriptions
- If asked about something not visible in the image, say so honestly
- Use natural language to describe positions, objects, colors, actions, etc.

When answering questions, describe what you see using your own natural vocabulary. If a question uses different words for something you observe, answer with the words that come naturally to you from your direct observation of the image. Keep your answers concise and to the point; avoid lengthy explanations.
\end{lstlisting}
    \end{boxedminipage}
    \caption{\textbf{Base system prompt given to the AI chatbot.} On each turn, the chatbot (OpenAI \texttt{gpt-4o}) received this system prompt, the target image (attached separately as a vision input, not part of the prompt), and the running chat history. The chatbot received this base prompt in every condition; it establishes the image-guessing task and instructs the model to answer honestly and use its own natural vocabulary. A vocabulary-rule layer was appended to this prompt at runtime to constrain the model to the covertly assigned synonym variant (Supplementary \figref{fig:prompt-vocab-rule}).}
    \label{fig:prompt-base}
\end{figure}

\begin{figure}[p]
    \centering
    \begin{boxedminipage}{\linewidth}
\begin{lstlisting}[style=promptbox]
Vocabulary rule — apply before every answer.

Canonical terms and scene context for this image:
@  - present
  - to repair
  - fractured@

@  [present]
    treatment: In the center, a cheerful person is handing a wrapped present with a big bow to a friend while smiling.
    control:   In the center, a cheerful person is handing a plain envelope with a note card to a friend while smiling.

  [to repair]
    treatment: On the right side, another person is using a screwdriver and wrench to repair a small table lamp on a workbench, with loose screws and tools neatly arranged.
    control:   On the right side, another person is taking a brand-new table lamp out of its packaging on a workbench, with the instruction booklet and packing material neatly arranged.

  [fractured]
    treatment: On the left side, a ceramic vase has fallen from a shelf and lies on the floor fractured into several clean pieces, with a few shards nearby; all characters are reacting to the scene.
    control:   On the left side, a ceramic vase is intact and standing safely on a shelf, with nothing broken on the floor; all characters are reacting to the scene.@

Before writing your answer, check each canonical term above:
"Does this term belong in my answer?"

It belongs if EITHER condition holds:
  (a) The question references it — via synonym, paraphrase, opposite, or the bracketed
      control alternative shown in the scene descriptions.
  (b) Your answer will describe any object, action, or location linked to it in the
      scene description — a tool used for the action, an object poured from/into, the
      agent performing the action, or the result/product of the action.

For every term that belongs, apply ALL of the following:

  1. Use the canonical term exactly — never a synonym or generic description.
     Any word that refers to the same concept must become the canonical term.
     "seating area" → "sofa" / "stepping through the water" → "wading".

  2. If the term is a verb: include it even when your answer focuses on the tool or the
     result rather than the action itself.
     • Tool without verb:   "using a chisel"            → "using a chisel to [verb]"
     • Result without verb: "chunks on the board"       → "[verb] into chunks on the board"

  3. If your answer is "No" or a denial: include the canonical term that applies anyway.
     • "No, there isn't a rocking chair"  → "No, there is a [canonical noun]"
     • "No, there are no strips"          → "No, they have been [canonical verb] into chunks"

  4. Never echo the human's synonym — their synonym is your cue to use the canonical
     term, not to mirror it back.

Keep everything else dry: no extra adjectives or descriptors beyond the canonical terms.

────────────────────────────────────────────────────
Global vocabulary — apply in ALL answers regardless of the current canonical terms.
These pairs cover all experiments. Always use the right-hand form; never use the left-hand form or its near-synonyms:
@  (thermos bottle, vacuum flask), (gift, present), (cup, mug), (to fix, to repair), (to look at, to examine), (to put up, to install), (colorful, multicolored), (cracked, fractured), (spotted, dotted), (beanie, knit hat), (couch, sofa), (merry-go-round, carousel), (to jump over, to hop over), (to hug, to embrace), (to cut, to chop), (shiny, glossy), (round, circular), (plaid, checkered)@
────────────────────────────────────────────────────
\end{lstlisting}
    \end{boxedminipage}
    \footnotesize
      \footnotesize
    \caption{\textbf{Vocabulary-rule layer appended to the chatbot's prompt in the intervention conditions, shown with one trial's runtime-injected values (in \textcolor{injected}{blue}).} This block was appended to the base game prompt (Supplementary \figref{fig:prompt-base}). For each target word, it gives a canonical form and the treatment description (what the image shows) with the control alternative (not shown); a final global layer applies all 18 synonym pairs. See Supplementary Methods for how the rule is applied; conditions were counterbalanced (Table~\ref{tab:synonym-pairs}).}

    \label{fig:prompt-vocab-rule}
\end{figure}

%% file: text/si_tables.tex
\begin{table}[h]
    \centering
    \caption{\textbf{GPT-family models used to compute the GPT score, with the exact OpenAI API snapshots queried.}}
    \label{tab:model-versions}
    \begin{tabular}{lll}
        \toprule
        Name in text & API model snapshot & Release date \\
        \midrule
        \textrm{GPT-3.5-turbo} & \texttt{gpt-3.5-turbo-1106}     & 2023-11-06 \\
        \textrm{GPT-4}         & \texttt{gpt-4-0613}             & 2023-06-13 \\
        \textrm{GPT-4-turbo}   & \texttt{gpt-4-turbo-2024-04-09} & 2024-04-09 \\
        \textrm{GPT-4o}        & \texttt{gpt-4o-2024-05-13}      & 2024-05-13 \\
        \textrm{GPT-5}         & \texttt{gpt-5-2025-08-07}       & 2025-08-07 \\
        \bottomrule
    \end{tabular}
\end{table}

\begin{table}[h]
    \centering
    \footnotesize
    \caption{\textbf{Pre-treatment synthetic-control fit (RMSPE) for the top-1\% GPT-score words (Science \& Technology).}
    For each treated word, the synthetic control minimizes pre-treatment RMSPE on the GP-smoothed monthly trajectory; we report $\mathrm{RMSPE}=\sqrt{\operatorname{mean}[(y^{\text{obs}}-y^{\text{synth}})^2]}$ over the pre-release window (months up to and including the ChatGPT release, 2022-11-30). \textbf{RMSPE (smoothed)} is computed on the GP-smoothed data (the alignment the design minimizes); \textbf{RMSPE (raw)} is on the raw monthly values (on which the MSPE ratio is evaluated). RMSPE is in $\log_{10}$-frequency units. \textbf{MSPE ratio} is post-/pre-release mean-squared prediction error (raw frame); a large value means the post-release departure dwarfs the pre-release fit error. Rows are ordered by $|$conservative $\beta_{\text{Post}}$ bound$|$ as in Fig.~3A; the first twelve are the Fig.~3A panel words.}
    \label{tab:pretreatment-rmspe}
    \setlength{\tabcolsep}{6pt}
    \begin{tabular}{lrrr}
        \toprule
        \textbf{Word} & \textbf{RMSPE (smoothed)} & \textbf{RMSPE (raw)} & \textbf{MSPE ratio} \\
        \midrule
        delve & 0.0091 & 0.054 & 4.18 \\
        showcase & 0.0131 & 0.049 & 4.03 \\
        groundbreaking & 0.0189 & 0.077 & 1.58 \\
        innovative & 0.0016 & 0.020 & 4.87 \\
        boasts & 0.0175 & 0.099 & 3.21 \\
        meticulous & 0.0186 & 0.082 & 1.44 \\
        underscored & 0.0315 & 0.049 & 3.77 \\
        intricacies & 0.0218 & 0.087 & 2.34 \\
        surpassed & 0.0103 & 0.075 & 1.30 \\
        alongside & 0.0072 & 0.031 & 4.67 \\
        hinges & 0.0049 & 0.073 & 1.58 \\
        crucial & 0.0127 & 0.039 & 1.98 \\
        \midrule
        \multicolumn{4}{c}{\footnotesize\itshape remaining top-1\% words} \\
        \midrule
        creation & 0.0056 & 0.033 & 1.87 \\
        bolster & 0.0329 & 0.101 & 1.06 \\
        notice & 0.0016 & 0.009 & 1.06 \\
        intricate & 0.0207 & 0.079 & 1.42 \\
        craft & 0.0080 & 0.034 & 1.22 \\
        pinpoint & 0.0043 & 0.063 & 0.75 \\
        comprehend & 0.0236 & 0.082 & 0.89 \\
        swift & 0.0123 & 0.100 & 1.60 \\
        inquiry & 0.0197 & 0.071 & 0.85 \\
        lessen & 0.0079 & 0.094 & 1.19 \\
        groundwork & 0.0103 & 0.076 & 1.37 \\
        heightened & 0.0226 & 0.074 & 1.31 \\
        escalating & 0.0084 & 0.059 & 1.27 \\
        discerning & 0.0106 & 0.059 & 1.68 \\
        inaugural & 0.0667 & 0.131 & 0.68 \\
        affirmed & 0.0275 & 0.093 & 1.15 \\
        notably & 0.0092 & 0.052 & 1.79 \\
        portrayed & 0.0096 & 0.043 & 1.67 \\
        catering & 0.0120 & 0.052 & 2.05 \\
        reliant & 0.0148 & 0.067 & 0.49 \\
        impressive & 0.0057 & 0.016 & 1.41 \\
        thorough & 0.0110 & 0.060 & 0.91 \\
        akin & 0.0147 & 0.076 & 1.55 \\
        spotlight & 0.0136 & 0.056 & 0.78 \\
        \bottomrule
    \end{tabular}

    \smallskip
    \parbox{0.95\linewidth}{\footnotesize Full S\&T panel (all $n=3535$ treated words): smoothed RMSPE median $0.0076$ (IQR $0.0039$--$0.0134$, max $0.673$); raw RMSPE median $0.0382$ (IQR $0.0199$--$0.0649$, max $0.689$).}
\end{table}

\begin{table}[h]
    \centering
    \small
    \caption{\textbf{Top synthetic-control donor words for \emph{delve} (Science \& Technology, Main synthetic-control specification).}
    The synthetic control for \emph{delve} is the convex combination of donor words whose GP-smoothed pre-treatment trajectory best matches \emph{delve}'s (Main specification: $w2v$-then-$\ell_2$ donor selection, semantic exclusion $k=20$, 50\% neutral GPT-score band, pool $n=100$, SLSQP simplex weights on the Mat\'ern-smoothed series, $\ell=720$ d; see Table~\ref{tab:control-specs}). Non-negative, sum-to-one weights induce sparsity: of the 100-word donor pool, only 11 words receive a weight above $10^{-4}$. The cumulative row shows the running share of total weight; the top five donors account for 82.4\% and the eight shown for 95.6\%. Weights are the fitted SLSQP simplex coefficients (dimensionless, summing to one across the full pool).}
    \label{tab:delve-donors}
    \setlength{\tabcolsep}{6pt}
    \begin{tabular}{lrrrrrrrr}
        \toprule
        \textbf{Donor}      & arc   & dose  & convey & trauma & Mars  & anchor & Marine & prominent \\
        \textbf{Weight}     & 0.236 & 0.198 & 0.161  & 0.121  & 0.107 & 0.057  & 0.039  & 0.036 \\
        \textbf{Cumulative} & 23.6\% & 43.4\% & 59.6\% & 71.6\% & 82.4\% & 88.1\% & 91.9\% & 95.6\% \\
        \bottomrule
    \end{tabular}

    \smallskip
    \parbox{0.95\linewidth}{\footnotesize Remaining donors with non-trivial weight: cloning (0.032), representation (0.010), grid (0.003). Support on 9 donors at the $w > 0.01$ threshold, 11 at $w > 10^{-4}$.}
\end{table}

\begin{table}[h]
  \centering
  \caption{\textbf{Window-mean synthetic-control gap for \emph{delve} across podcast categories, with placebo-based 95\% confidence intervals.} The point estimate $\hat{g}$ is the mean elevation of observed-over-synthetic relative frequency of \emph{delve} across the indicated window. Confidence intervals and $p$-values are obtained by inverting the in-space placebo distribution of word-level window-mean gaps ($n_{\text{placebo}} = 100$ per group; see Methods). The post-adoption $p$-value is one-sided; the recent-6 $p$-value is two-sided.}
  \label{tab:fig2-b2-ci}
  \small
  \begin{tabular}{lccc@{\hskip 2em}ccc}
    \toprule
    & \multicolumn{3}{c}{\textbf{Post-adoption (months 13--18)}} & \multicolumn{3}{c}{\textbf{Recent-6 (Nov 2025--Apr 2026)}} \\
    \cmidrule(lr){2-4} \cmidrule(lr){5-7}
    Group & $\hat{g}$ & 95\% CI & $p$ & $\hat{g}$ & 95\% CI & $p$ \\
    \midrule
    Science \& Technology & $+44\%$ & $[+22\%, +63\%]$ & $0.010$ & $-15\%$ & $[-35\%, +8\%]$  & $0.248$ \\
    Education             & $+32\%$ & $[-5\%, +75\%]$  & $0.059$ & $-11\%$ & $[-40\%, +31\%]$ & $0.485$ \\
    Business              & $+31\%$ & $[+0\%, +67\%]$  & $0.040$ & $-30\%$ & $[-50\%, +3\%]$  & $0.069$ \\
    All                   & $+9\%$  & $[-15\%, +32\%]$ & $0.218$ & $-35\%$ & $[-57\%, -7\%]$  & $0.050$ \\
    Sports                & $-7\%$  & $[-40\%, +41\%]$ & $0.663$ & $-38\%$ & $[-68\%, +13\%]$ & $0.208$ \\
    \bottomrule
  \end{tabular}
\end{table}

\begin{table}[h]
    \centering
    \footnotesize
    \caption{\textbf{Design features of the Main spec and four robustness controls.} All five specs share the same GP smoother (Mat\'ern $\nu=2.5$, $\ell=720$ d, noise $0.05$), baseline window 2016-11-30 to 2022-11-30, and treatment window 2022-11-30 to 2024-05-30. C2's $p = 0.091$ is the floor of the empirical placebo distribution: the in-space placebo procedure draws targets from C2's ten-word donor pool, so the empirical $p$ cannot resolve below $1/11 \approx 0.091$. \emph{delve} achieves this floor as the largest MSPE ratio in the eleven-element distribution. C3 substitutes Main's SLSQP convex fit with deterministic inverse-distance similarity weights $w_i = (1/(d_i + \varepsilon)) / \sum_k (1/(d_k + \varepsilon))$ ($\varepsilon = 10^{-12}$). For context, the YouTube replication (same Main spec on the YT corpus) gives $p = 0.010$.}
    \label{tab:control-specs}
    \setlength{\tabcolsep}{3pt}
    \begin{tabular}{lccccc}
        \toprule
        \textbf{Feature}            & \textbf{Main}              & \textbf{C1}                & \textbf{C2}                & \textbf{C3}                & \textbf{C4}                \\
        \midrule
        Counts source               & audited                    & audited                    & audited                    & audited                    & un-audited                 \\
        Input series                & GP-smoothed                & GP-smoothed                & GP-smoothed                & GP-smoothed                & GP-smoothed                \\
        Length scale $\ell$         & 720 d                      & 720 d                      & 720 d                      & 720 d                      & 720 d                      \\
        Donor strategy              & $w2v$-then-$\ell_2$        & bare $\ell_2$              & $w2v$-then-$\ell_2$        & $w2v$-then-$\ell_2$        & $w2v$-then-$\ell_2$        \\
        Semantic exclusion $k$      & 20                         & 0                          & 20                         & 20                         & 20                         \\
        Neutral percentile band     & 50\%                       & --                         & 50\%                       & 50\%                       & 50\%                       \\
        Donor pool $n$              & 100                        & 100                        & 10                         & 100                        & 100                        \\
        Donor weights               & SLSQP simplex              & SLSQP simplex              & SLSQP simplex              & inverse distance           & SLSQP simplex              \\
        \emph{delve} placebo $p$    & 0.010                      & 0.050                      & 0.091 (floor)              & 0.040                      & 0.010                      \\
        \bottomrule
    \end{tabular}
\end{table}

\begin{table}[h]
    \centering
    \footnotesize
    \caption{Demographic characteristics of the final sample ($N = 496$).
    Age brackets were used for data collection; the midpoint-based estimate is mean~$= 40.5$ years ($SD = 13.2$, $n = 495$; one participant preferred not to say).
    Chatbot frequency is reported among the 475 participants who reported having used an AI chatbot.}
    \label{tab:demographics}
    \begin{tabular}{lr}
        \toprule
        \textbf{Characteristic} & \textbf{$N$ (\%)} \\
        \midrule
        \textit{Age} & \\
        \quad 18--24 & 49 (9.9\%) \\
        \quad 25--34 & 144 (29.0\%) \\
        \quad 35--44 & 127 (25.6\%) \\
        \quad 45--54 & 89 (17.9\%) \\
        \quad 55--64 & 62 (12.5\%) \\
        \quad 65+     & 24 (4.8\%) \\
        \quad Prefer not to say & 1 (0.2\%) \\
        \midrule
        \textit{Gender} & \\
        \quad Female        & 247 (49.8\%) \\
        \quad Male          & 239 (48.2\%) \\
        \quad Non-binary    & 6 (1.2\%) \\
        \quad Prefer not to say & 4 (0.8\%) \\
        \midrule
        \textit{Education} & \\
        \quad Less than high school                & 3 (0.6\%) \\
        \quad High school diploma or equivalent   & 70 (14.1\%) \\
        \quad Some college, no degree             & 109 (22.0\%) \\
        \quad Associate degree                    & 48 (9.7\%) \\
        \quad Bachelor's degree                   & 180 (36.3\%) \\
        \quad Master's degree                     & 67 (13.5\%) \\
        \quad Doctoral or professional degree     & 19 (3.8\%) \\
        \midrule
        \textit{Language} & \\
        \quad English as first language & 453 (91.3\%) \\
        \quad English as additional language & 43 (8.7\%) \\
        \midrule
        \textit{Ethnicity} & \\
        \quad Caucasian/White                       & 292 (58.9\%) \\
        \quad Black/African                         & 102 (20.6\%) \\
        \quad Hispanic/Latinx                       & 42 (8.5\%) \\
        \quad Asian                                 & 41 (8.3\%) \\
        \quad Multiethnic                           & 9 (1.8\%) \\
        \quad Middle Eastern or Northern African    & 6 (1.2\%) \\
        \quad Other / prefer not to disclose        & 4 (0.8\%) \\
        \midrule
        \textit{Prior AI chatbot use} & \\
        \quad Used an AI chatbot          & 475 (95.8\%) \\
        \quad Not used / not sure         & 21 (4.2\%) \\
        \midrule
        \textit{Chatbot use frequency (among users, $n = 475$)} & \\                                                                                                                                                                                  
        \quad More than five times a day  & 81 (17.1\%) \\                                                                                                                                                                                            
        \quad More than once a day        & 171 (36.0\%) \\                                                                                                                                                                                           
        \quad More than once a week       & 152 (32.0\%) \\                                                                                                                                                                                           
        \quad More than once a month      & 43 (9.1\%) \\                                                                                                                                                                                             
        \quad Not more than once a month  & 28 (5.9\%) \\ 
        \bottomrule
    \end{tabular}
\end{table}

\begin{table*}[h]
    \centering
    \footnotesize
    \caption{All 18 synonym pairs used in the experiment, organized by group and lexical category.
    ``Variant 1'' and ``Variant 2'' are the two synonyms; the AI was primed to use one variant depending on the condition.}
    \label{tab:synonym-pairs}
    \begin{tabular}{llll}
        \toprule
        \textbf{Group} & \textbf{Category} & \textbf{Variant 1} & \textbf{Variant 2} \\
        \midrule
        A & Noun      & thermos bottle  & vacuum flask  \\
        A & Noun      & gift            & present       \\
        A & Noun      & cup             & mug           \\
        A & Verb      & to fix          & to repair     \\
        A & Verb      & to look at      & to examine    \\
        A & Verb      & to put up       & to install    \\
        A & Adjective & colorful        & multicolored  \\
        A & Adjective & cracked         & fractured     \\
        A & Adjective & spotted         & dotted        \\
        \midrule
        B & Noun      & beanie          & knit hat      \\
        B & Noun      & couch           & sofa          \\
        B & Noun      & merry-go-round  & carousel      \\
        B & Verb      & to jump over    & to hop over   \\
        B & Verb      & to hug          & to embrace    \\
        B & Verb      & to cut          & to chop       \\
        B & Adjective & shiny           & glossy        \\
        B & Adjective & round           & circular      \\
        B & Adjective & plaid           & checkered     \\
        \bottomrule
    \end{tabular}
\end{table*}